



%
%
%
%
%
%
%

\input phyzzx
\message{For printing the figures, you need to have the
files epsf.tex, 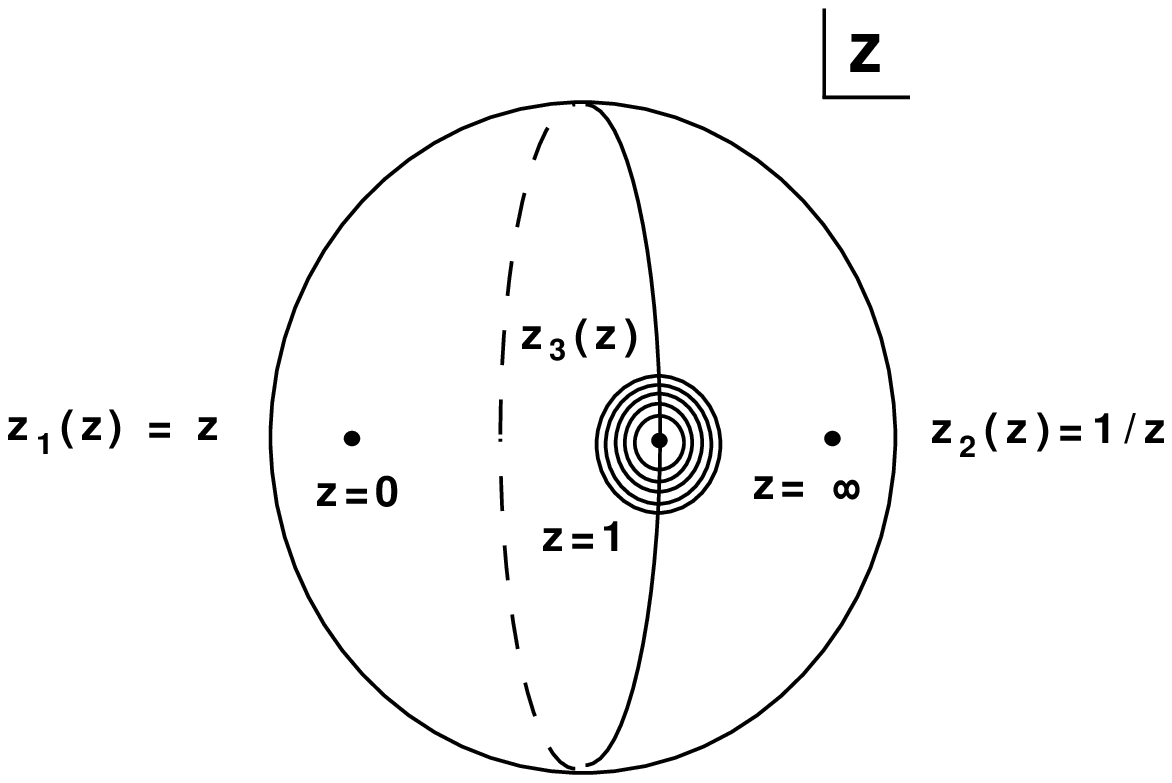, 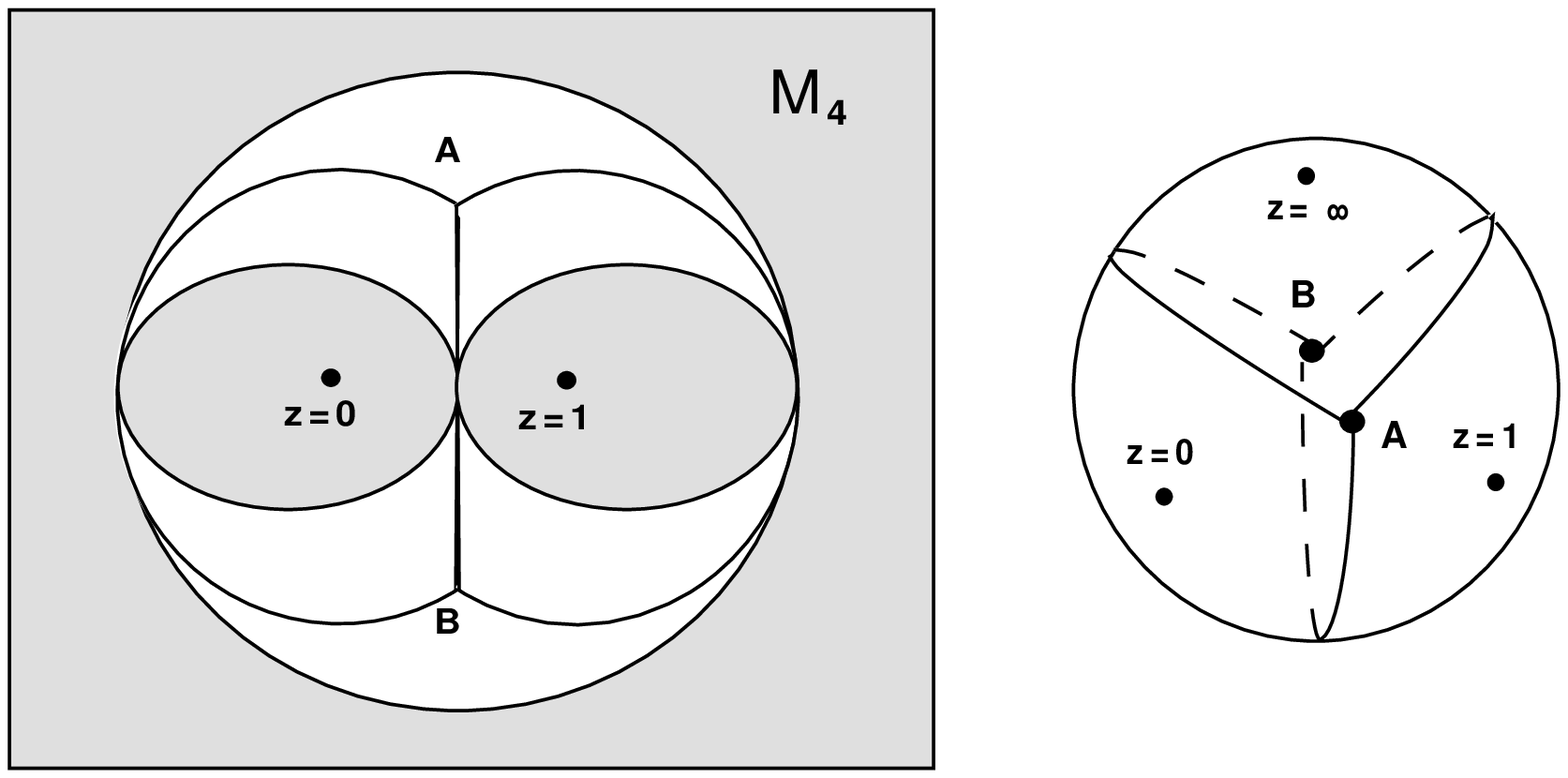 and szfig3.ps
in your area. Do you have these files? Enter y/n :    }
\read -1 to \figcount
\if y\figcount \message{This will come out with the figures}
\else \message{This will come out without the figures}\fi

\if y\figcount \input epsf \else\fi
\overfullrule 0pt
\catcode`\@=11 
\def\NEWrefmark#1{\step@ver{{\;#1}}}
\catcode`\@=12 
%

\def\square{\kern1pt\vbox{\hrule height 1.2pt\hbox{\vrule width 1.2pt\hskip 3pt
   \vbox{\vskip 6pt}\hskip 3pt\vrule width 0.6pt}\hrule height 0.6pt}\kern1pt}

\def\bra#1{\langle #1 |}
\def\ket#1{| #1 \rangle}

\def\vx{{x}}
\def\vy{{y}}
\def\ov{{\overline}}
\def\A{{\cal A}}
\def\B{{\cal B}}
\def\C{{\cal C}}
\def\D{{\cal D}}
\def\H{\widehat{\cal H}}
\def\HH{{\cal H}}
\def\F{{\cal F}}
\def\I{{\cal I}}

\def\W{{\cal W}}

\def\M{{\cal M}}

\def\O{{\cal O}}
\def\P{{\cal P}}

\def\s{{\cal S}}

\def\V{{\cal V}}
\def\PP{P}

\def\p{\partial}

\def\ov{\overline}

\def\wt{\widetilde}
\def\wh{\widehat}
\def\ss{\wt\s}
\overfullrule=0pt
\baselineskip 13pt plus 1pt minus 1pt
\nopubblock
{}~ \hfill \vbox{\hbox{MIT-CTP-2222}\hbox{TIFR-TH-93-30}
\hbox{hep-th/9307088} \hbox{July 1993}}\break
\titlepage
\title{A PROOF OF LOCAL BACKGROUND INDEPENDENCE}
\titlestyle{OF CLASSICAL CLOSED STRING FIELD THEORY}
\author{Ashoke Sen \foot{E-mail  address: sen@tifrvax.tifr.res.in,
sen@tifrvax.bitnet } }
\address{Tata Institute of Fundamental Research \break
Homi Bhabha Road, Bombay 400005, India}
\andauthor
{Barton Zwiebach \foot{E-mail address: zwiebach@irene.mit.edu,
zwiebach@mitlns.bitnet.\hfill\break Supported in part by D.O.E.
contract DE-AC02-76ER03069.}}
\address{Center for Theoretical Physics \break
LNS and Department of Physics\break
MIT, Cambridge, MA 02139, U.S.A.}

\abstract{ We give a complete proof of local background
independence of the classical master action for closed
strings by constructing explicitly, for any two nearby
conformal theories in a CFT theory space, a symplectic
diffeomorphism between their state spaces mapping the
corresponding non-polynomial string actions into each other.
We uncover a new family of string vertices, the lowest of
which is a three string vertex satisfying exact
Jacobi identities with respect to the original closed string
vertices. The homotopies between the two sets of string
vertices determine the diffeomorphism establishing background
independence. The linear part of the diffeomorphism is
implemented by a CFT theory-space connection determined
by the off-shell three closed string vertex, showing how string
field theory induces a natural interplay between Riemann surface
geometry and CFT theory
space geometry.}

\endpage

\def\mapdown#1{\Big\downarrow
   \rlap{$\vcenter{\hbox{$\scriptstyle#1$}}$}}
\def\mapup#1{\Big\uparrow
   \rlap{$\vcenter{\hbox{$\scriptstyle#1$}}$}}
\def\define#1#2\par{\def#1{\Ref#1{#2}\edef#1{\noexpand\refmark{#1}}}}
\def\con#1#2\noc{\let\?=\Ref\let\<=\refmark\let\Ref=\REFS
         \let\refmark=\undefined#1\let\Ref=\REFSCON#2
         \let\Ref=\?\let\refmark=\<\refsend}

\let\refmark=\NEWrefmark

\define\zwiebachlong{B. Zwiebach, `Closed string field fheory: Quantum
action and the Batalin-Vilkovisky master equation', Nucl. Phys {\bf B390}
(1993) 33, hep-th/9206084.}

\define\saadizwiebach{M. Saadi and B. Zwiebach, `Closed string
field theory from polyhedra', Ann. Phys. {\bf 192} (1989) 213.}

\define\kugokunitomo{T. Kugo,
H. Kunitomo and K. Suehiro, `Non-polynomial closed string
field theory', Phys. Lett. {\bf 226B} (1989) 48.}

\define\kugosuehiro{T. Kugo and K. Suehiro,  `Nonpolynomial closed
string field theory: action and gauge invariance', Nucl.
Phys. {\bf B337} (1990) 434.}

\define\kaku{M. Kaku, `Geometrical derivarion of string field theory from
first principles: closed strings and modular invariance. Phys. Rev.
{\bf D38} (1988) 3052;\hfill\break
M. Kaku and J. Lykken, `Modular Invariant closed string field theory',
Phys. Rev. {\bf D38} (1988) 3067.}

\define\zwiebachma{B. Zwiebach,  `How covariant closed string
theory solves a minimal area problem',
Comm. Math. Phys. {\bf 136} (1991) 83 ; `Consistency of closed string
polyhedra from minimal area', Phys. Lett. {\bf B241} (1990) 343.}

\define\sonodazwiebach{H. Sonoda and B. Zwiebach, `Closed string field theory
loops with symmetric factorizable quadratic differentials',
Nucl. Phys. {\bf B331} (1990) 592.}

\define\zwiebachqcs{B. Zwiebach, `Quantum closed strings from minimal
area', Mod. Phys. Lett. {\bf A5} (1990) 2753.}

\define\zwiebachtalk{B. Zwiebach, `Recursion Relations in Closed String
Field Theory', Proceedings of the ``Strings 90'' Superstring Workshop.
Eds. R. Arnowitt, et.al. (World Scientific, 1991) pp 266-275.}

\define\ranganathan{K. Ranganathan, `A criterion for flatness in minimal
area metrics that define string diagrams', Commun. Math. Phys.
{\bf 146} (1992) 429.}

\define\wolfzwiebach{M. Wolf and B. Zwiebach, `The plumbing of minimal
area surfaces', IASSNS-92/11, submitted to Jour. Geom. Phys. hep-th/9202062.}

\define\senone{A. Sen, `On the background independence of string
field theory', Nucl. Phys. {\bf B345} (1990) 551}

\define\sentwo{A. Sen,`On the background independence of
string field theory (II). Analysis of on-shell $S$-matrix elements',
Nucl. Phys. {\bf B347} (1990) 270}

\define\senthree{A. Sen, `On the background independence of string field
theory (III). Explicit field redefinitions',
Nucl. Phys. {\bf B391} (1993) 550, hep-th/9201041}

\define\senold{A. Sen, `Equations of motion in non-polynomial closed string
field theory and conformal invariance of two dimensional field theories',
Phys. Lett. {\bf B241} (1990) 350}

\define\mukherjisen{S. Mukherji and A. Sen, `Some all order classical
solutions in non-polynomial closed string field theory',
Nucl. Phys. {\bf B363} (1991) 639.}

\define\cubic{G.T. Horowitz, J. Lykken, R. Rohm and A. Strominger,
`A purely cubic action for string field theory',
Phys.
Rev. Lett. {\bf 57} (1986) 283.}

\define\hatazwiebach{H. Hata and B. Zwiebach, `Developing the covariant
Batalin-Vilkovisky approach to string theory', MIT-CTP-2184, to appear in
Annals of Physics. hep-th/9301097.}

\define\evansovrut{M. Evans and B. Ovrut, `Deformations of conformal
field theories and symmetries of the string', Phys. Rev.
{\bf D41} (1990) 3149.}

\define\sarojasen{R. Saroja and A. Sen, `Picture changing operators in closed
fermionic string field theory',
Phys. Lett. {\bf B286} (1992) 256, hep-th/9202087.}

\define\witten{E. Witten, `On background independent open-string field theory',
Phys. Rev. {\bf D46} (1992) 5467,
hep-th/9208027\hfill\break `Some computations in background independent
off-shell string field fheory', Phys. Rev. {\bf D47} (1993) 3405,
hep-th/9210065.}

\define\wittenqbi{E. Witten, `Quantum background independence in
string theory',
IASSNS-HEP-93/29, June 1993, hep-th/ 9306122.}

\define\liwitten{K. Li and E. Witten. `Role of short distance behavior in
off-shell open string field theory', IASSNS-HEP-93/7, hep-th/9303067;
\hfill\break
S. Shatashvili,`Comment on the background independent
open string theory', IASSNS-HEP-93/15, hep-th/9303143.}

\define\nelcam{M. Campbell, P. Nelson and E. Wong, `Stress tensor
perturbations in conformal field theory'
Int. Jour. Mod. Phys
{\bf A6} (1991) 4909}

\define\rangaconnection{K. Ranganathan, `Nearby CFT's in the operator
formalism: The role of a connection', to appear in Nucl. Phys. B.
hep-th/9210090.}

\define\kugozwiebach{T. Kugo and B. Zwiebach, `Target space duality
as a symmetry of string field theory',  Prog. Theor. Phys. {\bf 87}
(1992) 801, hep-th/9201040.}

\define\rangasonodazw{K. Ranganathan, H. Sonoda and B. Zwiebach,
`Connections on the state-space over conformal field theories',
MIT-preprint MIT-CTP-2193, April 1993, hep-th/9304053.}

\define\sonoda{H. Sonoda, Composite operators in QCD,
Nucl. Phys. {\bf B383} (1992) 173, hep-th/9205085;\hfill\break
``Operator Coefficients for Composite Operators in the $(\phi^4)_4$
Theory'', Nucl. Phys. {\bf B394} (1993) 302, hep-th/9205084. }

\define\nelson{P. Nelson, `Covariant insertion of general vertex operators',
Phys. Rev. Lett. {\bf 62} (1989) 993;\hfill\break
H. S. La and P. Nelson, `Effective field equations for fermionic strings',
Nucl. Phys. {\bf B332} (1990) 83;\hfill\break
J. Distler and P. Nelson, `Topological couplings and contact terms in
2-D field theory',
Comm. Math. Phys. {\bf 138} (1991) 273.}

\define\schwarz{A. Schwarz, `Geometry of Batalin-Vilkovisky quantization', UC
Davis preprint, hep-th/9205088, July 1992. }

\define\alvarez{L. Alvarez-Gaume, C. Gomez, G. Moore and C. Vafa,
`Strings in the operator formalism', Nucl.
Phys. {\bf B303} (1988) 455;\hfill\break
C. Vafa, `Operator formulation on Riemann surfaces',
Phys. Lett. {\bf B190} (1987) 47.}

\define\moore{G. Moore, `Finite in all directions', Yale University
preprint, YCTP-P12-93, hep-th/9305139.}

\define\stasheff{J. Stasheff, `Homotopy associativity of H-spaces, II.',
Trans. Amer. Math. Soc., {\bf 108} (1963) 293; `H-Spaces from a homotopy
point of view', Lecture Notes in Mathematics
{\bf 161}, Springer Verlag, 1970;\hfill\break
See also: T. Lada and J. Stasheff, `Introduction to sh Lie algebras for
physicists', hep-th/9209099.}

\define\sonodazw{H. Sonoda and B. Zwiebach, `Covariant closed string
theory cannot be cubic', Nucl. Phys. {\bf B336} (1990) 185.}

\define\wittenosft{E. Witten, `Noncommutative geometry and string field
theory', Nucl. Phys. {\bf B268} (1986) 253.}

\chapter{Introduction and Summary}

One of the most important open questions in string field theory is that
of finding a manifestly background independent formulation.
A consistent quantum closed string
field theory already exists
[\con\saadizwiebach\kugokunitomo\kugosuehiro\kaku\zwiebachma
\sonodazwiebach\zwiebachqcs\zwiebachtalk\ranganathan\wolfzwiebach
\zwiebachlong\noc].
It is written using the
Batalin-Vilkovisky
(BV) formalism which turned out to be remarkably efficient for string field
theory. This string field theory, however, requires for its formulation
a choice of a conformal field theory defining
a consistent background for string propagation. Such choice would not
be necessary in a manifestly background independent
formulation, where consistent backgrounds would
arise as classical solutions.
Since closed string field theory is not {\it manifestly} background
independent,
the obvious question is whether it is background independent {\it at all}.
In this paper we prove that closed string field theory is indeed
independent of the background in which it is formulated, as long as the
backgrounds are related by marginal deformations. Since our proof is
geometrical, we believe that it may
provide crucial insight for the construction of a manifestly
background independent closed string field theory.

A string field theory, in the BV formulation,
is defined by a master action $S$, which is a function on a subspace
$\H$ of the state space of the chosen CFT, and a symplectic structure
$\,\omega\,$, or BV antibracket, on $\H$.
The string field is just an arbitrary element of $\H$.
In writing down closed string field theory one has to make two types
of choices. The first one, as mentioned above, consists of choosing
a conformal theory from the space of two dimensional theories.
The second one, apparently on a totally different footing, is a choice
of string vertices for the string field action. This choice of vertices
determines how the Feynman diagrams of the resulting string field theory
decompose the moduli spaces of Riemann surfaces.
A canonical choice of string vertices arises from minimal area metrics,
but other choices are possible. The choice of an $n$-string vertex
for classical closed string theory, is the choice of a collection
of $n$-punctured spheres, each having specific choices
of local coordinates (defined up to phases) around each puncture.
This amounts to choosing a subspace of $\wh\P_n$, the space of all
inequivalent $n$-punctured spheres with all possible choices of local
coordinates on the punctures. Therefore, the choice of vertices is a
choice of subspaces from spaces of decorated Riemann surfaces.
It was shown recently [\hatazwiebach],
that string field theories corresponding to different
choices of string vertices, are, in fact, related by
field transformations canonical with respect to the BV antibracket.
This shows that these different string field theories
represent the same theory written in terms of different variables.

Given that we only know how to formulate closed string field theory
once we
choose a conformal field theory, the problem of background independence
is formulated as follows. Let $x$ and $y$ denote two different
conformal theories, and let $\H_x$ and $\H_y$ be their respective
state spaces. Let $(S_x ,\omega_x)$ and $(S_y ,\omega_y)$ be their
respective master actions and BV structures. Background independence
would mean that there is a string field transformation that establishes
the physical equivalence of the two theories. More precisely, we have
to find a diffeomorphism relating $\H_x$ to $\H_y$, such that under its
action the respective master actions and BV structures are taken into
each other. The main purpose of the present paper is to construct this
diffeomorphism explicitly for the case when we have nearby conformal
field theories related by an exactly marginal operator. We call this
the problem of {\it local} background independence. The natural
setting for this problem is therefore that of a space of
conformal field theories. The state spaces of the conformal field theories
then form a vector bundle over the space of conformal theories.
We will see in this paper how the choice of string vertices, necessary
for writing a string field theory, provides local geometrical structure
on this vector bundle. Thus string field theory is seen to induce
a natural interplay between Riemann surface theory and theory space geometry.
The geometrical structure induced on CFT theory-space is essential to
our proof of background independence.

Since the BV master action is not the gauge invariant classical
action nor the gauge fixed action (even though both
arise from the master action by simple operations), physical
background independence may not require background independence
of the master action. Our success in proving that the master
action is background independent provides further evidence of
the deep significance of the BV formulation of string theory.

The problem of local background independence was addressed earlier in
refs.[\con\senone\sentwo\senthree\noc] where
it was shown that, up to cubic
order in the string field, the classical actions of two string field
theories, formulated around two nearby conformal field theories, can be
related by a field redefinition. Due to various technical complications,
the result could not be extended to higher orders in the string field.
Moreover, there was no natural geometric construction of the field
redefinition that takes one string field theory action to the other.

Largely stimulated by this work, much progress has been made in
understanding deformations of conformal field theories
[\con\evansovrut\nelcam\kugozwiebach\rangaconnection\rangasonodazw\moore\noc].
It was understood
that having a space of conformal theories implies the existence of
connections on the vector bundle of state spaces over this theory space. A
connection is necessary to formulate precisely (and covariantly!) the
intuitive idea that correlation functions vary smoothly as we move in
theory space. A connection is also necessary to construct a conformal
theory using the state space of another conformal theory. There is no
unique connection on this vector bundle, and specific choices must be made
for specific purposes.  In [\rangasonodazw ]  a unified description of all
possible connections was given by generalizing the variational formula of
Sonoda [\sonoda]. A particularly useful connection $\wh\Gamma_\mu$, was seen
to satisfy the following variational formula:
$$D_\mu (\,\widehat\Gamma\,)
\bra{\,\Sigma\, } = -{1\over\pi} \hskip-6pt
\int_{\Sigma -
\cup_i D_i^{(1)}}\hskip-6pt d^2 z~\bra{\,\Sigma;z\,}\O_\mu\rangle.\eqn\xen$$
In here, $\bra{\, \Sigma\,}$ are the states of the operator formalism
encoding all the correlators of the punctured Riemann surface $\Sigma$.
This bra is a section on the vector bundle. In the right hand side we have
the integral, over the surface minus the {\it unit} disks around each puncture,
of the insertion of the exactly marginal operator $\O_\mu$ (in the operator
formalism this insertion is done by contracting a section $\bra{\Sigma;z}$,
corresponding to
the surface $\Sigma$ with an extra puncture at $z$, with the state
$\ket{\O_\mu}$).  Our whole input from the fact that we have a theory
space will be that a connection $\wh\Gamma_\mu$ satisfying the above
formula must exist.  String field theory contact interactions, such as
those defining the classical closed string field theory vertices $\V_n$, are
specified by punctured spheres $\Sigma$ whose unit disks around the
punctures cover fully and precisely the surfaces. It follows from the
above formula that, for such surfaces, there is no region to integrate over,
and therefore, the covariant derivative $D_\mu (\,\wh\Gamma \,)$ of the
closed string field theory vertices vanishes.

Coming back to the question of background independence,
it should be emphasized that, while we are discussing an infinitesimal
variation $\delta x^\mu$ in theory space, the diffeomorphism relating
the two relevant state spaces ({\it i.e.,} the redefinition
relating the corresponding string fields) is not linear. It is actually
nonpolynomial. The field independent part of it is a constant shift
corresponding to a perturbation by an exactly marginal operator.
The linear part of the map can be interpreted as defining a theory space
connection $\Gamma_\mu$. We find that background
independence, to quadratic order in the string field, requires that
the symplectic form be a covariantly constant
section (in theory space), and that the covariant derivative of the
BRST operator
must be given by $D_\mu (\,\Gamma \,) Q = \bra{V^{(3)}}
c\bar c\O_\mu\rangle$,
where
$\bra{V^{(3)}}$ is the off-shell three string vertex of closed string
field theory (with one of its state spaces turned into a ket). This formula
is remarkable in that the Riemann surface data encoding the three string
vertex of string field theory determines a particular connection in
theory space. The question of background independence to quadratic order
is therefore the question whether a connection $\Gamma_\mu$ satisfying
the two conditions
stated above exists. We find such a connection. This is done by first
showing that the canonical connection $\wh\Gamma_\mu$ satisfies  an equation
of the type $D_\mu (\,\wh\Gamma \,) Q =
\bra{V^{\prime(3)}}c\bar c\O_\mu\rangle$ where
$\bra{V^{\prime(3)}}$ is a new three string field vertex.
This vertex has an asymmetric
puncture, where $c\bar c\O_\mu$ is inserted, but is symmetric under
the exchange of the other two punctures (see Fig.1, in \S3.3).
It is then simple to show
that $\Gamma_\mu -\wh\Gamma_\mu$ can be expressed in terms of an
interpolating three string vertex $\B_3$ representing a deformation from
$\V_3$ to $\V'_3$ (the surfaces, or points in $\wh\P_3$,
corresponding to the string field vertices
$\bra{V^{(3)}}$ and $\bra{V^{\prime(3)}}$ respectively).

In proceeding to higher orders in the redefinition rather interesting
properties of the new vertex $\bra{V^{\prime(3)}}$ come into light.
If we
denote by $[\,,\,]$ the standard star product arising from the three
string vertex, and by $[ \,,\,]'$ the star product arising from the
new string vertex, we then find that
$$\big[ A_1 ,\, [A_2 ,A_3] \,\big]' \, \pm \hbox{cyclic}= 0\, .\eqn\newjac$$
The new three string vertex, together with the standard three string vertex,
satisfies a strict Jacobi like identity. An on-shell version of this
identity is sufficient to guarantee that the quadratic part of the
diffeomorphism exists. The new product $[\,,\,]'$ also satisfies consistency
conditions with respect to the higher products of closed string field theory.
We find that
$$\big[ A_1 ,\, [A_2 ,A_3, \cdots \,,A_N] \,\big]' \,\pm\hbox{cyclic}=
0\,,\quad
N\geq 3\, .\eqn\newhojac$$
An on-shell version of these identities is sufficient to guarantee
the existence of the desired diffeomorphism to all orders. We believe
the fact that these identities hold off-shell could prove necessary
for a general analysis of background independence where we must
consider shifts of the string field corresponding to arbitrary operators.
This new product could also be a useful tool in the understanding of
homotopy Lie Algebras [\stasheff ]. Moreover, as we will discuss in \S9,
it opens up the possibility of constructing string field
actions without using the BRST operator, or perhaps involving more
than one string field.  Such versions of string field theory could
represent progress towards a manifestly background independent formulation.

The pattern that emerges is as follows. If we denote the closed string
vertices by $\V_3,\V_4,\cdots$, and the new string vertex associated
to $\bra{V^{\prime(3)}}$ is denoted by $\V'_3$,
we first find a homotopy $\B_3$
between $\V'_3$ and $\V_3$. A new vertex $\V'_4$ is then constructed by
twist-sewing (sewing and integrating over twist)  $\V_3$ to all the
surfaces in $\B_3$. As a consequence of the above mentioned consistency
conditions, the boundaries of $\V_4$ and $\V'_4$ turn out to coincide.
This is
essential to be able to define a satisfactory interpolating  vertex $\B_4$
between them.  This process is continued recursively.  At every stage, a
new vertex $\V'_N$ is obtained by twist-sewing one lower dimensional
string vertex $\V$, with one interpolating vertex $\B$, in all possible
ways. The consistency conditions guarantee that the boundaries of $\V'_N$
and $\V_N$ coincide, and therefore, one can define the new interpolating
vertex $\B_N$.  The end result is an infinite family of vertices
$\V'_3,\V'_4 \cdots$, and an infinite family of interpolating vertices
$\B_3,\B_4\cdots\,$. These interpolating vertices define the full
diffeomorphism implementing background independence.

In a manifestly background independent formulation of string theory a
change of background should be implemented by a simple shift of the string
field.  A possible way to achieve this would  have the string field
be the coordinates labelling the space of two dimensional field theories.
For open string field theories such a manifestly background independent
approach has been proposed in Refs.[\witten,\liwitten], again based on the
BV formalism.  Other issues involving background independence have
been discussed in Ref.[\wittenqbi]. We feel intuitively, that a measure
of the degree of
background independence of a formulation is provided by the simplicity of
the field redefinition that takes us from string field theory in one
background to another. As we have sketched above, the relevant
diffeomorphism has a clear geometric interpretation.  This leads us to
believe that the present formulation of closed string field theory may not
be far from a manifestly background independent formalism. For the case of
the standard covariant open string field theory [\wittenosft], we show that
the redefinition is given
by a shift plus a linear transformation.

The plan of the paper is as follows. In \S2 we give all the background
material that is necessary for our analysis. In \S3 we develop
some preliminary results essential for our proof.  These include a
discussion of the canonical connection $\wh\Gamma_\mu$ in the presence of
a ghost conformal theory, a computation of the covariant derivative of the
BRST operator, and a study of the connectivity property of the spaces of
symmetric string vertices.  In \S4 we set up the general conditions for
background independence of closed string field theory, and explore
their explicit forms for the case of nearby conformal theories.
Since we work in the
Batalin-Vilkovisky formalism, background independence of the theory
requires existence of a field redefinition which maps not only the action,
but also the symplectic structure of the theory formulated around one
background to those in the theory formulated around a different
background.  In \S5 we prove background independence to quadratic and
cubic order in the string field.
This section develops most of the intuition necessary for
the later generalization to all orders.  In \S6 we discuss in
detail the new three string vertex $\V'_3$, and prove
\newjac , and \newhojac, in particular, we explain why they hold off-shell.
We also
find that $\V'_3$ can be used in conjuction with the standard closed string
vertices to find a new way to construct the moduli spaces $\M_n$ with the
use of fewer than usual Feynman diagrams.  \S7 gives the
construction of the symplectic diffeomorphism to all orders in the string
field.  In \S8 we turn to the question of existence of field
redefinitions that relate string field theories formulated around
backgrounds which are finite distance away, but are
related to each other by a set of marginal deformations.  We show that the
field redefinitions required in this case satisfy a set of differential
equations and prove that their integrability conditions are always
satisfied. Therefore the question of existence is reduced to proving that
in the process of integrating the diffeomorphism one does not encounter
infinities. We argue, but do not prove, that this should be the case.

We conclude this paper in \S9. There we present a proof of local
background independence for open string field theory.
We explore the possibility of extending our analysis to string
field theories based on general ({\it i.e.} non-overlap) vertices, and,
propose a setup for a proof of {\it quantum} background independence of closed
string field theory.
We speculate, on the basis of our results, on
formulations of closed string
theory with a higher degree of background independence.

\chapter{Review}

In this section we begin by reviewing the basics of closed string
field theory. We will describe the various moduli spaces
of surfaces relevant to off-shell amplitudes, and the properties of
differential forms in these moduli spaces, with particular attention
to the action of the BRST operator and to sewing properties. We then
give the precise definition of the symplectic structure
relevant to closed string field theory, as it arises from the symplectic
structure on the state space of a CFT including the reparametrization
ghosts. We explain
why, in the Batalin-Vilkovisky (BV) formalism, the symplectic structure
is necessary, in addition to the master action, to specify
the theory. Finally, we review the earlier work in background independence
of closed string field theory and discuss its relation to the present
work.

\section{Basics in String Field Theory}

The main geometrical input to the construction of the
classical closed string field theory is the set of string vertices $\V_n$.
The string vertices are properly thought as subspaces
of the space $\widehat\P_n=\P_n/\sim$, where $\P_n$ is the space
of $n$-punctured Riemann spheres equipped with local coordinates
around each
of the $n$ punctures, and $/\sim$ indicates that
two identical punctured spheres, with local coordinate systems that
differ by a constant phase around each puncture, should be identified.
Local coordinates up to phases are defined by ``coordinate
curves'', Jordan closed curves homotopic to the punctures that correspond
to the locus $|z| =1$ of the local coordinate $z$, with $z=0$ the puncture.

Given a point in $\P_n$, there
is an obvious projection to $\wh\P_n$, which consists in forgetting
about the phase of the local coordinate. There is another projection
$\pi$ from $\wh\P_n$ to $\M_n$, the moduli space of $n$-punctured
spheres, consisting of
forgetting about the local coordinates.
This allows us to regard $\wh\P_n$
as a fiber bundle, with $\M_n$ as the base space, and the space of the
local coordinate systems at the punctures modulo phases, as the fiber.
We denote by $\sigma$ a section of this fiber bundle.
This can be summarized in the
diagram
$$
\matrix{\P_n\cr
\mapdown{}\cr
\wh\P_n \cr
\sigma\,\mapup{} \,\,\mapdown{\pi} \,\, \cr
\M_n \cr} \eqn\projectmap$$

In classical closed string field theory, points in the
subspace $\V_n$ correspond to ``restricted polyhedra'' [\saadizwiebach ,
\kugokunitomo]. They
represent contact interactions, which means that for each $n$-punctured sphere
(with local coordinates) corresponding to a point in $\V_n$, the disks
determined by the coordinate curves cover fully the surface.
Moreover the subspaces $\V_n$
satisfy the recursion relation
$$\partial\, \V_n\,\, =\, -\,\,{1\over 2}\,
\hskip-5pt\sum_{n_1,n_2\geq 3\atop n_1+n_2 = n+2}
\,{\bf S} \big( \, \V_{n_1} \,\times\, \V_{n_2} \, \big) \,, \eqn\geom$$
where $\times$ indicates the subspace obtained by sewing each surface of
the subspace $\V_{n_1}$ with each surface of the subspace $\V_{n_2}$,
with sewing parameter $t=\exp i\theta$, and
$\theta \in [0,2\pi]$.
{\bf S} denotes the sum over
different splittings of the $n$ labelled punctures into two (unordered)
sets, one with $n_1-1$ punctures, to be attached to the free puntures of
of $\V_{n_1}$, and the other with $n_2-1$ punctures, to be attached to
the free punctures of $\V_{n_2}$.
Each inequivalent
contribution is counted twice in the sum on the right hand side, due to
symmetry under the exchange of the two vertices. This is compensated for
by the explicit factor of $1/2$.
In the left hand side $\partial\V_n$ denotes
the boundary of $\V_n$. The spaces $\V_n$ are actually
subspaces of a globally defined section $\sigma$ over
$\M_n$. This is the section determined by minimal area metrics.

\subsection{Reflectors.}
In the operator formulation of conformal field theory, to every $n$-punctured
surface $\Sigma$ with local coordinates, one assigns a state $\bra{\Sigma}\in
\HH^{*\otimes n}$, where $\HH$ denotes the Hilbert space of states in the
conformal field theory. The basic requirement is that the states must give a
representation of the algebra of sewing of Riemann surfaces. A particularly
useful state is the ``reflector'' state $\bra{R_{12}}\in \HH^*\otimes\HH^*$,
representing a two punctured sphere with local coordinates
$z_1=z$, and $z_2=1/z$, about
$z=0$, and $z=\infty$, respectively. Since there is a globally defined
conformal map exchanging the punctures and the local coordinates
(the map $z\to 1/z$), this state is symmetric under the exchange of the
state space labels: $\bra{R_{12}}=\bra{R_{21}}$. This state defines
a bilinear form (a metric) on $\HH$. Choose a basis $\ket{\Phi_i}$
of states in $\HH$, and let $\epsilon (\Phi_i) \equiv i$, denote the
grassmanality of the operator $\Phi_i$. We define the metric components
$$g_{ij}\,\equiv\,\bra{R_{12}}\Phi_i\rangle_1\ket{\Phi_j}_2  \,.\eqn\zmet$$
Here, $g_{ij}$ are numbers, and are non-vanishing only when $i+j$ is even.
This metric satisfies
$$g_{ij}= (-)^{ij}g_{ji}\,,\quad\epsilon (g_{ij})=i+j ={\rm even} \, ,
\eqn\conmet$$
One defines the inverse metric $g^{ij}$ satisfying
$g^{ik}g_{kj} = g_{jk}g^{ki} =\delta_j^i\,$, and, as a consequence
$$g^{ij}= -(-)^{(i+1)(j+1)}g^{ji}\,,\quad\epsilon (g^{ij})=i+j ={\rm even}
\,.\eqn\conimet$$
We define states with upper indices as $\ket{\Phi^i} \equiv
g^{ij}\ket{\Phi_j}$.
The reflector state is used to introduce a dual basis. One defines
$$\bra{\Phi^i} \equiv \bra{R\,}\Phi^i\rangle \, .\eqn\defdual$$
It then follows that $\bra{\Phi^i}\Phi_j\rangle = \delta_j^i$, and
$$\bra{R_{12}} = g_{ij} \,\, {}_2\bra{\Phi^j} {}_1\bra{\Phi^i} \, .
\eqn\bramet$$
One can also introduce the ket reflector $\ket{R_{12}}$
$$\ket{R_{12}} =\ket{\Phi_i}_1\, g^{ij} \, \ket{\Phi_j}_2  \, ,
\eqn\bramet$$
also symmetric
under the exchange of its state spaces. The contraction
$$\bra{R_{12}}R_{23}\rangle = {}_3{\bf 1}_1 \, ,\eqn\contracr$$
gives the relabeling operator.
The bra
$\bra{R_{12}}$ satisfies the following properties
$$\bra{R_{12}} (c_0^{(1)} + c_{0}^{(2)} ) =
\bra{R_{12}} (b_0^{(1)} - b_0^{(2)}) =\bra{R_{12}}(Q^{(1)}+Q^{(2)}\,) = 0.
\eqn\qonref$$
and similar properties with $b$, $c$ replaced by $\bar b$, $\bar c$.
The ket reflector satisfies analogous properties
$$
(c_0^{(1)} + c_0^{(2)} ) \ket{R_{12}} =
(b_0^{(1)} - b_0^{(2)} ) \ket{R_{12}} =
(Q^{(1)} + Q^{(2)} \,) \ket{R_{12}} = 0.
\eqn\qonconjref$$

The dynamical closed string field $\ket{\Psi}$ corresponds to
a state in the subspace $\H$ of $\HH$, spanned by the elements of
$\HH$ that are annihilated by $L_0-\bar L_0$, and
$b_0-\bar b_0\equiv b_0^-$,
$$\H = \Bigl\{ \ket{A} \,\,\Bigl\vert \,\ket{A}\in \HH,\, \hbox{and},\,\,
(L_0-\bar L_0)\ket{A}=
(b_0-\bar b_0)\ket{A}=0 \, \Bigr\}\, .\eqn\definehath$$
This makes it convenient to introduce the bra
$$\bra{R'_{12}}=\bra{R_{12}}\P_1\P_2 \, ,
\eqn\primref$$
where
$$
\P = \int_0^{2\pi}
{d\theta\over 2\pi} e^{i\theta(L_0 -\ov L_0)}\, ,
\eqn\edefprojxxx
$$
is the
projection to rotationally invariant ($L_0 = \bar L_0$) states.

\subsection{Differential Forms.}
The basic objects for the construction of the string field
interactions arise from differential
forms defined on the tangent space $T_{{}_\Sigma}\P_n$
based at the surface $\Sigma$.
We let $\Omega^{(k)n}_\Sigma$ denote a $(2n-6+k)$-form
labelled by $n$ arbitrary states $\ket{\Psi_1}, \ldots \ket{\Psi_n}$
in $\HH$.
We will generally omit
from the form the label $\Sigma$
corresponding to the surface. These forms are
explicitly given by [\zwiebachlong ,\alvarez ,\nelson ,\kugosuehiro]
$${\Omega_{}^{}}^{(k)n}_{\Psi_1\cdots\Psi_n}( V_1,\cdots , V_{2n -6+k} )
= (2\pi i)^{(3-n)}\bra{\Sigma}\,{\bf b}({\bf v}_1)\cdots
{\bf b}({\bf v}_{2n-6+k})\ket{\Psi_1}\cdots\ket{\Psi_n}.
\eqn\cdefform$$
The Schiffer vector ${\bf v}_r= (v_r^{(1)}(z),\cdots v_r^{(n)}(z))$
creates the deformation of the surface $\Sigma$ specified by the tangent
$V_r\in T_{{}_\Sigma}\P_n$,
and the antighost insertions are defined by
$${\bf b}({\bf v}) = \sum_{i=1}^n \biggl(
\oint b^{(i)}(z_i) v^{(i)}(z_i) {dz_i\over 2\pi i}
+\oint \overline b^{(i)}( \overline z_i)  \overline v^{(i)}
(\overline z_i) {d\overline z_i\over 2\pi i} \biggr).\eqn\kjhkjh$$
Here $\ointop$ is defined  such that
$\ointop dz/z=\ointop d\bar z/\bar z=2\pi i$.
Since there are no global sections in $\P_n$ we must work on $\wh\P_n$ where
there are global sections. It can be shown that for $\ket{\Psi_i}\in \H$
the above differential forms descend to well-defined forms on
$T_{{}_\Sigma}\wh\P_n$ [\nelson ,\zwiebachlong].

The above forms satisfy the basic identity (Ref.[\zwiebachlong],
Eqn.(7.49))
$${\Omega_{}^{}}^{(k+1)n}_{(\sum Q)\Psi_1\cdots\Psi_n}
= (-)^{k+1}\,\hbox{d}{\Omega_{}^{}}^{(k)n}_{\Psi_1\cdots\Psi_n},
\eqn\qivbu$$
which holds both for forms in $\P$ or $\wh\P$. Therefore, the BRST operator
$Q$ acts as an exterior derivative
on the extended moduli spaces. We will drop the off-shell
states from the formulas by writing the forms as bras in
$(\HH^*)^{\otimes n}$:
$${\Omega_{}^{}}^{(k)n}_{\Psi_1\cdots\Psi_n}
=\bra{\Omega^{(k)n}}\,\Psi_1\rangle\cdots\ket{\Psi_n}.
\eqn\corm$$
Then, Eqn.\qivbu\ reads
$$\bigl\langle\Omega^{(k+1)n}\,\bigl|\,\sum_{i=1}^n Q^{(i)}
= (-)^{k+1}\,\hbox{d}\,\bigl\langle\Omega^{(k)n}\bigr| \, ,\eqn\qixbu$$
The form
$\bra{\Omega^{(0)N}}$ (in $T_{{}_\Sigma}\wh\P_N$),
integrated over the subspace $\V_N$ of $\wh\P_N$, defines
the $N$-string interaction vertex\foot{Although the bosonic closed string
field theory action
can be constructed in terms of the vertices $\bra{\Omega^{(0)n}}$ only, the
$\bra{\Omega^{(k)n}}$ for $k<0$ are essential for construction of fermionic
string field theory[\sarojasen].}
$$
\bra{V^{(N)}} = \int_{\V_N} \bra{\Omega^{(0)N}} .\eqn\vexpression$$
More precisely, this equation should include the kets (in $\H$) that,
upon contraction, give a number. In terms of these vertices, the closed string
field theory master action is given by
$$S\, =\, {1\over 2}\,  \bra{R'_{12}}\, c_0^{-(2)} \, Q^{(2)}
\ket{\Psi}\ket{\Psi}
+\sum_{N=3}^\infty{1\over N!}
\langle\, V^{(N)}\ket{\Psi}_1\cdots\ket{\Psi}_N
\equiv \sum_{N=2}^\infty{1\over N!}
\langle\, V^{(N)}\ket{\Psi}_1\cdots\ket{\Psi}_N \eqn\emmeight$$
with $c_0^- =(c_0 -\ov c_0)/2$, and, with the master field $\ket{\Psi}$
an element of $\H$.
Here, for convenience, we have set the string coupling constant $g$ to 1.
The appropriate factors of $g$ can be easily recovered by a rescaling
$\ket{\Psi}\to g\ket{\Psi}$ in the final result.
The statistics of the expansion coefficients of $\ket{\Psi}$, along the
basis vectors of $\H$, are chosen in such a way that $\ket{\Psi}$ is always
even.

\subsection{Sewing Property.}
The forms $\Omega$ (in $T_\Sigma\P$) also satisfy a sewing property.
We introduce a ket
$$\ket{\ss(\theta)_{12}}={1\over 2\pi} b_0^{-(1)}
e^{i\theta(L^{(1)}_0-\ov L^{(1)}_0)}
\ket{R_{12}}\, , \eqn\edefnss$$
which, apart from the $b_0^{-}$ insertion, has the geometrical meaning
of sewing with a twist angle $\theta$. One can then prove (following
the methods of Ref.[\zwiebachlong], \S8) that
$$\bigl\langle\Omega^{(1)n_1+1}_{\Sigma_1}\,\bigl|\,
\bigl\langle\Omega^{(0)n_2+1}_{\Sigma_2}\,\bigl|\,\ss(\theta)\rangle \,=\,
\bigl\langle\Omega^{(0)n_1+n_2}_{\Sigma_1\cup_\theta\Sigma_2}\,
\bigl|\, , \eqn\ssewing$$
where the proper interpretation of this equation is that the left hand
side acts on
$2n_1-3$ tangent vectors of $T_{{}_{\Sigma_1}}\P_{n_1+1}$, and
$2n_2-4$ tangent vectors of $T_{{}_{\Sigma_2}}\P_{n_2+1}$, while the
right hand side acts on $2n_1+2n_2-6 = (2n_1 -3) + (2n_2-4) + 1$ vectors.
Of those, the first $(2n_1-3)$ are vectors in
$T_{{}_{\Sigma_1\cup_\theta\Sigma_2}}\P_{n_1+n_2}$, each of which creates the
deformation of the sewn surface that would be produced by deforming $\Sigma_1$
with the corresponding vector in
$T_{{}_{\Sigma_1}}\P_{n_1+1}$, and then sewing to $\Sigma_2$. The next
$(2n_2-4)$ vectors arise in a completely analogous fashion. The last
vector is $\p/\p\theta$, and is the generator of twist. It arises from
the $b_0^-$ insertion in the ket $\ket{\ss}$. We also define
$$\ket{\s_{12}} \equiv \int_0^{2\pi} d\theta
\ket{\ss(\theta)_{12}} =
b_0^{-(1)} \ket{R'_{12}}\, .\eqn\defsewket$$
The sewing ket $\ket{\s_{12}}$ is the familiar ket relevant to
``twist-sewing'',
that is, sewing with integration over the twist angle.  Moreover,
 it follows from
\qonconjref\ that the sewing ket $\ket{\s}$ is also symmetric.
The integrated version of \ssewing\ in $\wh\P$ will be useful for us.
Let $\B$ denote a $(2n_1-3)$ subspace of $\wh\P_{n_1+1}$ and
let $\V$
denote a $(2n_1-4)$ subspace of $\wh\P_{n_2+1}$. We then find
$$\int_\B\bigl\langle\Omega^{(1)n_1+1}\,\bigl|\,
\int_\V\bigl\langle\Omega^{(0)n_2+1}\,\bigl|\,\s\rangle \,=\,
\int_{\B\times\V}\bigl\langle\Omega^{(0)n_1+n_2}\,
\bigl|\, . \eqn\sewing$$
Here in the right hand side $\B\times\V$ is the (oriented) subspace of
$\wh\P_{n_1+n_2}$ obtained by twist-sewing every element of $\B$ to every
element of $\V$.\foot{The orientation of $\B\times \V$ is fixed by
an ordered basis $[ \cdots ]$ of tangent vectors at each point.
The induced orientation of $\B\times\V$ is $[[\B],[\V],\p/\p\theta ]$,
where the
vectors $[\B]$ and $[\V]$ were defined below eqn.\ssewing.}

\section{Batalin-Vilkovisky Structures}

We would like to review the BV structure that exists in a supermanifold,
and the BV structure that exists in the vector space $\H$. The BV structure is
nothing else than a symplectic structure. For a vector space we need
a bilinear odd-nondegenerate form. For the case of a manifold we need,
in addition, that the form be closed.

\subsection{Symplectic form on a Supermanifold}
We follow the conventions of Ref.[\hatazwiebach].
On a manifold the symplectic form $\omega$ reads
$$ \omega = - dz^i\, \omega_{ij} (z) \, dz^j\, .\eqn\sformm$$
The form $\omega$ is odd, nondegenerate and closed. Nondegeneracy
means that the matrix $\omega_{ij}$ is invertible, and the inverse matrix
is denoted by $\omega^{ij}$. One has
$$\omega^{ik}\,\omega_{kj} = \omega_{jk}\,\omega^{ki}
= \delta_j^i\, .\eqn\win$$
The following properties hold
$$\eqalign{
&\epsilon (\omega_{ij}) = \epsilon(\omega^{ij}) = i+j+1 \ , \cr
&\omega_{ij} = -(-)^{ij}\omega_{ji} \ , \cr
&\omega^{ij} = -(-)^{(i+1)(j+1)}\omega^{ji} \ . \cr}\eqn\PROPOMEGA$$
Using \sformm\ we derive the following transformation laws
$$\eqalign{
\omega_{pq} (\xi ) \,& = \, {\p_l z^i\over \p \xi^p}\, \omega_{ij} (z)\,
{\p_r z^j\over \p \xi^q}\, ,\cr
\omega^{pq} (\xi ) \,& = \, {\p_r \xi^p\over \p z^i}\, \omega^{ij} (z)\,
{\p_l \xi^q\over \p z^j}\, ,\cr}\eqn\tranfww$$
where $\p_l$ and $\p_r$ denote left and right derivatives respectively.

\subsection{Symplectic form on $\H$}
Let $A,B\in \H$, be vectors in the even-dimensional supervector space $\H$.
The symplectic form $\omega ( \cdot \,,\cdot )$ must have the
following exchange property:
$$\omega \, ( A , B ) = - (-)^{AB} \, \omega \, ( B,A ) \, . \eqn\excpr$$
In closed string field theory, $\H$ is the vector space
defined in \definehath,  and
the physically relevant choice of symplectic form reads
$$\omega \, ( A , B ) = \bra{R'_{12}\,}c_0^{-(2)}\,\ket{A}_1\,\ket{B}_2
\equiv \bra{\omega_{12}\,}\,A\rangle_1\,\ket{B}_2 \, .\eqn\ghty$$
The property \excpr\ is easily verified using \qonref.
We now introduce component notation as follows
$$\eqalign{
\bra{\omega_{12}}=\bra{R'_{12}} c_0^{-(2)} &\equiv
\,-\, ~_1\bra{\Phi^i}\,\,\omega_{ij}(x)\,\, ~_2\bra{\Phi^j},\cr
\ket{\s_{12}} = b_0^{-(1)} \ket{R'_{12}} &\equiv \,\,
\ket{\Phi_i}_1\,\,(-)^{j+1}\omega^{ij}(x)\,\,\,\ket{\Phi_j}_2 \,
,\cr}\eqn\emmten$$
where the sewing ket $\ket{\s_{12}}$ was introduced in Eqn.\defsewket.
It follows from Eqn.\emmten\  that
$$\omega_{ij}=(-)^{i+1}\bra{\Phi_i} c_0^- \ket{\Phi_j} \quad {\rm and} \quad
\omega^{ij}=-\bra{\Phi^i} b_0^- \ket{\Phi^j}\, ,\eqn\estatw$$
are real numbers which are non-vanishing only if the ghost numbers
of the states $\ket{\Phi_i}$ and $\ket{\Phi_j}$ add up to five.
Thus $\omega_{ij}$ defined this way automatically satisfies the first of
eqs.\PROPOMEGA.
Moreover, it is clear from the reflector properties that
$$\bra{\omega_{12}} = -\bra{\omega_{21}}\quad\hbox{and}\quad
\ket{\s_{12}}= \ket{\s_{21}}\, .\eqn\sprosf$$
These equations together with our definitions in \emmten\ imply the
expected exchange properties
$$\omega_{ij} = -(-)^{ij} \omega_{ji},\, \quad\hbox{and}\quad
\omega^{ij} = -(-)^{(i+1)(j+1)}\, \omega^{ji}, \eqn\exchsym$$
It follows from \contracr\ that $\bra{R'_{12}}R'_{23}\rangle= {}_3{\bf \P}_1$,
where the
operator on the right is an operator that changes the state space
label of states from one to three, and, at the same
time projects into the $L_0=\bar L_0$ subspace. We use this
to evaluate the contraction of  $\bra{\omega}$ with $\ket{\s}$. One
readily finds
$$\bra{\omega_{12}}\s_{23}\rangle = \bra{R'_{12}}c_0^{-(2)}
b_0^{-(2)}\ket{R'_{23}}=\bra{R'_{12}}
b_0^{-(3)}c_0^{-(3)}\ket{R'_{23}} = b_0^{-(3)}c_0^{-(3)}\,
{}_3{\P}_1 \, =\,{}_3{\P}_1 \,,\eqn\restrinv$$
where the last equality holds in the restricted space we
work on (where states are annihilated by $b_0^-$).
Equation \restrinv\ implies that our definitions in \emmten\ give, as expected,
$$\omega_{ik}\, \omega^{kj}\,=\, \delta^j_i\, .\eqn\emmfour$$

\subsection{Antibracket}
This structure arises from the symplectic structure in $\H$ as follows
[\zwiebachlong].
Consider a set of basis vectors $\ket{\wt\Phi^i}$ such that
$$\omega \, (\, \ket{\wt\Phi^i} , \ket{\Phi_j} ) = -\delta^i_j .\eqn\symbas$$
We can construct the tilde states as follows
$$\ket{\wt\Phi^i} \equiv (-)^i \bra{\Phi^i\,}\s\rangle\, .\eqn\tstate$$
It is straightforward to verify that \symbas\ is
satisfied by this definition.
It is also simple to see that $\ket{\wt\Phi^i}= b_0^-\ket{\Phi^i}$.
The string field is then expanded as
$$\ket{\Psi} = \sum_i \ket{\Phi_i}\,\psi^i =
\sum_{g(\Phi_s)\leq 2}\big(\, \ket{\Phi_s}\,\psi^s+ \ket{\wt\Phi^s}\,
\psi^*_s\,\big) \, ,
\eqn\fantf$$
where $\psi^s$ are fields, and $\psi^*_s$ are antifields. The second sum
is only over states of ghost number less than or equal to two. Since
the ghost number of $\ket{\wt\Phi^s}$
is five minus the ghost number of $\ket{\Phi_s}$,
the sum actually runs over a complete basis (in fact,
a symplectic basis of $\H$). The antibracket of two functions $A$ and $B$, of
the string field is defined as
$$\bigl\{ A \, , B \, \bigr\}= {\p_r A\over \p \psi^s}\,
{\p_lB\over \p \psi^*_s}\,-\,{\p_rA\over \p \psi^*_s}\,{\p_lB\over\p\psi^s}\,.
\eqn\antibr$$
It is a straightforward calculation to prove that
$$\bigl\{ A \, , B \, \bigr\}= {\partial_r A \over \partial \psi^i}
\omega^{ij} {\partial_lB\over \partial \psi^j} \,=\, (-)^{B+1}
{\p \,A \over \p \,\ket{\Psi}}\, {\p \,B \over \p \,\ket{\Psi}}\, \ket{\s}\, ,
\eqn\xantibr$$
where the sewing ket $\ket{\s}$ is gluing the two state spaces left open
by the differentiation with respect to the string field. There is no need
to specify left or right derivatives because the string field is even.

\section{Equivalence of Master Actions}

In a conventional field theory, two actions give classically equivalent
physics if they are related to each other via field redefinitions. In this
case the tree level $S$ matrices calculated from these two actions are
identical. In the Batalin-Vilkovisky formalism, however, specifying the
action does not completely specify the theory, even at the tree level. One
also needs the symplectic structure $\omega_{ij}$ to specify the theory
completely at the tree level.
It enters the theory in three different ways:

\pointbegin The master action $S$ must satisfy
$$\{S, S\}=0 \, ,\eqn\ebatvone $$
where $\{~,~\}$ is the antibracket defined with respect to the symplectic
structure $\omega$ (see \xantibr).
Thus, for a generic change of $\omega$, the master action $S$ will not even
remain a solution of the master equation.

\point The physical observables $\O$ must satisfy
$$ \{ S, \O\} =0  \, .\eqn\ebatvthree $$
Thus, even if we change $\omega$ in such a way that Eqn.\ebatvone\ is still
satisfied, the observables in the original theory do not remain
observables in the new theory.

\point Finally, given a set of observables $\O_i$, their correlation
function is calculated as,
$$ \langle \prod_i \O_i \rangle = \int_L d\psi e^{-S} \prod_i \O_i \, ,
\eqn\ebatvfour $$
where $L$ is a lagrangian submanifold of the full manifold $M$
of configurations of the master
fields. It has dimension equal to half of that of $M$, and
satisfies the property that, for any two tangent vectors $t^i{\p\over
\p\psi^i}$, $\tilde t^i{\p\over \p\psi^i}$ in the tangent space $T(L)$ of
$L$,
$$ \tilde t^i\, \omega_{ij}\, t^j=0 \, .\eqn\ebatvfive $$
Thus the definition of a lagrangian submanifold depends on the choice of
$\omega$.\foot{For a given choice of $\omega$, the right hand side of
Eqn.\ebatvfour\ is independent of the choice of lagrangian submanifold.
This is the main result of BV theory. For extensions, see  [\schwarz].}
This shows that even if we choose a set of observables and change $\omega$
in such a way that both
eqs.\ebatvone\ and \ebatvthree\ still hold with the new $\omega$,
their is no {\it a priori} guarantee that the correlation functions of
these observables will be the same in the two theories.

This shows that if we want to prove the equivalence of two theories,
it is not enough to find a field redefinition which maps one master action
to the other. If such field redefinition, in addition, maps the
symplectic structure of one theory to the other, the theories are clearly
equivalent.\foot{It seems likely that, under reasonable assumptions,
this mapping of both action and symplectic structure, is not only sufficient,
but is also necessary to prove the equivalence of two theories.}
In the next section we shall show
how this can be done for string field master actions and symplectic
structures arising from two different conformal field theories.

\section{Review of Earlier Work}

Background independence of closed string field theory has been analyzed
earlier in refs.[\senone,\sentwo,\senthree]. In this subsection we shall
briefly review
these results, and discuss their relationship with present analysis.

We begin with a review of
ref.[\senone ], which
analyzes background independence of the quadratic part of the
string field theory action.
Let us consider two conformal field theories
CFT and CFT$'$ related by an infinitesimal marginal deformation
$(\delta\lambda/\pi)\int d^2 z\, \O$.
Let $Q$ and $Q'$ denote the BRST charges of the two
string theories formulated around these two conformal field theories, and
$\bra{~}~\rangle$ and $\bra{~}~\rangle'$ be the BPZ inner products in
these two theories. Then, in a certain choice of basis, one finds that
$$\eqalign{
Q'- Q= &{\delta\lambda\over 2\pi i} \ointop_{|z|=\epsilon} \big (dz \bar
c(\bar z) \O(z,\bar z) + d \bar z c(z) \O(z, \bar z)\big) \, ,\cr
\bra{A} B\rangle' -\bra{A} B\rangle = & -{\delta\lambda\over \pi}
\int_{\epsilon\le |z|\le 1/\epsilon} d^2 z \bra{A}
\O(z, \bar z) \ket{B}\, .\cr}\eqn\edeltaqb$$
for some number $\epsilon$.
Also let $\delta\lambda\ket{\wh\O}=
\delta\lambda\ket{c\bar c\O}$
denote the classical  solution in string field theory formulated around
CFT that represents CFT$'$, and
$$ \hat Q = Q + \delta\lambda \,[\wh O, \,\,]\,\, ,\eqn\erevone  $$
be the nilpotent operator [\senold ] that appears in the kinetic term of the
string field theory action formulated around CFT after we shift fields by
an amount $\delta\lambda \ket{\wh\O}$.
It was shown that there is a transformation $S\equiv e^{\delta\lambda K}$
such that acting on states in $\H$
$$
\hat Q= S Q' S^{-1}, \quad \bra{S\Phi_1}c_0^- \ket{S\Phi_2} =\bra{\Phi_1}
c_0^- \ket{\Phi_2}', \quad \ket{\Phi_i}\in \H
\eqn\erevtwo
$$
to order $\delta\lambda$.

To compare this result to the results of the present paper we have to
express Eqn.\erevtwo\ in a different language.
Writing $Q'= Q+\delta\lambda \p_\lambda Q$, and noting that,
$$\eqalign{
\bra{\Phi_1}c_0^- \ket{\Phi_2}' - \bra{\Phi_1}
c_0^- \ket{\Phi_2} = & (-)^{\Phi_1} \,\,
({}'\bra{\omega_{12}}-\bra{\omega_{12}}) \ket{\Phi_1}_1 \ket{\Phi_2}_2\cr
= & (-)^{\Phi_1} \delta\lambda (\p_\lambda \bra{\omega_{12}})
\ket{\Phi_1}_1 \ket{\Phi_2}_2\,\, ,\cr }\eqn\erevthree$$
we can rewrite  eqs.\erevtwo, to order $\delta\lambda$, as
$$\p_\lambda Q + [K, Q] = [\wh\O \, , ~\, ]\,\,,
\quad \p_\lambda \bra{\omega_{12}}
-\bra{\omega_{12}} (K^{(1)} + K^{(2)}) =0\, .\eqn\erevfour$$
As we shall see in \S4, by demanding the
background independence of the  quadratic terms
of the master action, we arrive precisely at
equations of the form \erevfour, which are conditions on the covariant
derivatives of $Q$ and $\bra{\omega_{12}}$ with respect to the connection
$K$. Thus the proof of existence of $K$ given in ref.[\senone ] may be
taken as the proof of background independence of the quadratic part of the
master action.\foot{The analysis of ref.[\senone ] was for the classical
action, and not the master action. In this case only a combination of the
two equations in \erevfour, specifying the covariant derivative of
$\bra{\omega_{12}} (Q^{(1)}+Q^{(2)})$, is necessary for background
independence. However, a proof of existence of $K$ satisfying both
equations separately, was given in ref.[\senone]. As we shall see in
\S4, covariant constancy of $\bra{\omega_{12}}$ is necessary for
the symplectic structure of the theory formulated around CFT to get mapped
to the symplectic structure of the theory formulated around CFT$'$.} The
proof given in [\senone] was based on showing how to construct matrix
elements of $K$
satisfying Eqn.\erevfour.  However,
no closed expression for $K$ as an operator was obtained.  In this paper
we shall obtain a closed expression for this operator, which we shall call
$\Gamma_\mu$, by expressing it in terms of geometric
objects in the moduli space $\wh\P_3$ of three punctured spheres
(with local coordinates at the punctures).

In refs. [\sentwo,\senthree]  an attempt was made to prove the background
independence of
cubic and higher order terms of the string field theory. In particular, it
was shown that there is an explicit field redefinition
which transforms the {\it classical} string field theory action
formalated around CFT, to the {\it classical} string field theory action
formulated around CFT$'$, up to cubic
terms. Again, the proof involved explicit construction of the
different coefficients that appear in the field redefinition.
In this paper we find a field redefinition that relates the {\it master}
actions formulated around these two conformal field theories {\it to all
orders} in the string field. Moreover, the objects which describe the
the field redefinition are expressed in terms of geometric objects in
the moduli spaces $\wh\P_n$ of $n$-punctured Riemann spheres (with local
coordinates at the punctures).

\chapter{Connections and Symmetric Vertices}

In this section we develop some of the basic results that will be
necessary to carry out
our proof of background independence. After defining
covariant derivatives on (super) vector bundles, we discuss how the
connection $\wh\Gamma_\mu$ [\kugozwiebach,\ranganathan,\nelcam,\rangasonodazw]
is defined in the presence of ghosts. We then compute the covariant
derivative of the string field vertices with respect to this connection.
This includes the covariant derivative of the BRST operator,
which appears together with the symplectic form in the definition
of the two string vertex. It is here that the asymmetric three string
vertex $\V'_3$ makes its appearance.
Finally, we prove that the space of symmetric
closed string vertices is connected.

\section{Covariant Derivatives on the Vector Bundle}

Let $\F_n$ denote the vector bundle with the space of conformal field
theories, labelled by the coordinates $\{x^\mu\}$,
as the base space, and fiber $(\H_x)^{\otimes n}$ for
$n\ge 0$ and $(\H_x^*)^{\otimes (-n)}$ for
$n<0$.
We begin by defining the connection matrix
$$\Gamma_\mu \equiv \ket{\Phi_j}\, \Gamma_{\mu i}^{~~j} \, \bra{\Phi^i} \, ,
\eqn\mconn$$
with $\epsilon(\Gamma_{\mu i}^{~~j})
=(-)^{i+j}$. The connections that will
be of relevance for us in this paper will all have the property that
$\Gamma_{\mu i}^{~~j}$ are real numbers and vanish unless $\ket{\Phi_i}$
and $\ket{\Phi_j}$ have the same ghost numbers.
If $\ket{A(x)}$ and $\bra{B(x)}$ denote sections of $\F_1$ and $\F_{-1}$
respectively, then, the covariant derivatives acting on these sections are
defined to be
$$\eqalign{
D_\mu \, (\, \Gamma\, ) \, \ket{A\,} &\equiv \,\,\,
(\p_\mu+\Gamma_\mu) \, \ket{A\,}
\cr
D_\mu \, (\, \Gamma\, ) \, \bra{B\,} &\equiv \,\,\, \p_\mu \bra{B} \,-
\bra{B\,}\Gamma_\mu \,
\cr}\eqn\covbasis$$
It is clear that this definition preserves contraction of state
spaces, namely $D_\mu (\Gamma ) \bra{A}B\rangle =
\p_\mu\bra{A}B\rangle$.
The covariant derivative of general sections is obtained using
the above derivatives and keeping in mind that the derivatives
act tensorially.
The covariant derivative of the symplectic section is given by
$$\eqalign{
D_\mu \, (\Gamma \, ) \bra{\omega\,} &= -\,D_\mu \, (\Gamma \, )
\bigl(\,\, _1\bra{\Phi^i}\,\,\omega_{ij}(x)\, ~_2\bra{\Phi^j} \bigr) \cr
&= -\,\, _1\bra{\Phi^i}\,\Bigl( \partial_\mu \,\omega_{ij}
-(-)^{i(i'+1)}\, \Gamma_{\mu i}^{~i'}\,\,\omega_{i'j} - \omega_{ij'}\,
\Gamma_{\mu j}^{~j'} \,
\Bigr)~_2\bra{\Phi^j} \, . \cr }\eqn\covsym$$
This is sometimes written conventionally as
$$\eqalign{
\bigl( \D_\mu (\Gamma ) \, \omega\,\bigr)_{ij}&=\partial_\mu \,\omega_{ij}
-(-)^{i(i'+1)}\, \Gamma_{\mu i}^{~i'}\,\,\omega_{i'j} - \omega_{ij'}\,
\Gamma_{\mu j}^{~j'} \, .\cr } \eqn\sdervb$$

It is useful to introduce another kind of covariant derivative, one
relevant to functions on the whole vector bundle $\F_1$. Let $S(\psi^i , x)$
be such a function, with $\psi^i$ denoting the coordinates of $\H$.
We then define
$$\hbox{D}_\mu (\Gamma )\, S \equiv \partial_\mu S -
{\partial_r S\over\partial \psi^i}\,
\Gamma_{\mu j}^{~~i} \psi^j \, .\eqn\newcov$$
This covariant derivative examines the variation of the function as
we move in the base along $\delta x$, and, on the fiber by parallel transport.
For functions that arise naturally from sections, such as
$$F\, (\, \ket{\Psi} \,, x\, ) = \bra{\,\Sigma\, (x)\,}\Psi\rangle
\ket{\Psi}\cdots\ket{\Psi}\, ,\eqn\specfun$$
with $\bra{\,\Sigma (x)\,}$ a section
of $\F_{-n}\,$, and, $\ket{\Psi} = \ket{\Phi_i}\,\psi^i\,$ a
grassman even ket,
one can readily verify that
$$\hbox{D}_\mu (\Gamma )\, F\, (\, \ket{\Psi} \,, x\, ) =
\Bigl( D_\mu (\Gamma )\bra{\,\Sigma\, (x)\,} \Bigr)\,
\ket{\Psi}\cdots\ket{\Psi}\, .\eqn\specfun$$

\section{The connection $\widehat \Gamma$ upon inclusion of ghosts}

In this subsection we discuss the
extension of some of the theory-space geometry results of [\rangasonodazw ]
to the case when the space of CFT's is made of theories each of which is
the tensor product of a matter CFT times the standard CFT of
reparametrization ghosts. We discuss, in
particular, the canonical connection $\wh\Gamma_\mu$.
These results will be used in the next subsection for
the computation of the covariant derivative, with respect
to $\wh\Gamma_\mu$, of the string field vertices.

One of the main results of [\rangasonodazw ] was that the variational
formula of ref.[\sonoda] can be generalized to allow a unified
description of all possible connections. It was argued that such
a variational formula could, in fact, be taken as a definition of what
we mean by having a theory space. The formula reads
$$D_\mu (\Gamma\,) \bra{\Sigma\, } = -{1\over\pi} \hskip-6pt
\int_{\Sigma -
\cup_i D_i}\hskip-6pt d^2 z~ \bra{\,\Sigma;z\,} \O_\mu \rangle
-  \sum_{i=1}^n \bra{\,\Sigma\,}\,\omega_\mu^{(i)}.\eqn\xgen$$
where the surface sections $\bra{\, \Sigma\,}$ of the operator formalism
encode all correlators on the punctured surface $\Sigma$. The state
$\ket{\O_\mu}$ is exactly marginal, and is integrated over the surface
minus some disks $D_i$ around the punctures. Finally, the operator one forms
$\omega_\mu$ represent similarity transformations acting on each state
space of $\bra{\, \Sigma\,}$.
Given a domain $D$, and a one form $\omega_\mu\,$,
there must exist a connection $\Gamma_\mu$ such that the above equation holds.
In particular, taking $\omega_\mu=0$, and $D_i=D_i^{(1)}$ to be unit disks,
we are guaranteed the existence of the corresponding connection
$\widehat\Gamma$ satisfying
$$D_\mu (\,\widehat\Gamma\,) \bra{\,\Sigma\, } = -{1\over\pi} \hskip-6pt
\int_{\Sigma -
\cup_i D_i^{(1)}}\hskip-6pt d^2 z~\bra{\,\Sigma;z\,}\O_\mu\rangle.\eqn\xen$$

For string field theory, the relevant CFT theory-space
is made of theories each of which is a matter theory CFT$_M$, tensored
with the reparametrization ghost theory CFT$_G$. The ghost CFT is never
changed, and therefore, the
coordinates that parametrize the total
space
are the coordinates
arising from the specification of the matter theory CFT$_M$.
When we consider this theory space, the variational
formula \xen\
holds with $\ket{\O_\mu}$ the ghost number
zero state created by the action of $\O_\mu (z)\otimes {\bf 1}$ on the
SL(2,C) vacuum, where $\O_\mu(z)$ is constructed purely out of the matter
fields. It follows that the covariant derivative of $\bra{\Sigma}$
does not change the ghost number of the bra. Indeed, in the convention
where both the in and out vacuum have
ghost number zero, the bra $\bra{\Sigma}$,
corresponding to a surface of genus $g$ with $n$ punctures, has ghost
number $6g-6+6n$. Upon contraction, one loses six units of ghost number, and
therefore, $\ket{\O_\mu}$ must be of ghost number zero for
$\bra{\Sigma ;z}\O_\mu\rangle$ to be of the same ghost number
as $\bra{\Sigma}$.

Our argument of background independence will assume the existence
of $\widehat\Gamma_\mu$, as this is conceptually equivalent to the assumption
of having a theory space. This connection will be taken to be known, and this
will be our basic input from theory space geometry. It is useful, however,
to give a construction of $\widehat\Gamma_\mu$ in terms of the connection
$\wh\Gamma_\mu^{\hskip 1pt M}$ relevant to the theory space of CFT$_M$
(without the ghosts). This latter connection satisfies the variational formula
$$D_\mu (\,\widehat\Gamma^{\hskip 1pt M}\,) \bra{\,\Sigma^{\hskip 1pt M}\, }
= -{1\over\pi} \hskip-6pt
\int_{\Sigma -
\cup_i D_i^{(1)}}\hskip-6pt d^2 z~\bra{\,\Sigma^{\hskip 1pt M};z\,}
\O_\mu^{\hskip 1pt M}\rangle \, ,\eqn\mxen$$
where we have added the superscripts $M$
to remind ourselves that we are dealing
with the matter theory alone. We also denote by $\ket{\phi_i^{\hskip 1pt M}}$
a basis in $\HH_M$, and the connection coefficients read
$\wh\Gamma_{\mu i}^{\hskip 1pt Mj}$. Now consider the ghost CFT, and
choose as a basis of states the Fock space states formed, by acting
on the vacuum, with the usual ghost and antighost oscillators
$(c_n,\ov c_n , b_n, \ov b_n)$. Denote such basis states by
$\ket{\phi_I^{\hskip 1pt G}}$. It then follows that the basis states
of CFT$_M\otimes$CFT$_G$ are given by
$\ket{\phi_i^{\hskip 1pt M}}\otimes\ket{\phi_I^{\hskip 1pt G}}$.
We now claim that the connection $\wh\Gamma_\mu$ on the full (tensored)
theory space is given by
$${{\wh \Gamma}}_{\mu\,(i,I)}^{~~(j,J)}\,=\,
\wh\Gamma_{\mu i}^{\hskip 1pt Mj}\, \delta_I^J \, .\eqn\formcon$$
The Kronecker delta in the ghost labels implies that the connection
essentially ignores the ghosts. More precisely, we have
$$D_\mu (\, \wh\Gamma\,)
\Bigl(\, \ket{\phi_i^{\hskip 1pt M}}\otimes\ket{\phi_I^{\hskip 1pt G}}\,\Bigr)
= \Bigl(\,D_\mu (\, \wh\Gamma^{\hskip 1pt M}\,)
\ket{\phi_i^{\hskip 1pt M}}\, \Bigr)
\otimes\ket{\phi_I^{\hskip 1pt G}}\, .\eqn\split$$
In each tensored theory, the surface states are given by
$\bra{\Sigma} = \bra{\,\Sigma^{\hskip 1pt M}\, }\otimes
\bra{\,\Sigma^{\hskip 1pt G}\,}$, with the $M$ and $G$ superscripts
denoting the matter theory and the ghost theory respectively. Moreover,
$\ket{\O_\mu}=\ket{\O_\mu^{\hskip 1pt M}}\otimes\ket{0^{\hskip 1pt G}}$.
Let us now show that the ansatz \formcon, together with the
matter variational formula \mxen, indeed give us the variational
formula \xen\ for the tensored theory space.
We begin with the left hand side of \xen\
$$\eqalign{
D_\mu (\,\widehat\Gamma\,) \bra{\,\Sigma\, } &=
D_\mu (\,\widehat\Gamma\,) \Bigl(\bra{\,\Sigma^{\hskip 1pt M}\, }\otimes
\bra{\,\Sigma^{\hskip 1pt G}\, }\,\Bigr)\, =
\Bigl( \,D_\mu (\,\widehat\Gamma^{\hskip 1pt M}\,)
\bra{\,\Sigma^{\hskip 1pt M}\, }\Bigr)
\otimes\,\bra{\,\Sigma^{\hskip 1pt G}\, } \cr
&= -{1\over\pi} \hskip-6pt
\int_{\Sigma - \cup_i D_i^{(1)}}
\hskip-6pt d^2 z~\bra{\,\Sigma^{\hskip 1pt M};z\,}
\O_\mu^{\hskip 1pt M}\rangle \, \otimes \bra{\,\Sigma^{\hskip 1pt G}\, }\,,
\cr}\,\eqn\gettt$$
where in the first step we used \split, and in the second step we used \mxen.
Since the vacuum state deletes punctures we can write
$ \bra{\,\Sigma^{\hskip 1pt G}\, } =  \bra{\,\Sigma^{\hskip 1pt G};z}
0^{\hskip 1pt G}\rangle$, and back in \gettt\ we find
$$\eqalign{
D_\mu (\,\widehat\Gamma\,) \bra{\,\Sigma\, } &=
-{1\over\pi} \hskip-6pt
\int_{\Sigma - \cup_i D_i^{(1)}}
\hskip-6pt d^2 z\,\Bigl( \,\bra{\,\Sigma^{\hskip 1pt M};z\,}\otimes
 \bra{\,\Sigma^{\hskip 1pt G};z}\, \Bigr)
\Bigl( \ket{\O_\mu^{\hskip 1pt M}} \otimes\ket{0^{\hskip 1pt G}\, }\,\Bigr)\cr
&= -{1\over\pi} \hskip-6pt \int_{\Sigma - \cup_i D_i^{(1)}}
\hskip-6pt d^2 z~\bra{\,\Sigma;z\,}\O_\mu\rangle\, ,\cr}\eqn\gextt$$
which is the desired relation.

Let $\A$ denote an operator constructed purely from
the ghost fields. It can then be written explicitly as follows
$$\A \,=\, A_I^J (x) \, \Bigl[ {\bf 1}^M \otimes\Bigl( \,
\ket{\phi_J^G} \bra{\phi^{IG}}\, \bigr) \,\Bigr] \, .\eqn\explgh$$
The covariant derivative, computed with the help of \formcon, gives
$$D_\mu(\wh\Gamma) \A =\,(\p_\mu  A_I^J) (x) \,
\Bigl[ {\bf 1}^M \otimes\Bigl( \,
\ket{\phi_J^G} \bra{\phi^{IG}}\, \bigr) \,\Bigr] \,\equiv\, \p_\mu\A\, .
\eqn\eghostcov$$
Consider now the ghost operators $\{ c_n,\ov c_n,b_n,\ov b_n \}$. Given that
we have defined the basis states of CFT$_G$ in terms of these operators acting
on the vacuum,
their matrix elements (the analogs of $\A_I^J$) are all constants
throughout theory space. It then follows that the covariant derivative
of each of these operators vanishes
$$D_\mu(\,\wh\Gamma\,) \{ c_n,\ov c_n,b_n,\ov b_n \} = 0\,.\eqn\covdosc$$

\section{Covariant Derivatives $D_\mu(\, \wh\Gamma\,)$ of String Field
Vertices}

We shall now compute covariant derivatives of $N$-string field vertices
$\bra{V^{(N)}}$ with the connection $\wh\Gamma$.
We note that the string field vertices, for $N\geq 3$ all arise from punctured
spheres whose local coordinates cover fully the surfaces. Therefore, for
each sphere $\Sigma$ in $\V_N$, $\Sigma-\cup_i D_i^{(1)}$ vanishes,
and as a consequence $D_\mu(\,\wh\Gamma\,) \bra{\Sigma} = 0$ (see \xen).
Moreover, recall (\S2.1) that the string field vertex takes the form
$\bra{V^{(N)}}=\int_{\V_N} \bra{\Sigma}{\bf b}\cdots {\bf b}\,,$ where
${\bf b}$ are antighost insertions. Such insertions have nothing whatsoever
to do with CFT$_M$, they simply construct a measure on $\wh\P_N$.
Therefore $D_\mu(\,\wh\Gamma\,)\, {\bf b} = 0$. All this implies that
the covariant derivative of the string field vertex $\bra{V^{(N)}}$
vanishes
$$D_\mu (\, {\wh\Gamma}\, )\bra{V^{(N)}} =0\, ~~ {\rm for}~ N\ge 3.
\eqn\dervnzero$$

We would like to compute now the covariant derivative of the two
string vertex $\bra{V^{(2)}} = \bra{\omega_{12}}Q^{(2)}$. To this
end we will first calculate the covariant derivative of the symplectic
form, and then the covariant derivative of the BRST operator.
The symplectic form is given by $\bra{\omega_{12}} = \bra{R'_{12}}c_0^{-(2)}$,
where, $\bra{R'_{12}}=\bra{R_{12}}\P_1\P_2$ (\primref). Since $\bra{R_{12}}$
is an overlap two string vertex,
$D_\mu (\, {\wh\Gamma}\, )\bra{R_{12}} =0$. Furthermore,
$D_\mu (\, {\wh\Gamma}\, )(L_0-\ov L_0) =0$ (Ref.[\rangasonodazw], Eqn.5.5)
\foot{
This can be easily seen using the definition of $D_\mu
(\, {\wh\Gamma}\, ) L_n$, and $D_\mu (\, {\wh\Gamma}\, )\ov L_n$,
given below, and,
noting that with our definition of $\oint$, $\oint_{|z|=1} zd\bar z f=
\oint_{|z|=1}\bar z d z f$ for any function $f$ of $z$ and $\bar z$.}
implies that $D_\mu (\, {\wh\Gamma}\, )\P =0$. These results combine
to give
$$D_\mu (\, {\wh\Gamma}\, )\bra{R'_{12}}=0\, . \eqn\edmurpot$$
This equation, expressing the metric compatibility of $\widehat\Gamma$,
together with $D_\mu(\,\wh\Gamma\,)\, c_0^-=0$ (\covdosc), implies
that the covariant
derivative of the symplectic section vanishes
$$D_\mu(\, \widehat\Gamma\, )\bra{\, \omega\,}=0 .\eqn\ethreetwo$$

We now turn to the computation of $D_\mu (\, {\wh\Gamma}\,)\,Q$.
It was found in [\nelcam,\rangasonodazw ] that the covariant derivative
of the Virasoro operators, with respect to $\widehat\Gamma$ is given by
$$D_\mu (\widehat\Gamma \,)\, L_n
= \oint_{|z|=1} {d \bar{z}\over 2\pi i}~ z^{n+1}\, \bra{0,z, \infty^*}
\O_\mu\rangle = \oint_{|z|=1} {d \bar{z}\over 2\pi i }~ z^{n+1}\,
\O_\mu (z, \bar z) ,\eqn\dln$$
with $D_\mu(\wh\Gamma)\ov L_n$ given by a similar expression (recall
$\ointop dz/z=\ointop d\bar z/\bar z=2\pi i$). It then follows from
$Q =\sum (c_{-n} L_n + \ov c_{-n} \ov L_n )$, and Eqn.\covdosc, that
$$D_\mu(\, \widehat\Gamma\,)\, Q =
{1\over 2\pi i}\ointop_{|z|=1}(dz \bar c(\bar z)
\O_\mu(z,\bar z) + d\bar z c(z) \O_\mu(z,\bar z)).\eqn\ethreeone$$
This resembles Eqn.\edeltaqb\ for $\epsilon=1$. We are now set for
the computation of $D_\mu(\, \widehat\Gamma\,)\,\bra{V^{(2)}}$.

\subsection{Claim:} The covariant derivative of $\bra{V^{(2)}}$ is given by
$$D_\mu(\,\widehat\Gamma\,)\,\bra{V^{(2)}}=
D_\mu(\,\widehat\Gamma\,)\,\big( \bra{\,\omega_{12}\,}\, Q^{(2)}\big) =
\,  \bra{\, V^{\prime(3)}_{123}\, }
\,\widehat\O_\mu\rangle_3  \, ,\eqn\ethreenewa$$
where $\ket{\wh\O_\mu}=\ket{c\bar c \O_\mu}$, and the bra
$\bra{V^{\prime(3)}_{123}}$ is the surface state
corresponding to
a three punctured sphere $\V'_3$, shown in Fig.~1, and described as follows.
In the uniformizing coordinate $z$, it is
punctured at $z=0$, with a local coordinate $z_1(z) = z$, at
$z=\infty$, with local coordinate $z_2 (z)= 1/z$, and at $z=1$,
with local coordinate $z_3 (z) = h(z)$ left arbitrary.
The right hand side of
Eqn.\ethreenewa\ is independent of the choice of $z_3$, since
$\ket{\widehat \O_\mu}$,
which is inserted at $z=1$, is primary and of dimension zero. It should be
emphazised that this equation holds only upon contraction with states in
$\H$, namely, it is a strict identity if we multiply from
the right by $b_0^-\P$ for each state space.

\if y\figcount
\midinsert
$$\epsffile{szfig1.ps}$$
\nobreak
\narrower
\singlespace
\noindent
Figure 1. Here we show the asymmetric three punctured sphere
$\V'_3$. The local coordinates $z_1(z)=z$, based at $z=0$, and $z_2(z)=1/z$,
based at $z=\infty$, cover the sphere fully. The coordinate $z_3(z)$, based
at $z=1$, is undetermined.
\medskip
\endinsert

\else\fi

\subsection{Proof.}
Using \ethreetwo\ and \ethreeone, we get,
$$D_\mu(\,\hat\Gamma\,)\,(\bra{\omega_{12}}Q^{(2)})
=\, \bra{\omega_{12}} {1\over 2\pi i}
\ointop_{|z|=1}\Bigl( \,dz \bar c(\bar z)
\O_\mu(z,\bar z)+d\bar z c(z) \O_\mu(z,\bar z)\Bigr)^{(2)},\eqn\ethreeoneaa$$
where the operator inside the contour integral is an operator on the
state space $(2)$. We therefore need to show that
$$\bra{\omega_{12}}
{1\over 2\pi i}\ointop_{|z|=1}\Bigl(\, dz \bar c(\bar z)
\O_\mu(z,\bar z)+d\bar z c(z) \O_\mu(z,\bar z)\, \Bigr)^{(2)}
=\bra{V^{\prime(3)}_{123}}\widehat \O_\mu\rangle_3
\eqn\ethreeonea$$
This can be done following a similar analysis in ref.[\senone].
Consider the left hand side of the equation, and
separate out the ghost zero mode from
the bra $\bra{\omega\,}$
$$\bra{{R'}_{12}}{1\over 2\pi i} \ointop_{|z|=1} \Big( dz\, c_0^-\, \bar c
(\bar z)
\O_\mu(z,\bar z)
+d\bar z\, c_0^-\, c(z) \O_\mu(z,\bar z) \Big)^{(2)}\, .\eqn\ethreeonec$$
Using Virasoro Ward identities we find
$$\eqalign{
\bar c\O_\mu (z=e^{i\theta}) &= e^{-i\theta} e^{i(L_0-\ov L_0)\theta}
\,\,\bar c\O_\mu (z=1)\, e^{-i(L_0-\ov L_0)\theta}\,\, ,\cr
c\O_\mu (z=e^{i\theta})& = e^{i\theta}\,\, e^{i(L_0-\ov L_0)\theta}
\,\, c\O_\mu (z=1)\, e^{-i(L_0-\ov L_0)\theta} .\cr}\eqn\vward$$
The operator $e^{i(L_0-\overline L_0)\theta}$ on the left
commutes with $c_0^-$, and gives one
acting on the primed reflector, and, the operator
$e^{-i(L_0-\ov L_0)\theta}$ on the right, gives one
acting on states in $\H$.
Using these relations we can explicitly perform the $z$, $\bar z$
integrals in Eqn.\ethreeonec, and bring it to the form:
$$\bra{{R'}_{12}}\,\,\Bigl( \,  c_0^- (c(1)+\bar c(1))\O_\mu(1)
\,\Bigr)^{(2)} \,\eqn\ethreeoned$$
On the other hand, from the geometrical description of $\bra{V^{\prime(3)}}$
we have that
$$\bra{V^{\prime (3)}_{123}}\widehat\O_\mu\rangle_3
=\bra{{R'}_{12}} \, \Bigl( \, c(1)\bar c(1) \O_\mu(1) \, \Bigr)^{(2)} \, ,
\eqn\ethreeonee$$
since the third puncture sits at $z_2 =1$.
We now must show that \ethreeoned\ and the right hand side of
\ethreeonee\ agree upon contraction with states annihilated by $b_0^-$.
The simplest way to verify this is to multiply both expressions from
the right by the factor $b_0^{-(2)}b_0^{-(1)}$. One must then move
$b_0^{-(2)}b_0^{-(1)}$ all the way to the left and use
$\bra{R'_{12}}b_0^{-(2)}b_0^{-(1)}=0$, which follows from the properties
of the reflector and $(b_0^-)^2=0$.
This establishes the desired result.\hfill\square

\section{Symmetric Vertices and Their Deformations}

Our analysis of symplectic (or antibracket preserving) diffeomophisms
requires careful consideration of the meaning of symmetric vertices
and their deformations.
In this section we will develop the necessary
results on symmetric string vertices. The basic result that we need is
that given two symmetric string vertices, there is a continuous
deformation of one vertex into the other via symmetric string vertices.
In other words, the space of symmetric string vertices is connected.
We will prove this result by using the methods of Ref. [\sonodazw ].
We discuss explicitly, because of their special features, the cases of
two and three string vertices. We then consider all higher string vertices.

\subsection{Two String Vertices.}
A two string vertex is
a two punctured sphere with a coordinate curve $\C$ (\S2.1) surrounding each
puncture.
If we consider the punctures to be at $z=0$, and at $z=\infty$,
with $z$ the uniformizing coordinate, there is a one complex parameter
family of conformal maps taking the punctured sphere into itself, namely,
the maps $z\to az$, with $a$ constant. Two two-string vertices are
identical if their corresponding coordinate curves $(\C_1 ,\C_2)$,
and $(\C'_1 ,\C'_2)$ are mapped into each other
by the map $T_a: z\to az$, for some $a$:
$$ (T_a \,\C_1 ,T_a \,\C_2)= (\C'_1 ,\C'_2)\, . \eqn\tpv$$
A two string vertex is defined to
be {\it symmetric} if any well defined map on the sphere exchanging the
two punctures exchanges the coordinate curves up to the above equivalence.
The map $I_b : z\to b/z$ is the most general map exchanging the
punctures (at zero and infinity). Thus a vertex is symmetric if
$$(I_b \,\C_1 ,I_b \,\C_2)= (T_a\,\C_2 , T_a\,\C_1)\, . \eqn\stev$$
It follows from the above that
$$(I_1 \,\C_1 ,I_1 \,\C_2)= (T_{c}\,\C_2 , T_{c}\,\C_1)\, . \eqn\stev$$
for $c=a/b$. Thus, a two string vertex is symmetric
if there is a constant
$c$ such that the above relation holds. It follows from \stev\ that a symmetric
two string vertex is always of the form
$$( \C_1\, ,\, I_1 \,T_c \, \C_1 \, )\, \,, \,\, (c\not= 0).\eqn\stsvs$$

We now want to show that given two symmetric
vertices $(\C_1 ,I_1 \,T_c \, \C_1 \,)$, and $(\C'_1 ,I_1 \,T_{c'} \, \C'_1)$,
with $c$ and $c'$ two constants different from zero,
there is a continuous deformation taking one into
the other, such that, at every stage we have a symmetric two string
vertex. To this end, we introduce a homotopy $c(t)$
interpolating between the two constants: $c(0)=c$, $c(1) =c'$,
and $c(t)\not= 0$ for all $t\in [0,1]$. Since the coordinate curves
$\C_1$ and $\C'_1$ are Jordan closed curves surrounding $z=0$, they are
homotopic, and therefore, there is a homotopy
$\C_1(t)$ such that $\C_1(0)=\C_1$, and  $\C_1(1)=\C'_1$. It is then
clear that
$\bigl( \, \C_1 (t)\,\, ,\,\, I_1 \circ T_{c(t)} \C_1 (t)\,\,
\bigr) \,$
provides the desired homotopy between the two string vertices.
What we did was elementary, we deformed
arbitrarily one of the coordinate curves, and defined the other coordinate
curve to be such that we obtain a symmetric two string vertex at every
stage of the deformation.

\subsection{Three string vertices.}
The maps taking a three punctured sphere into itself arise from
a map $T$ that cycles the three punctures, and a map $E$ that
exchanges two punctures leaving the other fixed. A three
string vertex is said to be symmetric if $T$
cycles the coordinate curves, and, $E$ exchanges two coordinate curves
leaving the other fixed.  The requirement of invariance of this coordinate
curve under $E$ implies that given an arbitrary coordinate curve around one
puncture, it is not always possible to obtain a symmetric vertex.
It was shown in
[\sonodazw], however, that given a coordinate curve $\C_1$
satisfying $E \C_1 = \C_1$, the vertex $( \C_1 , T \C_1 , T^2\C_1 )$ is
symmetric. Moreover, all symmetric three-string vertices
can be written in this way.
Thus, given two symmetric vertices $( \C_1 , T \C_1 , T^2\C_1 )$
and $( \C'_1 , T \C'_1 , T^2\C'_1 )$, we must find a homotopy
$\C_1(t)$ between $\C_1$ and $\C'_1$ with $E\C_1(t) = \C_1 (t)$.
This requirement  is
easily visualized if $\C_1 (t)$ is chosen to surround the puncture
at $z=0$,  and the other two punctures are placed at $z=1$ and $z=-1$. Then,
the map $E$ takes the form
$E: z\to -z$, and it acts on $\C_1(t)$ by reflection around the origin.
This means that $\C_1(t)$ can be broken into two open curves
$\C^u(t)$ and $\C^l(t)$ (for upper
and lower) whose endpoints, one on the positive real
axis, and the other on the negative real axis, coincide, with the map $E$
exchanging the open curves. The curves $\C^u(t)$ and $\C^l(t)$ are homotopic
to open curves lying fully on the upper and lower half plane respectively.
Thus, the open curves $\C_1^u$ and ${\C'}_1^u$ are homotopic, and any
arbitrary homotopy between them can be extended by reflection to a consistent
homotopy of the curves $\C_1$ and $\C'_1$. The vertex
$(\C_1(t), T\C_1(t), T^2 \C_1(t))$ then defines a
homotopy, via symmetric closed string vertices, between the two symmetric
vertices.

In our analysis we shall also need to construct homotopies between three
string vertices which are not fully symmetric, but symmetric under the
exchange of two legs. Such vertices are characterized by the coordinate
curves $(\C_1, \C_2, E\C_2)$ with $\C_1$ satisfying $\C_1=E\C_1$, but
$\C_2$ arbitrary. The homotopy between two such vertices maintaining the
exchange symmetry $2\leftrightarrow 3$ is given by $(\C_1(t), \C_2(t),
E\C_2(t))$, where $\C_1(t)$ is the homotopy between $\C_1$ and $\C_1'$
satisfying $\C_1(t)=E\C_1(t)$, and $\C_2(t)$ is any arbitrary homotopy
between $\C_2$ and $\C_2'$.

\subsection{Higher String Vertices}
A vertex $\V_n\, (n\geq 4)$ is a subspace of $\widehat \P_n$, typically,
of dimensionality $2n-6$. It is said to
be symmetric if $\V_n$ includes, for each punctured
surface with local coordinates, all the
copies of this surface that differ only by the assignment of the
labels $\{1,2\cdots , n\}$ to the underlying punctures.
Two string vertices $\V_n$ and $\V'_n$ will be
said to be in the same {\it class} if their boundaries coincide as
punctured Riemann surfaces without local coordinates.
We claim that given two
symmetric string vertices $\V_n$ and $\V'_n$ in the same class, there is
a homotopy $\V_n (t)$, such that $\V_n (0)= \V_n\,$, $\V_n (1)=\V'_n\,$,
and for all $t\in [0,1]$, $\V_n(t)$ is a symmetric vertex in the same class.

This homotopy is simple to build when both $\V_n$, and $\V'_n$, define
sections $\sigma$, and $\sigma'$ respectively, over $\M_n$.
Then both vertices determine a common space
$D_n= \pi (\V_n)=\pi(\V'_n)\subset\M_n$,
of labeled punctured surfaces (without local coordinates), and
$\V_n = \sigma (D_n)$ and $\V'_n = \sigma' (D_n)$.
Our aim is to define
a homotopy $\sigma (t)\bigl( \D_n\bigr)\in \widehat\P_n$,
such that $\sigma (0) = \sigma$, $\sigma (1)= \sigma '$, and
$\sigma (t)\bigl( \D_n\bigr)$is symmetric and in the same class for
all $t\in [0,1]$.
First consider a homotopy taking each local coordinate $z_i$ for each
surface on $\sigma (\D_n)$  into the coordinate $k z_i$, with
$k$ a sufficiently large constant so that for each surface
$\sigma(\PP)$ ($\PP\in \D_n$),
the new coordinate curves lie completely
within the corresponding
coordinate curves of the surface $\sigma'(\PP)$.
(The constant $k<\infty$ is guaranteed to exist because $D_n$ is compact).
Since this deformation is independent of the labelling of the punctures
it defines a symmetric deformation manifestly preserving the class of the
vertex.
We can now imagine each surface $\sigma' (\PP)$ as equipped with
two coordinate curves around each puncture; the one arising from
$\sigma (\PP)$ by the above deformation, completely inside the
one defined by the section $\sigma'$. Let $(\C_1 ,\cdots \C_n)$
denote the small curves and $(\C'_1 ,\cdots \C'_n)$ denote the
big curves. We now define the homotopy $(\C_1(t) ,\cdots \C_n(t))$
as follows. Let $m$ be the map taking the annulus determined by the curves
$\C_i$ and $\C'_i$ to the standard annulus $1\leq |z| \leq 2$.
The homotopy is provided by inverse images of the
circles $|z| = 1+t$, for $t\in [0,1]$, namely $\C_i(t) = m^{-1} \{ |z|=1+t\}$.
This clearly gives us a continuous path on the fiber over
$\PP$ from $\sigma (\PP)$ up to $\sigma' (\PP)$. We now define
$\sigma(t)$ to act in precisely this way for any surface on $D_n$.
We claim that $\sigma(t)(D_n)$ is a section, that is, a continuous map
from $D_n$ to $\widehat\P_n$. This should be clear, since nearby surfaces
$\PP ,\PP' \in D_n$ must be mapped to nearby surfaces
$\sigma(t)(\PP ) \,,\sigma(t)(\PP')\in \widehat\P_n$ for any fixed
value of $t$. The sections $\sigma(t)(D_n)$ must be symmetric since
the deformations are done without reference to the labels of the
punctures, it only involves the punctured surfaces and their local
coordinates. It is also clear that for exceptional surfaces with
automorphisms, that is, conformal maps that
exchange punctures, there is no complication. \foot{This implies that
the present argument also applies to two-punctured and three-punctured
spheres, as particular cases.}
Any automorphism of a surface
must correspond to conformal maps that take the annuli considered above
into themselves or into each other. This is because the maps must take
the inner circles into each other and the outer circles into each other.
But any such map must take, in the standard picture of the annulus
as $1\leq |z|\leq 2$, the constant $|z|$ circles into each other.
This shows that our homotopy must respect the automorphisms.

A final point concerns the case when the vertices are not sections.
This is the case, for example, when the projection from (some
subspace of) $\V_n$ to $\M_n$ is many to one.
It is enough to discuss the
case when one of the vertices is a section and the
other is not, since once we know how to construct such symmetric homotopy,
we can find a homotopy between each non-section vertex and a common
(section) vertex of the same class, and by composition we find the
desired homotopy between the non-section vertices.
The idea, for taking a non-section vertex
into a section one goes as follows.
We extend the section vertex arbitrarily  (not even keeping symmetry)
so that the projection of the non-section vertex into $\M_n$ is inside
the projection of the extended section into $\M_n$. We then do exactly
as we did above for every surface in the nonsection vertex, we produce
the two curves around each puncture and construct the homotopy. This
homotopy flattens the non-section vertex over the section vertex. In
this process we obtain folds due to the surfaces that are contained more
than once in the non-section. The
folds now cancel out, since they represent identical spaces with
opposite orientation. This gives us the desired homotopy.

During our analysis, we shall encounter vertices
which are symmetric in all but one of the external legs. Such vertices
can also be deformed into each other maintaining their symmetry. This is
done by following exactly the same procedure as above of deforming the
coordinate curves around all the punctures, including the special one.

\chapter{Formulating the Problem of Background Independence of CSFT}

In this section we shall discuss the issues involved
in proving background independence
of the  string field theory action.
We denote by $x^\mu$ the set of coordinates labelling the moduli space of the
conformal field theories. For each point $x^\mu$ in the moduli space, the
state space $\H_x$ contains the states of all ghost numbers of CFT$_x$
annihilated by $b_0^-$ and $L_0^-$. The state space $\H_x$ is a
symplectic vector space, namely it is endowed with an odd nondegenerate
two-form $\omega_x$. Acting on two vectors $A,B \in \H_x$, we have
$\omega_x (A,B) \sim
\omega_{x\,ji} A^iB^j$,
with $\omega_{x\,ij}$ constants.
The string field $|\Psi_x\rangle$ is an
element of $\H_x$, and the string field master action
$S_x\,(\,\ket{\Psi_x})$ is a map from
$\H_x$ to the space of real numbers\foot{More precisely, since
we are dealing with the master action, it is a map to the space of
even elements of a grassmann algebra, which we will continue to
denote as $R$.}. We have included an extra
explicit dependence on the conformal theory at $x^\mu$ as a subscript
of the action. This takes into account the fact that the construction
of the action makes use of ingredients of the conformal theory in
question, such as the BRST operator and correlators.
As has already been emphasized, in the BV formalism, both, the master
action $S_x$ and the symplectic form $\omega_x$ are crucial in
specifying the theory.

An issue that will play a role at various points of our
discussion is whether the state space $\H_x$ should be thought of
as a vector space or as a manifold \foot{Throughout this paper all vector
spaces and manifolds,
are actually supervector spaces and supermanifolds, respectively.}.
The string field, by definition, is a vector in the state space
$\H_x$. Any point in this state space represents a configuration of the
string field. In conventional field theory, however, the set
of field configurations naturally define a manifold, for example, the space
of metrics in gravity. Therefore it is sometimes convenient to think of
$\H_x$ as a manifold, with the string field defining coordinates on it.
When a vector space is viewed as a manifold, the tangent space at any
point $p$ on the manifold is naturally identified with the vector space
itself. This allows us to define a symplectic form on the {\it manifold}
$\H_x$, from the symplectic form $\omega_x$ on the {\it vector space} $\H_x$.
This symplectic form on the manifold is necessary to be able
to define the antibracket of two functions on the manifold.
Using the natural coordinates induced by the basis vectors of $\H_x$,
we see that the components of the symplectic form that we have obtained
on the manifold $\H_x$ are constants.
The action $S_x$ may now be regarded as a map from the manifold $\H_x$ to $R$.
A general invertible string field redefinition is then naturally thought of as
a diffeomorphism of the manifold $\H_x$ into itself. A particularly relevant
subclass of diffeomorphisms are those that preserve the symplectic structure
on the manifold $\H_x$. Such diffeomorphisms
have featured in the proof that two string field theories formulated on
$\H_x$ but using different string field vertices are physically equivalent
[\hatazwiebach]. We will sometimes
separate out linear maps arising from the general diffeomorphisms and then
it will be useful to use the picture of $\H_x$ as a linear vector space.
\foot{On general grounds, we can expect that the correspondance between
the vector space $\H$ and the manifold of field configurations
holds only locally. It may happen that some points in $\H$ far away from
the origin do not represent allowed configurations, for example, a fluctuation
$h_{\mu\nu}$ of a background metric $\hat g_{\mu\nu}$ making the total
metric negative.  Or it could be that $\H$ actually represents only
a patch in the space of all allowed field configurations. }

\section{The General Conditions for Background Independence}

The question of background independence of  string
field theory may now be formulated as follows. Given two string field
actions
$S_x : \H_x \to R$
and $S_y : \H_y \to R$, formulated around two different conformal
theories $x$ and $y$,  we demand the
existence of a diffeomorphism
$$F_{y,x} :\H_x\to \H_y\, ,\eqn\diffeo$$
between the corresponding spaces $\H_x$ and $\H_y$
such that
$$\eqalign{\omega_x &={F_{y,x}}^*\,\omega_y\,\, ,\cr
\quad S_x &={F_{y,x}}^*\, S_y\,\, ,\cr}\eqn\econstraint$$
with $F_{y,x}^{~~~*}$ denoting the pullback performed using the diffeomorphism
$F_{y,x}$. These equations imply that the diffeomorphism maps
{\it both} the symplectic structure and master action on $\H_y$ to
those in $\H_x$.
The question of background independence is simply the question whether such
symplectic, or antibracket preserving, diffeomorphism exists.

In order to make our discussion more concrete, let us choose a complete set of
basis states $\ket{\Phi_i}$ in $\H_x$ for all values of $x$. The target space
fields $\psi^i_\vx$ are defined to be the
components of the string field along the basis vectors
$$\ket{\Psi_x}=\sum_i \ket{\Phi_i\,} \, \psi^i_x\, .\eqn\efour$$
The string field action $S (  \psi_x ,x)$ is a
function of the string field coordinates $\psi^i_\vx$ and the coordinate $x$
labelling the space of conformal field theories.
Eqs.\diffeo\ and \econstraint\ may now be rewritten as
$$\psi^i_{\vy}=F^i\,(\psi_\vx, \vx, \vy)\, , \eqn\ashoke$$
together with the background independence conditions
$$\eqalign{
S\, (\, \psi_x\, , x\,) &= S\, (\, \psi_y\, , y\,)\, ,\crr
\omega_{ i' j'}(\vx)\,\, &=
{\p_l F^i\over \p \psi_\vx^{i'}}\,\,\omega_{ij}(\vy)\,
 \,{\p_r F^j\over \p \psi_\vx^{j'}} \, ,\cr}\eqn\efive$$
where $\p_r$ and $\p_l$, as usual, denote derivatives from the
right and from the left respectively.
We have not included a $\psi$ dependence in the $\omega_{ij}$'s,
since, as argued above, they are constants in this coordinate system.

\section{Background Independence for Nearby Backgrounds}

Let us now consider the case of nearby conformal field theories corresponding
to the points $\vx$ and $\vx+\delta\vx$. Since we have a vector
bundle there is a notion of smoothness in the choice of basis vectors
throughout theory space. Therefore, an infinitesimal
shift in theory space must require an infinitesimal transformation $F^i$
$$\psi^i_{x+\delta x} = F^i\,(\psi_\vx\, ,\, \vx, \vx+\delta\vx)=
\,\psi^i_\vx\, + \,\delta x^\mu\,\cdot
f^i_\mu\,(\psi_\vx\, , \vx)+ \O (\delta x^2)\,\, ,\eqn\eseven$$
for some function $f^i_\mu$. For $y=x+\delta x$,  equations \efive\ and
\eseven\ demand that
$$\eqalign{
&{\p\omega_{i'j'}(\vx)\over\p x^\mu}+
{\p_l f^i_\mu\over\p\psi^{i'}_\vx}\,\omega_{ij'}(\vx)
+\omega_{i'j}(x){\p_r f^j_\mu\over\p\psi^{j'}_\vx}=0\, .
\cr&{\p S(\psi_\vx, \vx)\over \p x^\mu} +
{\p_r S(\psi_\vx, \vx)\over \p \psi^i_\vx} f^i_\mu  =0,\cr}\eqn\eeight$$
The question of background independence of string field theory under
infinitesimal change of background reduces to the question of existence of
$f^i_\mu(\psi_x\, ,x)$ satisfying equations \eeight.

Let us now define objects $\Gamma_\mu^{~i}$, $\Gamma_{\mu j}^{~~i}$,
and  $h^i_\mu$ by separating out of $f^i_\mu$ the $\psi$ independent part
$-\Gamma_\mu^{~i}$, and the part linear in $\psi$:
$$f^i_\mu\,(\psi_\vx\, , \vx)\,= - \Gamma_\mu^{~i} \, (x) -
 \Gamma_{\mu j}^{~~i} \,(x)\, \psi^j_x\,
\,  - \,  h^i_\mu\,(\psi_x\, , x)\, ,\eqn\exseven$$
where $h^i_\mu$ contains quadratic and higher order terms in $\psi_x$.
As the notation suggests, the linear part of $f^i_\mu$ defines
a connection $\Gamma_{\mu i}^{~~j}(x)$ on the vector bundle of state
spaces over CFT theory space.
We shall restrict to field redefinitions that preserve ghost
number.\foot{Here ghost number refers to the ghost number in the string
field theory in the BV formalism. This is equal to the ghost number of the
state in the first quantised theory minus 2.}
This will imply, among other things, that $\Gamma_{\mu j}^{~~i}$ is
non-vanishing only if the states $\ket{\Phi_i}$ and $\ket{\Phi_j}$ carry
the same ghost number. This also shows that $\epsilon(\Gamma_{\mu
j}^{~~i})=(-1)^{i+j}=1$, which is consistent with the fact the
$\Gamma_{\mu j}^{~~i}$ are real numbers.
Needless to say, covariant derivatives
involving the connection $\Gamma_{\mu j}^{~~i}$ appear when we analyze the
content of our background independence equations.
Upon partial expansion, equations \eeight\ become the equations
$$\bigl( \, \D_\mu (\Gamma )\, \omega \bigr)_{i'j'} \, - \,
{\p_l h^i_\mu\over\p\psi^{i'}_\vx}\,\omega_{ij'}(\vx)\,\,-\,
\omega_{i'j}(x)\, {\p_r h^j_\mu\over\p\psi^{j'}_\vx}=0\, ,\eqn\fpart$$
$$\hbox{D}_\mu \, S- \, {\p_r S\over \p \psi^i}
\, (\, \Gamma_\mu^{~i} + h^i_\mu\, )\, =0,\eqn\yyeight$$
The covariant derivative
$\hbox{D}_\mu$ was defined in \newcov , and
the covariant derivative of $\omega$, is the
standard covariant derivative of sections on a vector bundle defined in
Eqn.\sdervb.
Indeed, being
$\psi$ independent, the symplectic form can
be viewed as a section on the vector bundle.

We shall first analyze the consequences of Eqn.\fpart.
Since $h^i_\mu$ is quadratic and higher orders
in $\psi$, the $\psi^i_x$ independent part of this equation gives
$$ D_\mu(\Gamma)\bra{\omega}=0, \eqn\eanother $$
whereas the $\psi^i_x$ dependent part of this equation gives
$${\p_l h^i_\mu\over\p\psi^{i'}_\vx}\,\omega_{ij'}(\vx)\,\,+\,
\omega_{i'j}(x)\, {\p_r h^j_\mu\over\p\psi^{j'}_\vx}=0\, .\eqn\getsign$$
Since $\omega$ is $\psi$ independent, we can write the above
equation in the following form
$$(-)^{i'j'}\,{\p_r( h^i_\mu\,\omega_{ij'})\,\over\p\psi^{i'}_\vx}\,\,-\,
 {\p_r (h^j_\mu\,\omega_{ji'})\,\over\p\psi^{j'}_\vx}=0\, \eqn\pain$$

It is convenient to use index free notation to appreciate the meaning
of the above equations. We take
$$\ket{h_\mu}=\ket{\Phi_i}\, h^i_\mu = \sum_{N=2}^\infty \,\,
{1\over N!} \, \, \, {}_{(01\cdots N)}
\bra{\,\Gamma_\mu^{(N+1)}}\s_{0e}\rangle\,
\ket{\Psi}_1\cdots \ket{\Psi}_N \, ,
\eqn\emmnine$$
where the object $\bra{\,\Gamma_\mu^{(N+1)}}$ introduced in the
above expansion is a bra in $(\H^*)^{\otimes (N+1)}$. Since $\ket{h_\mu}$
and $\ket{\Psi}$
are even, and $\ket{\s}$ is odd, $\bra{\,\Gamma_\mu^{(N+1)}}$ must be odd.
Its first state space, denoted as `0', has been
contracted with $\ket{\s_{0e}}$, where $e$, for external, is the label
of the resulting state in the left hand side of the equation. By definition,
$\bra{\,\Gamma_\mu^{(N+1)}}$ is symmetric on the state space labels
$1$ to $N$. It is a simple calculation using \emmten\ and \restrinv\
to show that
$$h^i_\mu\, \omega_{ij} = -\sum_{N=2}^\infty \,\,{1\over N!} \, \, \,
{}_{(01\cdots N)}
\bra{\,\Gamma_\mu^{(N+1)}}\Psi\rangle_1\cdots \ket{\Psi}_N\ket{\Phi_j}_0 \, .
\eqn\eine$$
Back in \pain\ we find, that for each value of $N\geq 2$
$$(-)^{i'j'}\,{\p_r\over\p\psi^{i'}}\,\Bigl(
\bra{\,\Gamma_\mu^{(N+1)}}\Psi\rangle_1\cdots \ket{\Psi}_N\ket{\Phi_{j'}}_0
\Bigr)\,\,-\,{\p_r \over\p\psi^{j'}}
\Bigl(\bra{\,\Gamma_\mu^{(N+1)}}
\Psi\rangle_1\cdots \ket{\Psi}_N\ket{\Phi_{i'}}_0 \Bigr)=0\,, \eqn\axin$$
which gives
$$\bra{\,\Gamma_\mu^{(N+1)}}\,{\Psi\rangle}_1\cdots\ket{\Psi}_{N-1}\,\Bigl(
\ket{\Phi_{i'}}_N\ket{\Phi_{j'}}_0
-\ket{\Phi_{i'}}_0\ket{\Phi_{j'}}_N \Bigr)=
0\,. \eqn\xaxin$$
This equation implies that the bra $\bra{\,\Gamma_\mu^{(N+1)}}$ must
be symmetric between its zeroth state space, which was, a priori, on
a different footing, and any other state space. Therefore
$\bra{\,\Gamma_\mu^{(N+1)}}$ must be a {\it totally} symmetric vertex.

To summarize, the conditions that the symplectic form is preserved are
simply
$$D_\mu\, ( \Gamma\, )\, \bra{\,\omega\, }=0 \, ,\quad\hbox{and,}\quad
\bra{\,\Gamma_\mu^{(N+1)}} \in {\bf S}\, \bigl(\, \H^{*\otimes(N+1)}
\bigr) \, .\eqn\emmtwelve$$

We now analyze the consequences of Eqn.\yyeight.
The $\psi$ independent part of the diffeomorphism is given by
$$\ket{\Phi_i}\,\Gamma_\mu^{\, i}\,=\ket{\, \widehat\O_\mu\, }
\equiv\, {}_0\bra{\Gamma_\mu^{(1)}}\s_{0e}\rangle  \, ,\eqn\emmthirteen$$
where we have introduced, in analogy to \emmnine\ a bra
$\bra{\Gamma_\mu^{(1)}}\in \H^*$.
This equation indicates that
the $\psi$ independent part of the diffeomorphism
is the classical
solution representing the theory at $x+\delta x$,
in the string field theory formulated around the background $x$.
We can now rewrite Eqn.\yyeight\  more clearly as
$$\hbox{D}_\mu \, S- \, {\p_r\, S\over \p\, \ket{\Psi}}
\, \bigl(\, \ket{\widehat\O_\mu} + \ket{h_\mu\,} \bigr) \,
=0,\eqn\xzyeight$$

Let us now derive the explicit conditions arising from the above equation.
Making use of \specfun\ and \emmeight\ it follows that
$${\rm D}_\mu S=\sum_{N=2}^\infty {1\over N!}(D_\mu\bra{V^{(N)}})
\ket{\Psi}_1\cdots\ket{\Psi}_N \, .\eqn\ezzone$$
We then find that the terms quadratic in $\ket{\Psi}$ in Eqn.\xzyeight\ give,
$$D_\mu(\Gamma)\, \bigl( \bra{\omega_{12}}\, Q^{(2)} \, \bigr)
= \, \bra{V^{(3)}_{123}\,}\,\widehat\O_\mu\rangle_3\, .
\eqn\emmfourteen$$
and terms involving higher powers of $\ket{\Psi}$ give,
$$\bra{\Gamma_\mu^{(N)}} \sum_{i=1}^N Q^{(i)} \, = \, -\, \sum_{m=3}^{N-1}
{\bf S}\Big(\bra{\Gamma_\mu^{(N-m+2)}}\, \bra{V^{(m)}}\s\rangle\Big) \,+\,
D_\mu(\Gamma )\bra{V^{(N)}}\,
- \bra{V^{(N+1)}}\widehat O_\mu\rangle \, .\eqn\emmfifteen$$
In the first term on the right hand side of the above equation $\ket{\s}$
sews any one of the $(N-m+2)$ legs of $\bra{\Gamma^{(N-m+2)}}$ to any one
of the $m$ legs of $\bra{V^{(m)}}$.
This gives a bra in $(\H^*)^{\otimes N}$ which is symmetric in its
first $N-m+1$ legs and also in the last $m-1$ legs.
As it was the case below Eqn.\geom,
${\bf S}$ denotes complete symmetrization of this bra. The total
number of terms in ${\bf S} ( \cdots ) $ is  ${N\choose m-1}$, which
is the number of ways of splitting the $N$ external labels into two sets,
one to be attached to $\bra{V}$, and one to be attached to $\bra{\Gamma_\mu}$.

In the next three sections
we shall see how to obtain the connection $\Gamma_\mu$
and symmetric $\bra{\Gamma_\mu^{(N)}}$'s for $N\ge 3$
satisfying the conditions of background independence expressed
in equations \emmtwelve, \emmfourteen\ and \emmfifteen. The diffeomorphism
implementing background independence will be given by
$$\ket{\Psi}_{x+\delta x} = {}_{x+\delta x}\,{{\cal I}_{}}_x \, \Bigl[
\,\ket{\Psi} -\delta x^\mu \Bigl(\,
\Gamma_\mu \ket{\Psi}+ \sum_{N\geq 0\atop N\not= 1}\,\,
{1\over N!} \, \, {}_{(01\cdots N)}
\bra{\,\Gamma_\mu^{(N+1)}}\s_{0e}\rangle\,
\ket{\Psi}_1\cdots \ket{\Psi}_N \, \Bigr)\,\Bigr] \, . \eqn\emxe$$
This equation follows from our definitions \exseven, \emmnine, and
\emmthirteen . Note that we have incorporated the $\psi$ independent
shift into the sum (the $N=0$ term) but not the linear term. As
we will see later, part of the connection $\Gamma_\mu$ will be
incorporated into the sum.
Since the left hand side is
a string field at $x+\delta x$, while the input $\ket{\Psi}$ in the
right hand side is a string field at $x$, we have included the
``copying'' operator  ${}_x{\cal I}_y =\sum_i
\ket{\Phi_i(x)}\bra{\Phi^i(y)}$, which is the operator that copies a state
from one state space to another one. The copying operator, acting on
a vector in one state space, gives a vector in another state space, with
the same value for all of its components.

\chapter{Background Independence to Quadratic and Cubic Order}

We have derived in the previous section the explicit conditions
for background independence. The diffeomorphism relating the
two theories must be symplectic, that is, it should preserve the
BV structure, and we have observed that the linear part of the
diffeomorphism has the index structure of a connection. In the
present section we will study the first three conditions for
background independence, namely
$$\eqalignno{
D_\mu\, ( \Gamma\, )\, \bra{\,\omega\, }&=\,\,0 , &\eqname\one\cr
D_\mu(\Gamma)\, \big( \bra{\omega_{12}}\, Q^{(2)} \big) \,
&= \, \bra{V^{(3)}_{123}\,}\,\widehat\O_\mu\rangle_3\, , &\eqname\two\cr
\bra{\Gamma_\mu^{(3)}} \sum_{i=1}^3 Q^{(i)} \, -\,
D_\mu(\Gamma )\bra{V^{(3)}}\,
&= -\bra{V^{(4)}}\widehat O_\mu\rangle \, .&\eqname\xteen\cr}$$
We will show how to solve for the connection $\Gamma_\mu$, and for the
three string vertex $\bra{\Gamma_\mu^{(3)}}$. It will then be
a simple matter to extend the discussion to all orders. This will
be done in the next two sections.

\section{Finding the String Field Theory Connection $\Gamma$}

We now show how to obtain a connection $\Gamma_\mu$
satisfying Eqns.\one--\xteen. The middle equation, Eqn.\two , that fixes the
covariant derivative of the BRST operator in terms of the three
string vertex contracted with the marginal operator, will be
our main input. As we will see, this equation does not determine
the connection completely.  Nevertheless, we will write a geometrical
expression that solves equation \two . It will then be
straightforward to see that equation \one\ is satisfied. The
third equation, which in fact, determines part of the connection
could give rise to an inconsistency. We explain why this does not happen,
and how this last equation can be used to solve for
$\bra{\Gamma_\mu^{(3)}}$.

The main observation that leads to the solution for $\Gamma$
is that the canonical connection $\widehat\Gamma$ satisfies rather
similar equations. We have (see \ethreetwo, \ethreenewa\ and \dervnzero )
$$\eqalignno{
D_\mu\, (\,\widehat\Gamma\, )\, \bra{\,\omega\, }&=\,\,0 , &\eqname\oone\cr
D_\mu(\,\widehat\Gamma\,)\, \bra{\omega_{12}}\, Q^{(2)} \,
&= \, \bra{V'^{(3)}_{123}\,}\,\widehat\O_\mu\rangle_3\, , &\eqname\ttwo\cr
D_\mu(\,\widehat\Gamma \,)\bra{V^{(3)}}\,
&= 0. &\eqname\xxteen\cr}$$
These equations indicate that it should be simpler to try to find the
difference between the two connections. We therefore introduce
the operator one form $\Delta\Gamma_\mu$
as the difference between the connection $\Gamma_\mu$ and the
canonical connection $\widehat\Gamma_\mu$
$$\Delta\Gamma_\mu=\Gamma_\mu -\widehat\Gamma_\mu .\eqn\ethreefour$$
Associated to the operator $\Delta\Gamma_\mu$ it is convenient
to introduce the bra
$\bra{\Delta\Gamma_\mu}\in \H^*\otimes\H^*$ as
$$\bra{\Delta\Gamma_\mu\,(1,2)}=\bra{\,\omega_{12}\, }\,
(\Delta\Gamma_\mu)^{(1)}\, .
\eqn\ethreefive$$
The difference between Eqns.\one\ and \oone\ gives
$$\bra{\omega_{12}}\Bigl( \Delta\Gamma_\mu^{(1)}+\Delta\Gamma_\mu^{(2)}\,\Bigr)
=0 \quad\to\quad\bra{\Delta\Gamma_\mu\,(1,2)}-
\bra{\Delta\Gamma_\mu\,(2,1)}=0\, ,\eqn\ethreesix$$
which is simply the condition that $\bra{\Delta\Gamma_\mu}$ is symmetric.
The covariant constancy of $\bra{\omega}$ implies that
equation \two\ can be written as
$$\bra{\omega_{12}}\,\Bigl( \partial_\mu Q  +
[ \Gamma_\mu \, , Q ] \Bigr)^{(2)}\,
= \, \bra{V^{(3)}_{123}\,}\,\widehat\O_\mu\rangle_3\, , \eqn\tttwo$$
Subtraction of the similar equation following from \ttwo\ gives
$$\bra{\omega_{12}}\,\bigl[ \Delta\Gamma_\mu \, , Q \bigr]^{(2)}
=\,-\,\bigl(\,\bra{V_{123}^{\prime(3)}} -\bra{V_{123}^{(3)}}\, \bigr)
\,\,\ket{\widehat\O_\mu}_3\,\,.\eqn\esevenaaa$$
Making use of $\bra{\omega_{12}}(Q^{(2)}+Q^{(1)})=0$, and \ethreefive\  we find
$$\bra{\,\Delta\Gamma_\mu\,}\, (\, Q^{(1)}+Q^{(2)}\, )
=\bigl(\,\bra{V_{123}^{\prime(3)}} -\bra{V_{123}^{(3)}}\, \bigr)
\,\,\ket{\widehat\O_\mu}_3\,\,.\eqn\ethreeseven$$
We shall now show how to find a $\bra{\Delta \Gamma_\mu}$ satisfying
eqs.\ethreesix\ and \ethreeseven.

The right hand side of \ethreeseven\ shows the difference between
two three string vertices. The two three-string-vertices represent
two different points $\V_3$ and $\V'_3$ in the space $\wh\P_3$.
Since the vertices
are contracted with the marginal operator, they effectively behave
as two-string vertices, {\it i.e.} the right hand side of \ethreeseven\
belongs to $\H^*\otimes\H^*$. Both,
$\bra{V_{123}^{\prime(3)}}\widehat\O_\mu\rangle_3$ and
$\bra{V_{123}^{(3)}}\widehat\O_\mu\rangle_3$ are symmetric two string
vertices. Let $\B_3$ be a path in $\wh\P_3$ representing a symmetric
homotopy between $\V_3$ and $\V'_3$, namely, every point of $\B_3$
is a three string vertex with a special puncture, and symmetric under
the exchange of the other two punctures. This homotopy can be constructed
as explained in \S3.4. We therefore have
$$\p\B_3 = \V'_3 - \V_3 .\eqn\ekakatwo$$
We now claim that
$$\bra{\Delta\Gamma_\mu}=\,-\,\int_{\B_3}\bra{\,
\Omega^{(1)3}\,}\wh\O_\mu\rangle_3
\,,\eqn\ethreetwentyone$$
is the solution to Eqs.\ethreesix\ and \ethreeseven . Here
$\bra{\Omega^{(1)3}}$ (see \S2.1) is a one form in $\wh\P_3$. Since
the interpolation path is a symmetric homotopy, $\bra{\Delta\Gamma_\mu}$
satisfies \ethreesix . We now verify \ethreeseven\ is also satisfied
$$\eqalign{
\bra{\Delta\Gamma_\mu}\,(Q^{(1)}+Q^{(2)}\,)
&= -\int_{\B_3}\bra{\,\Omega^{(1)3}\,}\sum_{i=1}^3 Q^{(i)}\ket{\wh\O_\mu}_3
= \,\int_{\B_3} \hbox{d}\,\bra{\Omega^{(0)3}}\, \wh\O_\mu\, \rangle_3\, \cr
&=\,\int_{\p\B_3} \bra{\Omega^{(0)3}}\, \wh\O_\mu\, \rangle_3\,
=\,\int_{\V'_3} \bra{\Omega^{(0)3}}\, \wh\O_\mu\, \rangle_3\,
-\int_{\V_3} \bra{\Omega^{(0)3}}\, \wh\O_\mu\, \rangle_3\,\cr
&=\bra{\,V_{123}^{\prime(3)}\,}\widehat\O_\mu\rangle_3
-\bra{\,V_{123}^{(3)}\,}\widehat\O_\mu\rangle_3\, .\cr}\eqn\ethreenine$$
where use was made of \qixbu\ and of \ekakatwo . This proves
that \ethreeseven\ is satisfied.
(Since $\V_3$ ($\V'_3$) refers to a single point in $\wh\P_3$,
$\int_{\V_3}$ ($\int_{\V'_3}$) in the
second line of the equation simply denotes that the integrand needs to
be evaluated at that point.)
We shall call $\B_3$
(and the corresponding state $\bra{B^{(3)}}\equiv
\int_{\B_3}\bra{\Omega^{(1)3}}$)
the interpolating three-string vertex.

\subsection{Ambiguities in $\Gamma_\mu$}
The result in \ethreetwentyone\ implies an obvious ambiguity in the
connection $\Gamma_\mu$ arising from the possibility of using two
different homotopies $\B_3$ and $\B'_3$ between the initial and
final three string vertices. Given two such homotopies we can find
a two dimensional region ${\cal D}_3$ in $\wh\P_3$ such that
$\partial {\cal D}_3 = \B'_3 - \B_3$.  Therefore, if we let
$$\bra{\Delta\Gamma_\mu}=-\int_{\B_3}\bra{\,\Omega^{(1)3}\,}\wh\O_\mu\rangle_3
\,,\quad\hbox{and}\quad\bra{\Delta'\Gamma_\mu}=
-\int_{\B'_3}\bra{\,\Omega^{(1)3}\,}\wh\O_\mu\rangle_3 \,,\eqn\yone$$
we then have that
$$\eqalign{
\bra{\Delta'\Gamma_\mu}-\bra{\Delta\Gamma_\mu}&=
\Bigl( -\,\int_{\B'_3}+\int_{\B_3}\,\Bigr)
\bra{\,\Omega^{(1)3}\,}\wh\O_\mu\rangle_3
=\,-\,\int_{{\cal D}_3} \hbox{d}\,\bra{\,\Omega^{(1)3}\,}\wh\O_\mu\rangle_3\cr
&=\,-\,\int_{{\cal D}_3}\,\bra{\,\Omega^{(2)3}\,}\wh\O_\mu\rangle_3\,\,
(Q^{(1)}+Q^{(2)}) \,,\cr}\eqn\yone$$
where use was made of \qixbu\ in the last step. This ambiguity, of the
form $\bra{\eta\,}(Q^{(1)}+Q^{(2)})$, could be expected from \ethreeseven\
and reflects the fact that the field redefinition that maps $S_x$ to
$S_{x+\delta x}$ is determined only up to a gauge transformation.
\foot{The ambiguity is not itself ambiguous, the right hand side of
\yone , by virtue of $Q^2=0$, does not depend on the chosen homotopy
${\cal D}_3$.}

\section{Constraints on $\Gamma_\mu$ and solving for $\bra{\Gamma^{(3)}_\mu}$}

Let us now turn to equation \xteen. Consider the second term in the
left hand side and use \xxteen\ to write it as
$$
D_\mu(\,\Gamma\,)\bra{V^{(3)}} =
D_\mu (\,\widehat\Gamma\, +\Delta\Gamma_\mu\,)\bra{V^{(3)}}=
 - \sum_{i=1}^3 \bra{V^{(3)}} \Delta\Gamma_\mu^{(i)}\, . \eqn\elalaone$$
Consider any one term in the final right hand side, for example,
$\bra{V^{(3)}_{123}\,} \Delta\Gamma_\mu^{(1)}$. Using \ethreefive\ and
\restrinv , this term can be rewritten as
$\bra{V^{(3)}_{1'23}\,}\bra{\Delta\Gamma_\mu (1,0)}\s_{01'}\rangle$,
where we get the geometrical picture of twist-sewing the three
string vertex to the vertex $\bra{\Delta\Gamma_\mu}$.
Back in the \elalaone\ we obtain
$$D_\mu(\,\Gamma\,)\bra{V^{(3)}} = -{\bf S}
\Big( \bra{\Delta\Gamma_\mu}\bra{V^{(3)}} \s\rangle \Bigr)\,, \eqn\ozone$$
where, as usual, {\bf S} denotes symmetrization on the external legs
(one out of $\bra{\Delta\Gamma_\mu}$ and two out of $\bra{V^{(3)}}$) requiring
a total of three terms. Using this result in \xteen\ we obtain
$$\bra{\Gamma_\mu^{(3)}} \sum_{i=1}^3 Q^{(i)} =-
{\bf S}\Big( \bra{\Delta\Gamma_\mu}\bra{V^{(3)}} \s\rangle \,\Bigr)
 -\bra{V^{(4)}}\widehat O_\mu\rangle \, .\eqn\elalatwo$$
We now notice that, if we contract both sides of the above equation with
three BRST invariant states, the left hand
side of the equation vanishes identically. The equation then
imposes a constraint on $\bra{\Delta\Gamma_\mu}$. We must show
that this constraint is satisfied identically by the expression for
$\bra{\Delta\Gamma_\mu}$ given in Eqn.\ethreetwentyone. In fact, understanding
why this constraint is satisfied, holds the key to solving the equation
for the unknown $\bra{\Gamma_\mu^{(3)}}$.
Using the expression \ethreetwentyone\ we may rewrite Eqn.\elalatwo\ as
$$\eqalign{
\bra{\Gamma_\mu^{(3)}} \sum_{i=1}^3 Q^{(i)} \, =  &  \,
{\bf S} \Big(\int_{\B_3}\bra{\Omega^{(1)3}}
\int_{\V_3}\bra{\Omega^{(0)3}} \s\rangle \,\ket{\wh\O_\mu}\Big)
-\bra{V^{(4)}}\widehat \O_\mu\rangle \cr
= & \, \int_{\V'_4} \bra{\Omega^{(0)4}}
\wh \O_\mu\rangle_4 -
\int_{\V_4} \bra{\Omega^{(0)4}}
\wh\O_\mu\rangle  \,, \cr } \eqn\elalathree$$
where the region  $\V'_4$ is given by
$$\V'_4 = {\bf S}(\B_3\times \V_3) \, .\eqn\elalafour$$
In deriving the second line of Eqn.\elalathree\ we have used the sewing
property \sewing . The first term in the right hand side
of \elalathree\  denotes an integral identical to the second one, but over
a different region $\V'_4$. Note that in \elalafour\ the special puncture,
where $\wh\O_\mu$ is inserted, is always on $\B_3$
(which we define to be the fourth leg of the vertex $\V'_4$);
the symmetrization involves
only the other legs (yielding three terms). The set of surfaces in
$\V'_4$ corresponds to three Feynman diagrams built by twist-sewing the
interpolating three string vertex $\B_3$ and the three string vertex $\V_3$.
We shall show that, remarkably, apart from the local coordinates,
the set of surfaces $\V'_4$ coincides exactly with the set of surfaces
in $\V_4$. This guarantees that, when we contract \elalathree\ with
three physical states, represented by dimension zero primaries, the right hand
side vanishes, as desired. When the representatives for the physical states
are arbitrary, differing from dimension zero primaries by BRST
exact states, the right hand side vanishes because of the additional
feature that the boundaries of $\V'_4$ and $\V_4$ agree precisely (except for
the local coordinate at the special puncture, $-$ more on this later).
These facts
will allow us to calculate $\bra{\Gamma_\mu^{(3)}}$.
Let us therefore explain how $\V'_4$ turns out to be so special.

\if y\figcount

\midinsert
$$\epsffile{szfig2.ps}$$
\nobreak
\narrower
\singlespace
\noindent
Figure 2. To the left we show the moduli space $\M_4$ of four
punctured spheres. The three shaded regions correspond to the
surfaces generated by the three Feynman graphs contributing to
this amplitude. The three closed string vertex $\V_3$ is shown
to the right. Since it is a three punctured sphere it can be
identified with $\M_4$.
\medskip
\endinsert

\else\fi

\subsection{Restoring Jacobi Identity} It is useful to recall how
the standard vertex $\V_4$ arises. If we take two three-string-vertices $\V_3$,
and form the three standard Feynman graphs suitable for four-string
amplitudes, we do not cover $\M_4$. With $\M_4$ thought itself as
the complex plane (compactified at infinity) and punctured at $z=0,1$,
and $\infty$, the three Feynman diagrams cover three nonoverlapping
disks around $0,1,$ and $\infty$, as shown in Fig.~2. In this
representation, any point $z$ in the plane represents a four-punctured
sphere, punctured at $0,1,z,$ and $\infty$. The region missed, not
shaded in the figure, represents the surfaces in $\V_4$. This diagram
does not tell, however, how to choose local coordinates on the punctures
of the missing surfaces {\it inside} the region $\V_4$,
although it does tell us how to choose them on the boundary of $\V_4$.
The boundaries of the three disks, which
coincide with the boundary of $\V_4$, correspond to twist-sewing of
two three string vertices.

Let us now figure out what region of
moduli space is obtained with the three diagrams of ${\bf S}(\B_3\times
\V_3)$.
To this end it is easiest to examine the boundaries of the regions.
Since twist sewing does not introduce boundaries, and $\V_3$ is a point,
the boundaries  arise from
$$\p \bigl( \, {\bf S}(\B_3\times \V_3 )\bigr) = {\bf S}\,\bigl(
(\p\B_3)\times\V_3\bigr)  =
{\bf S}\,(\V'_3\times \V_3 )-
{\bf S} (\V_3\times \V_3 )\, ,\eqn\checkb$$
where use was made of \ekakatwo .
The boundary $-{\bf S}(\V_3\times\V_3)$
corresponds to the configurations arising from twist-sewing of two
three-string vertices. For a given Feynman diagram, they
coincide with the boundary of a shaded disk in the figure. This
boundary coincides precisely with $\p\V_4$ as the recursion
relation \geom\ indicates (the factor of one-half is absent because the
special puncture breaks the symmetry leading to double counting in \geom ).
How about
the boundary ${\bf S}(\V'_3\times\V_3)$? In each of the three Feynman
diagrams contributing to this boundary, the special puncture of
$\V'_3$ is not sewn. The sewn configuration can be viewed as
a new copy of $\V_3$ coupling the
two free legs of $\V_3$, and the free leg of $\V_3'$, with the special
puncture of $\V'_3$ landing on the coordinate curve of
the puncture that comes from  $\V_3'$. The twist makes the special puncture
travel around that coordinate curve.  For the other two Feynman graphs,
the special puncture will land on the coordinate curves of the other two
punctures of the final vertex $\V_3$.
Since the three string vertex $\V_3$ is
an overlap (the coordinate curves coincide two at a time),
the three boundaries cancel out as the special puncture
travels each piece of the coordinate curves in opposite directions.
It is fun to see where in $\M_4$ this cancellation is taking place.
For this purpose, we identify the three punctured
sphere $\V_3$ with $\M_4$, by thinking of $z=0,1,$ and $\infty$,
as the three punctures of $\V_3$ (see Fig.~2).
The coordinate curves are then nothing else but the familiar lines
joining points $A$, and $B$ of the figure. Therefore, each Feynman
diagram of ${\bf S}(\B_3\times\V_3)$ covers the region interpolating from
the boundary of a shaded disk up to the coordinate curve. The three
diagrams together cover the missing region $\V_4$. The cancellation
of the boundaries ${\bf S} (\V_3'\times \V_3)$
is due to the tight fit of the three regions. Clearly
the vertex $\V'_3$ has nice properties with respect to the vertex
$\V_3$, the different channels agree. We refer to this as $\V'_3$
restoring a Jacobi identity to $\V_3$.

The above  cancellation of boundaries, if it is to happen in $\widehat\P_4$,
requires careful consideration of the local coordinate on the
special puncture. The above argument proves that the boundaries
cancel if we ignore the local coordinate on the special puncture.
This is really all we need for the present application, since
$\wh\O_\mu$ is inserted there. Thus $\V'_4$ and $\V_4$ have the same
boundaries, if we ignore the local coordinate at the special puncture.
This shows that their projections to $\M_4$ have the same boundaries,
and are therefore identical.\foot{Since $\V'_4$ need not be a section,
when projecting down to $\M_4$ one should remember that surfaces
produced more than once cancel out in pairs.} We will show in \S6
that remarkably, it is simple
to choose the coordinate on the special puncture in $\V'_3$ such that
the above cancellation takes place fully off-shell (on $\wh\P_4$). We
will therefore have that the boundaries of $\V'_4$ and $\V_4$ agree
strictly.

\subsection{Determination of $\bra{\Gamma^{(3)}_\mu}$}
We shall now show how to construct $\bra{\Gamma^{(3)}_\mu}$ satisfying
\elalathree. We give the construction for the case where local coordinates
on the special puncture have been chosen so that, as a result,
$\p\V'_4=\p\V_4$ strictly (see \S6). The case when we do not choose
coordinates on the special puncture is treated exactly analogously.

Since $\V'_4$ and $\V_4$ are symmetric (in three legs)
closed string vertices  with
common boundary, there is a symmetric homotopy between the two
vertices keeping the boundary fixed (see \S3.4). Let $\B_4$ denote the
subspace generated by the homotopy. It then follows that
$$\p\B_4=\V'_4 -\V_4 ,\eqn\elalasixteen$$
and then
$$\bra{\Gamma_\mu^{(3)}} = \,-\, \int_{\B_4}
\bra{\Omega^{(1)4}}\wh\O_\mu\rangle \, , \eqn\elalafifteen$$
provides the desired  solution of \elalathree. Since $\B_4$ is a
symmetric homotopy the bra $\bra{\Gamma_\mu^{(3)}}$ is symmetric in
its three state spaces, as required to have a symplectic diffeomorphism.
We refer to $\B_4$ (and the corresponding state $\bra{B^{(4)}}
\equiv\int_{\B_4}\bra{\Omega^{(1)4}}$) as the interpolating four-string vertex.

If no coordinate is chosen at the special puncture, one must introduce
a projection $\pi_f : \wh\P_n\to \wh\P'_n\,$, with $\wh\P'_n$ the space
where the local coordinate around the special puncture is forgotten.
As long as we integrate objects contracted with $\ket{\wh\O_\mu}$, our
integrals  can be thought as integrals on $\wh\P'$. Since
the $\pi_f$ projections of $\V'_4$ and $\V_4$ have the same boundary,
and define symmetric vertices, there is a symmetric homotopy
$\W_4\in \wh\P'_4$ interpolating between them, and
$\bra{\Gamma_\mu^{(3)}} = \int_{\W_4}\bra{\Omega^{(1)4}}\wh\O_\mu\rangle$
provides the desired solution.

\chapter{New Vertices and Identities}

In this section we will study the new geometrical structures
relevant to the problem of background independence. We will
develop the results beyond what is strictly necessary for
the problem of background independence, as studied in the
present paper. Our results suggest that
the new family of vertices are relevant for studying
shifts of the string field that are completely general, namely,
do not correspond to dimension zero primary fields.
We begin by a detailed analysis of the asymmetric
three punctured sphere $\V'_3$.
We show that, the condition of off-shell exchange symmetry
between the two symmetric punctures of $\V'_3$,
constrains the local coordinate on the
asymmetric puncture enough, to guarantee
{\it off-shell} consistency properties for $\V'_3$, with
respect to the polyhedral closed string vertices. We then
explain how $\V'_3$ can be used, in conjuction with the standard
closed string vertices, to produce a good cover of the moduli
spaces $\M_n$ of punctured spheres.

\section{The Three Punctured Sphere ${\V\,}'_3$}

In the previous section we made use of a special three punctured
sphere $\V'_3$. This sphere, using the uniformizing coordinate $z$, is
punctured at $z=0$, with a local coordinate $z_1(z) = z$, and punctured at
$z=\infty$, with local coordinate $z_2 = 1/z$. The asymmetric
puncture is located at $z=1$. This
is the puncture where the marginal field is inserted, making
the local coordinate $z_3$ at this puncture irrelevant. We now want to
fix this coordinate in order to achieve the strongest possible
identities.

In the same way as the three punctured sphere $\V_3$ is used to define
the string field product $[\,\, ,\,\, ]$, we can use the yet-to-be
fully specified
sphere $\V'_3$ to define a new product $[\,\,,\,\,]'$ as follows
$$[ A , B ]' \equiv \bra{{V'}^{(3)}_{123}}\, \s_{\,3e}\rangle\,\ket{A}_1\,
\ket B_2  \, .\eqn\dsymprod$$
Here the states $A$, and $B$, are inserted on the punctures at
zero, and at infinity, respectively, and the product comes
out of the asymmetric (third) puncture. The state space of the product
is labelled by $e$, for external.
It is natural to demand that this product be symmetric, or graded commutative,
$$ [ A , B]' = (-)^{AB} \,[ B , A]'. \eqn\symprod$$
This condition will restrict the choice of
coordinate at the third puncture.

We discussed in \S5 the nice interplay between the three string vertex $\V_3$
of closed string field theory and $\V'_3$. That interplay can be
summarized in the following relation
$$\Bigl[ \,A_1 , \,[ A_2 , A_3 \, ] \,\Bigr]'(-)^{A_1(1+A_3)} \,
+ \,\hbox{cyclic}\, =\,0 ,\eqn\miden$$
which was guaranteed by the pictures to hold on shell, that is, the
left hand side vanishes when contracted with dimension zero primary
states.
It is clearly
of interest to know if the coordinate $z_3$ of the sphere $\V'_3$
can be chosen so that the above identity holds strictly. The surprising
result we will prove is that, the condition on $z_3$ required by
graded commutativity (\symprod), is actually sufficient to guarantee
that \miden\ holds strictly. Equation \miden\ is a curious variation
on the usual Jacobi identity of a Lie bracket. In a homotopy Lie algebra
we have a bracket that satisfies a Jacobi identity weakly; here we
have found another bracket with a curious compatibility
with the original bracket. Furthermore, the product $[\,,\,]'$ also
satisfies remarkable properties with respect to the higher string
vertices. Again, if \symprod\ holds, we can show that
$$\Bigl[ \, A_1 ,[\, A_2 , \cdots  , A_n \,] \,\Bigr]' \,\pm  \,
\hbox{cyclic} =0 .\eqn\xiden$$
This identity (in fact, only its on-shell version)  will be necessary
to complete our proof of background independence to higher orders. The sign
factors can be written out explicitly using Eqn.(4.14)
of Ref.[\zwiebachlong]. If all $A$'s are even, all terms in \xiden\
have a plus sign.

The geometrical version of the above identities is summarized by
the expression
$${\bf S} \bigl( \V'_3\times\V_M \bigr) =0\, , \quad M\geq 3\,,\eqn\xgeom$$
which indicates that the various subspaces of $\wh\P_n$ arising from
twist-sewing $\V'_3$ to the surfaces in $\V_M$ add up to zero. As usual,
${\bf S}$ symmetrizes over the free puncture of $\V'_3$ and the free
punctures of $\V_M$ (the special puncture of $\V'_3$ is not symmetrized over).

As an aside, we observe that the $\V'_3$ sphere could  be used to define
another product
$$ A \circ B = \bra{{V'}^{(3)}_{123}}\,
\s_{\,1e}\rangle\,\ket{A}_2\,\ket B_3  \, , \eqn\otherp$$
where the first state in the product is inserted on the second puncture,
the second state is inserted on the third puncture (the asymmetric puncture),
and the product comes out of the first puncture. It is clear that this
product is not symmetric
$$ A \circ B \not= (-)^{AB} B \circ A . \eqn\nosymm$$
This product may be eventually useful to write some new interactions
for string fields (see \S9), but will not be explored systematically here.

\section{The Local Coordinate in the Asymmetric Puncture of $\V'_3$}

The sphere $\V'_3$, using the uniformizing coordinate $z$, is
punctured at $z=0$, with a local coordinate $z_1(z) = z$, and punctured at
$z=\infty$, with local coordinate $z_2 = 1/z$. The asymmetric
puncture, whose local coordinate $z_3$ has not been fixed, is located at $z=1$.
It is natural to demand that the manifest exchange symmetry between the
local coordinates at zero and infinity be respected by the asymmetric
puncture.  The map $z\to 1/z$,
which exchanges $z_1$ and $z_2$, indeed leaves the asymmetric puncture
at $z=1$ fixed, but we need more. Points near this puncture must be
mapped to points having the same $z_3$ coordinate,
up to a common constant phase.
Since the
exchange map squared is the identity, the phase must be simply $(\pm 1)$.
Moreover, since the map, near $z=1$, acts as a reflection, the
local coordinate cannot remain unchanged and we must choose
the minus sign. We therefore require
$$z_3 (z) = - z_3 (1/z) . \eqn\symconvp$$
It is not hard to parametrize the most general solution of the above equation.
Since $z_3$ must vanish at $z=1$, we can, without loss of
generality, write
$$z_3(z) = f_O\, \Bigl(\, {z-1\over z+1}\, \Bigr) \, \equiv h(z)\, ,
\eqn\solvez$$
where $f_O$ is a function that vanishes at zero, and, is one to one inside a
disk around the origin. Thus, for small $y$, $f_O(y)\propto y$.
Eqn.\symconvp\ implies that the function $f_O$ must be odd.
We see that the condition
of off-shell symmetry, under the exchange of the two symmetric punctures,
leaves the coordinate at the asymmetric puncture fairly unconstrained.

Nevertheless, it is not possible to make a choice
of $f_O$ such that the vertex $\V'_3$ becomes cyclic, as it is simple
to prove that the cyclic map $T(z) =1/(1-z)$ cannot map the local
coordinate $z_1$ into the local coordinate $z_3$. It should also be
no surprise that there is no choice of $f_O$ that makes this vertex
fully symmetric on the three punctures. The simplest possible choice
of $f_O$, namely, the identity function, makes $\V'_3$ a projective
vertex. It should be clear from our analysis of three punctured
spheres in \S3.4, that any coordinate around the asymmetric
puncture that could, in principle, be extended to a fully symmetric
vertex by choosing related coordinates at zero and at infinity, is
a consistent choice for the coordinate in the asymmetric puncture of
$\V'_3$. This means, for example, that the asymmetric puncture could
simply keep the local coordinate of the closed string vertex $\V_3$.

\if y\figcount

\midinsert
$$\epsffile{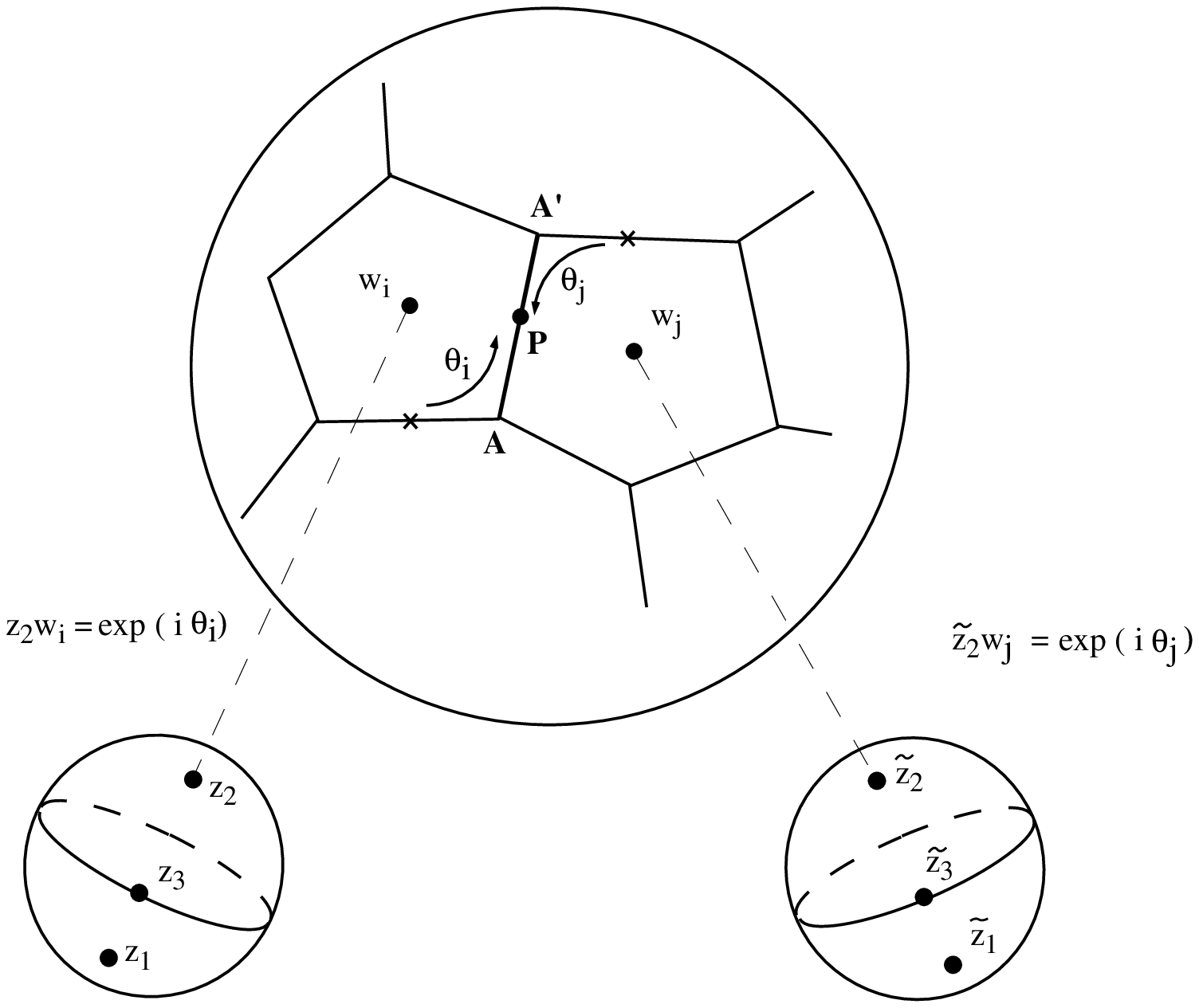}$$
\nobreak
\narrower
\singlespace
\noindent
Figure 3. This figure is used to derive the conditions on the local
coordinate at the asymmetric puncture of $\V'_3$ in order for equations
\miden, and \xiden\ to hold off-shell. We explore what happens when
the asymmetric puncture lands on the edge $AA'$ of the polyhedron in
two different ways.
\medskip
\endinsert

\else\fi

We now claim that off-shell exchange symmetry of $\V'_3$ (condition \symconvp),
is all we need to get equations \miden, and
\xiden\ to hold off-shell.
As discussed in \S5.2 the
overlap nature of the three string vertex is responsible for the
on-shell version of \miden . Since all the higher string vertices, the
restricted polyhedra, are also contact interactions, we will treat
the general situation. We first need to make a preliminary
observation. Consider an arbitrary polyhedron, as shown in Fig.~3 ,
and let $w_i$ and $w_j$ be the local coordinates in two adjacent faces
which share the edge $AA'$ of the polyhedron. Let us now show that,
in a neighborhood of this edge, the coordinates have the transition
function $w_i\,w_j = \exp (i\phi)$, for some phase $\phi$. This need not
be true for a generic contact interaction, but happens here because
the closed string theory contact interactions arise from Jenkins-Strebel
quadratic differentials [\saadizwiebach ]. We can think of the
interaction represented by the polyhedron, as having semiinfinite cylinders
(with metrics) of circumference $2\pi$, whose boundaries are glued
{\it isometrically} following the instructions of the polyhedron. This
means that every edge of the polyhedron is parametrized by length,
consistently, from the viewpoint of the two cylinders attaching to the edge.
Consider a point $P\in AA'$ with local coordinate $w_i(P)= \exp (i\theta_i)$
and $w_j(P) = \exp (i\theta_j)$.
{}From the point of view of the cylinder attached to the $i$th face,
$\theta_i$ measures the distance of $P$ from some fixed point along
the boundary of the cylinder, whereas from the point of view of the cylinder
attached to the $j$th face, $\theta_j$ measures the distance of $P$ from some
other fixed point along the boundary of the cylinder (see figure). Thus
$$\theta_i +\theta_j \equiv \phi\, ,\eqn\eththphi$$
is a constant, independent of the choice of $P$.
It then follows that the transition
function must be of the form
$$w_i(P')\, w_j (P') = \exp[i\phi] \,
,\eqn\trfunc$$
since it clearly works for $P'=P$;  it works for $P'\in AA'$ by our arguments
about the parametrization
of the edge, and therefore by analyticity must
work in a neighborhood of the edge.\foot{This is consistent with the fact
that the quadratic differentials $(dw_1)^2/w_1^2$ and $(dw_2)^2/w_2^2$ defined
for each disk, must agree on the edge.} This proves our statement about
transition functions.

Let us now address the issue of duality. We must therefore consider two
configurations. In the first one, a sphere $\V'_3$, with coordinates
($z_1,z_2,z_3$), is sewn to the polyhedron via the relation
$$z_2\, w_i = \exp (i\theta_i)\, ,\eqn\confone$$
so that the special puncture of $\V'_3$ lands at $P$ (since $z_2 =1$, for
the special puncture, and $w_i = \exp (i\theta_i)$, for $P$). In the
second configuration, another sphere  $\V'_3$, with coordinates
($\wt z_1,\wt z_2,\wt z_3$), is sewn to the polyhedron via the relation
$$\wt z_2\, w_j = \exp (i\theta_j)\, ,\eqn\conftwo$$
so that the special puncture of $\V'_3$ again lands at $P$
(since $\wt z_2 =1$, for
the special puncture, and $w_j = \exp (i\theta_j)$, for $P$). In doing this
we have made sure that the two configurations, corresponding to two
polyhedra with an extra puncture on an edge, are conformally equivalent
surfaces. We must now see if the local coordinates agree.
The local coordinates corresponding to the faces of the polyhedra agree,
as we verify next.
In the first configuration, the coordinate $z_1$ ends up
as the coordinate of the $i$-th face, while in the second configuration,
$w_i$ remains as the local coordinate. It follows from $z_2z_1=1$, and
Eqn.\confone, that
$$ w_i = z_1 \exp (i\theta_i)\, ,\eqn\xconfone$$
which shows that the two coordinates simply differ by a phase.
Conversely, in the second configuration, the coordinate $\wt z_1$ ends up
as the coordinate of the $j$-th face, while in the first configuration,
$w_j$ remains as the local coordinate. It follows from $\wt z_2\wt z_1=1$, and
Eqn.\conftwo, that
$$w_j = \wt z_1 \exp (i\theta_j)\, ,\eqn\xconftwo$$
again showing just a phase difference. The less clear point is that
the local coordinates arising in the first configuration from $z_3$
(located at $P$) agrees with the local coordinate arising in the
second configuration from $\wt z_3$. Consider a point $P'$ near
$P$ and let $z_1(P')$ and $\wt z_1 (P')$ be its coordinate in the
first and second configurations respectively. It follows by
multiplication of the last two equations that
$$w_i (P')\, w_j(P') = z_1 (P') \, \wt z_1(P') \exp [i(\theta_i+\theta_j)]
= z_1(P') \, \wt z_1(P') \exp[i\phi] \, ,
\eqn\relc$$
and, using \trfunc, it follows that
$$z_1 (P') \, \wt z_1(P')  = 1 \, .\eqn\xrelc$$
If we use the uniformizing coordinate $z$ on $\V'_3$ (such that $z_1=z$),
we can write $z_3 (p) = h(z(p)) = h(z_1(p))$. Using this language, it
follows that the duality condition, namely, the requirement that the $z_3$
and $\wt z_3$ coordinates of $P'$ agree, gives
$$z_3(P') = \pm \wt z_3 (P') \quad \longrightarrow \quad
h\,(z_1(P'))=\pm\, h\,(\wt z_1 (P')\, .\eqn\zzz$$
It therefore follows, using \xrelc, that we must have
$$ h\, (z_1(P'))=\pm\,  h(\, 1/ z_1 (P')\, )\, ,\eqn\zzz$$
which we recognize immediately as the condition \symconvp\ for off-shell
symmetry of $\V'_3$ under the exchange of the first and second punctures
(the plus sign cannot be realized). This completes our argument for
off-shell duality. In this way we have established that \miden\ holds
off-shell. Eqn.\xiden\ also holds off-shell, since off-shell duality
holds for any $n$-polyhedron, and therefore, for the full
$\V_n$ space defining the product of $(n-1)$ string fields.

\section{A Surprising Covering Result}

It is a simple consequence of our previous discussion that we can
produce a smooth covering of the moduli space $\M_4$
of four punctured spheres, with three string diagrams
built by sewing, with the standard propagator, a vertex $\V_3$, and a
vertex $\V'_3$. In the three string diagrams, the special puncture retains
its label, and the labels of the other punctures are exchanged, as usual,
to get a symmetric combination. The reason this covers smoothly moduli
space is that, when the propagators collapse, the three string diagrams,
by our duality argument, match precisely, and when the propagators
are infinitely long, we get the proper degenerations.

We claim the following generalization of this result:
A complete (smooth) cover of the moduli space of $n$-punctured
Riemann surfaces is generated with all the standard tree-level
Feynman diagrams that can be built using the vertices
$\{ \V_3 \cdots \V_{n-1}\}$, and the vertex $\V'_3$, with the condition
that in each diagram, $\V'_3$ appears once. The label of the special
puncture is always the same.

A proof by induction
would only require to show that all (non-degenerate) boundaries
of the Feynman graphs match, since, the good cover of the lower dimensional
moduli spaces guarantees we cannot miss any degeneration.
We will only sketch an argument for this matching of boundaries.
In building the moduli space of $n$-punctured spheres, the Feynman diagrams
with lowest number of propagators are  ${\bf S}(\V_{n-1}\cdot\V'_3)$ which
have  one propagator, indicated by the dot between the vertices.
The (non-degenerate)
boundaries of this graph arise from the collapsed propagator, yielding
${\bf S}(\V_{n-1}\times\V'_3)=0$ (Eqn.\xgeom), and, from the standard
boundary of the vertex $\V_{n-1}$, yielding terms of the form
$\V_p\times\V_q\cdot\V'_3$. These boundaries, cancel with
the boundaries of the Feynman diagrams $\V_p\cdot\V_q\cdot\V'_3$
arising from the collapse of the first propagator. When the second
propagator in this graph collapses, it yields the boundaries
$\V_p\cdot\V_q\times\V'_3$, which do not cancel out by themselves. They
would, if the term was of the form $\V_p\cdot(\V_q\times\V'_3)$, where the
parenthesis indicates that the propagator connects to either a leg in $\V_q$,
or a leg in $\V'_3$.  We therefore need to introduce the Feynman graphs of the
type $\V_p\cdot\V'_3\cdot\V_q$, which provide the missing boundaries,
when a propagator collapses. The graphs $\V_p\cdot\V_q\cdot\V'_3$,
together with $\V_p\cdot\V'_3\cdot\V_q$, amount to all graphs with two
propagators, and one $\V'_3$ vertex. In these graphs, the boundaries arising
from the vertices, would then be matched with boundaries from all graphs
with three propagators,
involving one $\V'_3$ vertex. This goes on until all boundaries are accounted
for.

In retrospect, we can argue that this result is a consequence of the
background independence of string field theory.
If $S(\ket{\Psi})$ denotes the string field theory action at the point
$x$, then,
$$\wh S(\,\ket{\Psi}\,)\, \equiv\, S(\,\ket{\Psi}\,)+\delta x^\mu
\bra{{\p S\over \p\ket{\Psi}}} \wh\O_\mu\rangle\,\,,
\eqn\ecovxxx$$
denotes the string field theory action obtained when the string field
is shifted by an amount $\delta x^\mu\ket{\wh\O_\mu}$.
In this action we use the state space $\H_x$. On the other, we can consider
the string field action based on $\H_{x+\delta x}$, which is of the form
$\wt S (\,\ket{\Psi}) = \sum_{N=2} {1\over N!}
\bra{V^{(N)}}\Psi\rangle\cdots\ket{\Psi}$,
where both the bras, and the kets, refer to objects in $\H_{x+\delta x}$.
Since contraction is invariant under parallel transport, we can
parallel transport both the bras, and the kets, from $\H_{x+\delta x}$
to $\H_x$ without changing the value of the action. This can be done
with any connection. Using the connection $\wh\Gamma_\mu$, we have
$$\wt S (\ket{\Psi}) = \sum_{N=2} {1\over N!}
\Bigl(\bra{V^{(N)}} + \delta x^\mu D_\mu (\,\wh\Gamma\,)\bra{V^{(N)}}\Bigr)
\bigl(\, \ket{\Psi} + \delta x^\mu \wh\Gamma_\mu \ket{\Psi} \,\bigr)^N \, ,
\eqn\moveac$$
where we made use of standard parallel
transport formulas (see [\rangasonodazw],
\S2.4). All objects in this right hand side refer to $\H_x$.
We now do the field redefinition
$\ket{\wt\Psi} \equiv \ket{\Psi} +
\delta x^\mu \wh\Gamma_\mu \ket{\Psi}$, and,
use \dervnzero\ and \ethreenewa, to find that
$\wt S$ takes the form
$$\wt S(\ket{\Psi}) = S(\,\ket{\wt\Psi}\,) +\half\, \delta x^\mu\,
\bra{V^{\prime(3)}_{123}} \wh\O_\mu\rangle_3 \ket{\wt\Psi}_2 \ket{\wt\Psi}_1\,.
\eqn\ecovyyy$$
While formulated on $\H_x$, it still represents string field theory
at the background $x+\delta x$.
If string field theory is
background independent, the actions in \ecovxxx\ and \ecovyyy\ must
be related via a
field redefinition involving linear and higher order terms. As a result,
the on-shell $S$-matrix elements calculated from the two theories must
be the same. The order $\delta x^\mu$ terms of the $S$-matrix elements
in the theory described by the first action can be computed
by treating the order $\delta x^\mu$ term in the action as perturbation,
and are given (including the combinatoric factors) by the ordinary Feynman
diagrams of the unperturbed theory with one of the legs being
$\ket{\wh\O_\mu}$.\foot{The calculation is more complicated if there
are divergences associated with the insertion of $\wh\O_\mu$ on the
external leg; in this case, one has to take into account wave-function
renormalization of the external state. For the purpose of this argument
we can choose $\wh\O_\mu$ and the external states in such a way that
no such divergence is present.}
This is known to cover the moduli space of punctured
spheres. On the other hand, using the same reasoning, we see that
the order $\delta x^\mu$ terms of the
$S$-matrix elements calculated from the second action are given by
Feynman diagrams constructed from the ordinary string vertices with
one and only one insertion of the vertex $\bra{V^{\prime(3)}}$, with
the state $\ket{\wh\O_\mu}$ inserted at the asymmetric puncture. Thus,
these diagrams must also cover the moduli space fully. This establishes
the covering result.

\chapter{Constructing The Complete Diffeomorphism}

We have constructed so far the first few pieces of the diffeomorphism
establishing background independence. Indeed, in \S5,
by defining $\Gamma_\mu = \Delta\Gamma_\mu + \wh\Gamma_\mu$, with
$\wh\Gamma_\mu$ the canonical connection, we were able to solve for
the one-form $\Delta\Gamma_\mu$, and hence for $\Gamma_\mu$. We also
solved for the one-form $\bra{\Gamma_\mu^{(3)}}$. The purpose of the
present section is to find the solution for all the higher bras
$\bra{\Gamma_\mu^{(N)}}$, with $N\geq 4$. It is useful to define
$$\bra{\Gamma_\mu^{(2)}}\equiv \bra{\Delta\Gamma_\mu}\, ,\eqn\ekkthree$$
since the one-form $\Delta\Gamma_\mu$ is really on the same footing as
any bra $\bra{\Gamma_\mu^{(N)}}$. With this definition, the diffeomorphism
implementing background independence, as written in \emxe , becomes
$$\ket{\Psi}_{x+\delta x} = {}_{x+\delta x}\,{{\cal I}_{}}_x \, \Bigl[
\,\ket{\Psi} -\delta x^\mu \Bigl(\,
\wh\Gamma_\mu \ket{\Psi}+ \sum_{N\geq 0}\,\,
{1\over N!} \, \, {}_{(01\cdots N)}
\bra{\,\Gamma_\mu^{(N+1)}}\s_{0e}\rangle\,
\ket{\Psi}_1\cdots \ket{\Psi}_N \, \Bigr)\,\Bigr] \, . \eqn\xemxe$$
It is instructive to interpret the above diffeomorphism as the result
of a canonical transformation,
followed by parallel transport.
It is clear from \xantibr,  and the symmetry of
$\bra{\Gamma_\mu^{(N)}}$ that
$$\sum_{N\geq 0}\,\,
{1\over N!} \, \, {}_{(01\cdots N)}
\bra{\,\Gamma_\mu^{(N+1)}}\s_{0e}\rangle\,
\ket{\Psi}_1\cdots \ket{\Psi}_N\,\, = \,-\,\Bigr\{ \,
{\bf U}_\mu\, , \, \ket{\Psi} \, \Bigr\}\, \, ,\eqn\cantra$$
where the generator ${\bf U}_\mu$ of the canonical transformation is given
by
$${\bf U}_\mu = \sum_{N\geq 1}\,\,
{1\over N!} \, \, {}_{(1\cdots N)}
\bra{\,\Gamma_\mu^{(N)}}\Psi\rangle_1\cdots \ket{\Psi}_N\, .\eqn\gencan$$
It then follows that
$$\ket{\Psi}_{x+\delta x} = {}_{x+\delta x}\,{{\cal I}_{}}_x \, \Bigl[
\,\,\ket{\Psi}\,+\, \delta x^\mu\,\Bigr\{ \,{\bf U}_\mu\,
,\,\ket{\Psi}\,  \Bigr\}\, -\delta x^\mu \cdot
\wh\Gamma_\mu\, \ket{\Psi}\,\,\,\Bigr] \, , \eqn\xxxe$$
which shows that the string field at $x+\delta x$ is obtained from
the string field at $x$ by first performing a canonical transformation
with generator ${\bf U}_\mu$, and then performing parallel transport with the
canonical flat connection $\wh\Gamma$.

Back to our central topic in this section, the background independence
conditions can also be written more clearly
with the help of \ekkthree.
Starting from $D_\mu(\wh\Gamma) \bra{V^{(N)}}=0$,
and following the same steps we performed at the beginning of \S5.2, we find
$$D_\mu(\,\Gamma\,)\bra{V^{(N)}} = -{\bf S}
\Big( \bra{\Gamma_\mu^{(2)}}\bra{V^{(N)}} \s\rangle \Bigr)\,. \eqn\ozone$$
This relation, back in the background independence condition
\emmfifteen, gives us
$$\bra{\Gamma_\mu^{(N)}} \sum_{i=1}^N Q^{(i)} \, = \, -\, \sum_{m=3}^N
{\bf S}\Big( \bra{\Gamma_\mu^{(N-m+2)}}\, \bra{V^{(m)}}\s\rangle\Big)
\, - \bra{V^{(N+1)}}\widehat O_\mu\rangle \, .\eqn\etwox$$
The solutions we have found so far can be written
in the form
$$\bra{\Gamma_\mu^{(2)}}=\,-\,\int_{\B_3}\bra{\,\Omega^{(1)3}\,}
\wh\O_\mu\rangle_3 ,\quad \p\B_3 = \V'_3 - \V_3 .\eqn\sptwo$$
$$\bra{\Gamma_\mu^{(3)}} = \,-\, \int_{\B_4}
\bra{\Omega^{(1)4}}\wh\O_\mu\rangle_4 \, ,
\quad\p\B_4=\V'_4 -\V_4 ,\eqn\spteen$$
where $\V'_4 = {\bf S}(\B_3\times\V_3)$ satisfies $\p\V'_4 = \p\V_4$.
This suggests that the higher order solutions take the form
$$\bra{\Gamma^{(N)}_\mu} =  -\,\int_{\B_{N+1}} \bra{\Omega^{(1)N+1}}
\O_\mu\rangle_{N+1} ,\quad \p\B_{N+1}=\V'_{N+1} -\V_{N+1} \, ,
\eqn\ebbb$$
where $\B_{N+1}$ must be a symmetric (in first $N$ legs) homotopy between
$\V_{N+1}$ and some vertex  $\V'_{N+1}$. The symmetric homotopy
is required in order to
to have a symplectic diffeomorphism. We can derive
what the vertex $\V'_{N+1}$ should be, by considering condition \etwox,
together with our ansatz \ebbb . We find
$$-\int_{\B_{N+1}}\bra{\Omega^{(1)N+1}} \sum_{i=1}^{N+1} Q^{(i)} \,
= \,  \sum_{m=3}^{N}
{\bf S}\int_{\B_{N-m+3}}\hskip-6pt\bra{\Omega^{(1)N-m+3}}\,
\int_{\V_m} \bra{\Omega^{(0)m}}\s\rangle\ \,-\,
\int_{\V_{N+1}} \bra{\Omega^{(0)N+1}} \, ,\eqn\zzfive$$
where we peeled off the common state $\ket{\O_\mu}$.\foot{Note that
\zzfive\ implies \etwox, but is a stronger constraint than \etwox.}
Here {\bf S} denotes symmetrization in all the free legs of
$\B_{N-m+3}$, except for the $(N-m+3)$-th leg (where $\O_\mu$ is to be
attached),
and all the free legs of $\V_m\,$.
$\ket{\s}$ sews one of the first $N-m+2$ legs of
$\B_{N-m+3}$, with one of the legs of $\V_m$.
Making use of \qixbu, and \sewing,  we rewrite the above equation as
$$\int_{\p\B_{N+1}}\bra{\Omega^{(0)N+1}}
= \, {\sum_{m=3}^{N}} \int_{\,\,{\bf S}(\B_{N-m+3}\times\V_m)}
\hskip-12pt\bra{\Omega^{(0)N+1}}\,
\,-\,\int_{\V_{N+1}} \bra{\Omega^{(0)N+1}} \, ,\eqn\zzsix$$
where the three integrals have a common integrand. It follows from this
equation that
$$\p\B_{N+1}=
\sum_{m=3}^N{\bf S}(\B_{N-m+3}\times\V_m)-\V_{N+1}\,.\eqn\ffdd$$
Upon comparison with \ebbb,  we conclude that $\V'_{N+1}$ must be given by
$$\V'_{N+1} = \,  \sum_{m=3}^{N}
{\bf S}\Big(\B_{N-m+3}\times \V_m\Big) = \,{\bf S}\Bigl(
\B_N \times\V_3 + \cdots +\B_3\times\V_N\Bigr) .\eqn\zzseven$$
This is a simple expression. It says that the new vertex $\V'$, to
any order, is obtained by twist-sewing an interpolating vertex $\B$ of lower
order, with an old vertex $\V$, in all possible ways. While this definition
always makes sense, Eqn.\ebbb\ implies a strong constraint. Since $\p^2=0$,
we must have that
$$\p\V'_{N+1}=\p\V_{N+1}\, .\eqn\zznine$$
Indeed, if this property holds, we can always find a symmetric interpolating
vertex $\B_{N+1}$. Therefore, our problem is to show that $\V'$, as defined
in \zzseven, satisfies \zznine .

\subsection{Proving the Coincidence of Boundaries.}
We shall carry out the proof via induction. Let us assume that
we have found new vertices $\V'_3 ,\cdots ,\V'_M$ and the corresponding
interpolating vertices $\B_3, \cdots , \B_M$, such that
$$\eqalign{\p\B_n &= \V'_n -\V_n \, ,\cr
\p\V'_{n}&=\p\V_{n}\, ,\cr
\V'_n \,\,&= \,  \sum_{m=3}^{n-1}
{\bf S}(\B_{n-m+2}\times \V_m) \, ,\cr}\eqn\begpro$$
for all $n$ in the interval $3\leq n\leq M$.
We then want to show that $\V'_{M+1}$, defined as
$$\V'_{M+1}=\sum_{m=3}^{M}{\bf S}(\B_{M-m+3}\times\V_m)\, ,\eqn\ekkfourteen$$
satisfies $\p\V'_{M+1} = \p\V_{M+1}$. This would allow us to define
$\B_{M+1}$, and continue the recursion procedure.

We must simply compute the boundary of $\V'_{M+1}$. From eqn.\ekkfourteen\
we get,
$$
\p\V'_{M+1}=\sum_{m=3}^{M} {\bf S} (\p\B_{M-m+3}\times \V_m) +
\sum_{m=4}^M {\bf S}(\B_{M-m+3}\times \p\V_m)
\eqn\ekkfifteen
$$
using $\p\V_3=0$.
With the help of the first equation in \begpro , and the expression
\geom\ for $\p\V_m$ we rewrite the above equation as
$$\eqalign{\p\V'_{M+1} &=
-\sum_{m=3}^M {\bf S} (\V_{M-m+3} \times \V_m) + \sum_{m=3}^M
{\bf S} (\V'_{M-m+3} \times \V_m)\cr
&\,\,\,-\sum_{m=4}^M \sum_{p=3}^{m-1} {\bf S} (\B_{M-m+3}
\times  \V_{m-p+2}\times\V_p ),\cr}\eqn\ekksixteen$$
where we have adopted the convention that $\A\times \B\times \C$ denotes
set of surfaces obtained by twist-sewing one puncture of $\A$ with one
puncture of $\B$, and another puncture of $\B$ with a puncture of $\C$.
According to this convention, the last term of the above equation
only contains terms where $\B_{M-m+3}$ is sewed to $\V_{m-p+2}$, but
not to $\V_p$. This is
responsible for the absence of a factor of $(1/2)$
present in Eqn.\geom.
The first term of the right hand side of
Eqn.\ekksixteen\ is recognized to be $\p\V_{M+1}$, keeping in mind that
{\bf S} symmetrizes all but the $(M-m+3)$-th leg of $\V_{M-m+3}$ in this
term, thereby accounting for the missing factor of (1/2).
We therefore have
$$\p\V'_{M+1}- \p \V_{M+1}
= \sum_{p=3}^M
{\bf S} (\V'_{M-p+3} \times \V_p)
-\sum_{m=4}^M \sum_{p=3}^{m-1} {\bf S} (\B_{M-m+3} \times  \V_{m-p+2}\times
\V_p ).\eqn\ekksixteen$$
We must now show that the right hand side of the above equation
vanishes. The $p=M$ term in the first sum on the right hand side gives
${\bf S} (\V'_3\times \V_M)$, and vanishes by Eqn.\xgeom. The other terms in
this sum can be rewritten using the third of Eqn.\begpro, for $n\le M$,
$$\sum_{p=3}^{M-1}\sum_{m=p+1}^M  {\bf S} \Bigl( \,\bigl(\B_{M-m+3} \times
\V_{m-p+2}\,\bigr)\times \V_p \Bigr)\,
=\sum_{m=4}^M \sum_{p=3}^{m-1} {\bf S} \Bigl( \,\bigl(\B_{M-m+3} \times
\V_{m-p+2}\,\bigr)\times \V_p \Bigr)\, ,\eqn\ifitwere$$
where the extra parenthesis indicate that $\V_p$ is sewn to both legs that
come out of $\B_{M-m+3}$, and legs that come out of $\V_{m-p+2}$.
Thus $\p\V'_{M+1}-\p\V_{M+1}$ is given by the difference between this term
and the last term of Eqn.\ekksixteen. This gives
$$\p\V'_{M+1}-\p\V_{M+1}=\sum_{m=4}^M \sum_{p=3}^{m-1} {\bf S} \bigl(
\V_{m-p+2}\times\B_{M-m+3} \times  \V_p\bigr) \, ,\eqn\wishaway$$
We claim this term vanishes, in fact, each term corresponding to a
fixed value of $m$ vanishes:
$$\sum_{p=3}^{m-1} {\bf S} \bigl(
\V_{m-p+2}\times\B_{M-m+3} \times  \V_p\bigr) =0 \, .\eqn\wishaway$$
Note that all terms in this equation involve the same vertex $\B$. This
vertex is symmetric under the exchange of any of its state spaces
(except the one where $\widehat\O_\mu$ is to be inserted, which
cannot be used to sew into the $\V$ vertices). The above relation
can be rewritten as
$${1\over 2} \sum_{p=3}^{m-1} {\bf S} \Bigl(\,
\V_{m-p+2}\times\B_{M-m+3} \times  \V_p
\,\,\,+ \,\,\,\V_p\times\B_{M-m+3} \times  \V_{m-p+2}\Bigr) =0 \, .\eqn\wishy$$
This equation holds because, for each value of $p$, each of
the two terms in the above expression produces
the same subspace of $\P_{M+1}$, (this is manifest due to the symmetry
of $\cal B$), but with opposite orientation. A way to show the orientations
are opposite is to write the expression for the corresponding string
amplitudes and to check they cancel.
The amplitude is written as follows
$$\eqalign{
{}&\bra{B^{(M-m+3)} (1,2,\cdots)}\,\,\Bigl(\,\, \bra{V^{(m-p+2)}(1',
\cdots )} \bra{V^{(p)} (2',\cdots)}\s_{11'}\rangle\ket{\s_{22'}}\cr
{}&\quad\quad \quad \qquad \qquad\qquad + \bra{V^{(p)}(1',\cdots )}
\bra{V^{(m-p+2)}(2',\cdots)}
\s_{11'}\rangle\ket{\s_{22'}}\,\Bigr)\cr}\eqn\asen$$
where
$$\bra{B^{(N)}}=\int_{\B_N}\bra{\Omega^{(1)N}}\, . \eqn\defbrabn$$
Eqn.\asen\ can be rewritten as
$$\eqalign{
{}&\bra{B^{(M-m+3)} (1,2,\cdots)}\,\,\Bigl(\,\, \bra{V^{(m-p+2)}(1',
\cdots )}  \bra{V^{(p)} (2',\cdots)}\cr
{}&\quad\qquad\qquad\qquad\qquad\,\,\, +
\bra{V^{(m-p+2)}(2',\cdots)} \bra{V^{(p)}(1',\cdots )}\,\, \Bigr)
 \,\,\ket{\s_{11'}}\ket{\s_{22'}}\, ,\cr}\eqn\sashoke$$
which vanishes identically since the product of sewing kets is
antisymmetric under the exchanges $1\leftrightarrow 2$,
$1'\leftrightarrow 2'$, while the rest of the expression is manifestly
symmetric (recall that all external legs, with the exception
of the last one in $\B$, are symmetrized).
This proves the desired result, and verifies the consistency of
our construction of the full nonlinear diffeomorphism implementing
background independence to all orders in the string coupling constant.

\chapter{Backgrounds a Finite Distance Apart}

We have proven in \S5--\S7  the existence of a fully nonlinear infinitesimal
diffeomorphism relating string field theories formulated around
infinitesimally close conformal field theories. This diffeomorphism
established local background independence of closed string field theory.
If we have a CFT theory space, it is natural to ask if this proof of
local background independence can be extended to the case when the two
conformal theories are a finite distance apart in theory space.
\foot{Distance can be defined using the Zamolodchikov metric.} The
finite distance diffeomorphism would be obtained by
integrating the infinitesimal diffeomorphism along a path in theory space
joining the two conformal theories [\senthree,\kugozwiebach].
There are two aspects
to the question of existence of a finite distance diffeomorphism.
The first is a formal one. Are there local integrability conditions that
must be satisfied in order for the diffeomorphism to be path independent?
We show here that there are no such integrability conditions.
The second hinges
on the fact that we are dealing with an infinite dimensional vector
bundle. Is it possible to integrate the diffeomorphism without getting
infinities? We will not deal with this question in detail,
but will argue that
finite distance diffeomorphisms are expected to exist.

\section{Composition Properties}

We begin by noting that the diffeomorphism $F_{y,x}: \H_x\to\H_y$ that
relates the string master actions $S_x$ and $S_y$,
and the symplectic forms $\omega_x$ and $\omega_y$, arising
from two different conformal field theories, is ambiguous due to the presence
of gauge (and possibly other global) symmetries of the action.
A symmetry ${\bf g}_x: \H_x\to \H_x$ of the string field theory at $x$,
is a transformation of the
string field leaving the action, and the symplectic form invariant, namely
$S_x = {\bf g}_x^* \,S_x$ and $\omega_x = {\bf g}_x^*\, \omega_x$.\foot{Note
that in the Batalin-Vilkovisky formalism only those transformations that
preserve the symplectic structure together with the action are genuine
symmetries of the (tree level) theory.}
It follows that whenever
$F_{x,y}$ is a diffeomorphism that relates string theory at $x$ and at $y$,
so is ${\bf g}_y \circ F_{y,x}\circ {\bf g}_x$. This gives us the equivalence
relation:
$$F_{y,x}\approx{\bf g}_y \circ\,F_{y,x}\,\circ\, {\bf g}_x \,.\eqn\ambgauge$$
In fact, it is sufficient to consider gauge transformations on the left, since
$$F_{y,x}\,\circ\, {\bf g}_x =
\bigl( F_{y,x}\,\circ\, {\bf g}_x \circ\, F_{x,y}\,\bigr)\circ\, F_{y,x}\, .
\eqn\ambgaugex$$
In the above equation the map in parenthesis is a symmetry transformation at
$y$ since it is a diffeomorphism from $\H_y$ preserving $S_y$ and $\omega_y$.
This implies that any symmetry transformation applied before performing
the diffeomorphism
can be written as a symmetry transformation applied after performing
the diffeomorphism.

By definition, the diffeomorphisms establishing the equivalence of
string field theories at different points in CFT theory space must
satisfy a composition law. Given three points $x,y$, and $z$, we must
have
$$F_{z,x} \approx F_{z,y}\, \circ\, F_{y,x}\, .\eqn\compos$$
The right hand side is a diffeomorphism from $\H_x$ to
$\H_z$ establishing the equivalence of the corresponding
string field theories, therefore uniqueness
(up to symmetry transformations)
of the diffeomorphism  relating
two state spaces implies the equality. Explicitly, in the
notation of \ashoke\ this equation reads
$$F\, (\, \psi_x\, ,  x,z \,) \approx F\, \bigl(\, F\,(\, \psi_x\, ,x,y\,)
\, , y\, , z \, \bigr),\eqn\expltransf$$
where we have supressed, for clarity, the vector indices on the string
field components and on $F$.

\section{Differential Equation for $F$ and its Integrability}

In order to find a differential equation for the diffeomorphism $F$, we
apply equation \expltransf\ for the case when $z=y+\delta y$, to find
$$\eqalign{
F^i\, (\, \psi_x\, ,  x,\, y+\delta y \,) &= F^i\, \bigl(\, F\,(\,
\psi_x\, ,x,y\,)
\, , y\, , y+\delta y \, \bigr)\cr
&= F^i\,(\, \psi_x\, ,x,y\,) + \delta y^\mu \,\cdot f^i_\mu \,\bigl(\,
F\,(\, \psi_x\, ,x,y\,)\, , y\, \bigr) + \O (\delta y^2) ,\cr}\eqn\extransf$$
where use was made of Eqn.\eseven. For the convenience of writing we have
replaced the $\approx$ symbol by $=$ in the above equation, but we should
always keep in mind that the equality in the above equation is true only
in the sense of equivalence defined in Eqn.\ambgauge. In particular, we are
allowed to add any infinitesimal symmetry transformation
to the right hand side of the above equation.  Eqn.\extransf\ then gives
$${\p F^i(\psi_\vx, x, y)\over \p  y^\mu}=f^i_\mu\, \bigl( \,F
(\,\psi_x,x,y), y\, \bigr) \, .\eqn\enine$$
Since the  existence of $f^i_\mu$ has already been proved, the proof of
existence of  $F$ reduces to showing the integrability of the set of
partial differential
equations \enine\ with the boundary condition
$$
F^i\, (\,\psi_\vx,\vx,\vx)=\psi^i_\vx.
\eqn\eextraone
$$
Since the infinitesimal
diffeomorphism $f_\mu^i$ preserves the symplectic structure, it is
guaranteed that the finite diffeomorphism $F(\psi_x, x, y)$ obtained by
integration, will also map the symplectic
structure at the point $x$ to the symplectic structure at the point $y$.

The integrability conditions for \enine\ arise by taking a further
derivative of the equation and antisymmetrizing:
$$ \Delta_{\mu\nu} F^i \equiv \,\Bigl(
{\p \over \p y^\mu} {\p\over \p y^\nu}\,-
{\p \over \p y^\nu} {\p\over \p y^\mu}\,\Bigr) \, F^i\,(\psi_x \, ,
x , y \, )\,= \, 0\,.\eqn\pdmc$$
Making use of \enine\ to evaluate the second derivatives we find
$$\Delta_{\mu\nu} F^i
=\biggl(\,\,{\partial f^i_\mu(\psi_{\vy}, \vy)\over \p y^{\nu}}
+{\partial_r f^i_\mu(\psi_{\vy}, \vy)\over \p\psi^j_{\vy}}f^j_\nu
(\psi_{\vy}\,, \vy)\biggr) -(\mu\leftrightarrow \nu)\, .\eqn\eten$$
If we can show that our solution for $f_\mu^i$, satisfying the
local background independence conditions \eeight, implies that
$\Delta_{\mu\nu} F^i =0$, then we would have proved
(formal) background independence of string field
theory for finite deformations of the background.

Actually the condition $\Delta_{\mu\nu} F^i =0$ is too strong.
This is due to the fact that
we are interested in obtaining
a solution $F^i(\psi_x, x, y)$ which is single
valued {\it only} when it is regarded as a point in the space of all
diffeomorphisms modulo the set of gauge transformations at $y$.
In other words, it is acceptable if integration of
Eqn.\enine\ along two
different paths gives different $F^i(\psi_x, x, y)$'s which are related by
a symmetry transformation of $S_y$.
Indeed, $(\delta_1x^\mu)(\delta_2x^\nu)\Delta_{\mu\nu} F^i$ gives the
difference between the diffeomorphisms obtained when going from $x$,
to $x+\delta_1x+\delta_2x = y$, along the two obvious paths.
Thus all we need is that $\Delta_{\mu\nu} F^i$
be a symmetry at $y$, namely $(\p_r S/ \p\psi^i) \Delta_{\mu\nu} F^i=0$.
This gives
$${\p_r S\,(\psi_\vy\, ,y)\over \p\psi^i_\vy}
\Big[\Big({\partial f^i_\mu(\psi_{\vy}, \vy)\over \p y^{\nu}}
+{\partial_r f^i_\mu(\psi_{\vy}, \vy)\over \p\psi^j_{\vy}}f^j_\nu
(\psi_{\vy}, \vy)\Big) -\Big(\mu\leftrightarrow \nu\Big)\Big]
=0\eqn\eeleven$$
We shall now show that this equation is automatically satisfied by the
solution of the local background independence conditions.
We start with the second equation of \eeight\ and differentiate it
with respect to $x^\nu$:
$$\eqalign{
{\p^2 S \over \p x^\nu \p x^\mu } &=\,
 -\,{\p_r\over\p \psi_x^i} \Bigl(
{\p S \over \p x^\nu }\Bigr) \, f^i_\mu\, - \,
{\p_r S \over \p \psi^i_x} { \p f_\mu^i\over \p x^\nu }  \cr
&=-{\p_r S \over \p \psi^i_x}\, \biggl[\,{ \p f_\mu^i\over \p x^\nu }\,
- \,{\p_r f^i_\nu\over \p \psi^j_x }\, f^j_\mu\biggr]  +
\Bigl(
{\p_r\over \p \psi^i_x} {\p_r \over \p\psi^j_x}S\Bigr)\,
f^i_\mu\,f^j_\nu  .\cr } \eqn\lkjh$$
Upon antisymmetrization in $\mu$ and $\nu$, the last term in the
right hand side drops out, and the remaining terms are seen to coincide
with the desired expression in Eqn.~\eeleven\ upon replacement of $x$ by
$y$.   We thus see that the integrability conditions required for obtaining
the finite field redefinitions $F^i$ from the infinitesimal field
redefinitions given by $f^i_\mu$ are automatically satisfied.
This is not surprising, however.
Since $f^i_\mu$ satisfies
Eqn.\eeight, we are
guaranteed that by integrating Eqn.\enine\ from $x$ to $y$ along any path
we must get a transformation $F(\psi_x, x, y)$ that maps $S_x$ to $S_y$.
Thus if we obtain different $F$'s by integrating along different paths,
they must differ by a symmetry transformation of $S_y$.

\section{Integrability without Divergences ?}

We have shown in the above paragraphs that there are no local
integrability conditions that ought to be satisfied, {\it i.e.} our
infinitesimal diffeomorphisms can be integrated and we are guaranteed not
to run into trouble unless we find infinite quantities.  In order to avoid
infinities to first approximation, products of the connection $\Gamma_\mu$
must be finite,
as is the case for the connection $c_\mu$ ( or $\bar
c_\mu$) of Refs.[\sonoda,\rangasonodazw].  This, of course, cannot be the
complete story since the diffeomorphism involves higher order bras
$\bra{\Gamma_\mu^{(N)}}$, and they must also enter in a full discussion.
In fact, a finite-distance field redefinition will involve the products of
all the $\bra{\Gamma^{(N)}}$'s, and the question of existence of
divergence free field redefinitions connecting two string field theories
reduces to the question of finiteness of these products.

While a complete
analysis ought to be done, it seems plausible that no infinities will
arise.  The products of $\bra{\Gamma^{(N)}}$'s are obtained by sewing
punctured spheres, and the only possible divergence in this procedure
comes from the configurations in $\bra{\Gamma^{(N)}}$ representing
surfaces where the special puncture lies on the coordinate curve of some
free puncture.  When two such
configurations are sewn, we may get divergences due to the collision of
the special punctures.  These dangerous configurations in
$\bra{\Gamma^{(N)}}$, can all be traced back to the sewing of the special
vertex $\V'_3$ to an ordinary string vertex.
Thus the only possible sources of divergences may be traced
to the introduction of $\bra{V^{\prime(3)}}$ in our analysis. This, in
turn, came from the connection $\wh\Gamma_\mu$.
The explicit presence of $\wh\Gamma_\mu$ in Eqn.\xemxe\ will also give
rise to divergences during the process
of integrating the equations for finite field redefinitions, since,
as was shown in ref.[\rangasonodazw],
the product
of two $\wh\Gamma_\mu$'s is divergent, and hence
the connection $\wh\Gamma_\mu$
cannot be used to parallel transport over a finite distance.
This divergence also appears due to the collision of $\wh\O_\mu$'s,
and must be related
to the divergence that arises in the process of sewing two
$\bra{V^{\prime(3)}}$ vertices due to the collision of the special punctures.
(No such divergences occur in the sewing of ordinary string vertices.)

It is clear from the above discussion that all
sources of divergence can finally be traced to the introduction of the
connection $\wh\Gamma_\mu$ in our analysis.
But this appearance
is purely fictitious, and is due to the fact that we have chosen to
express the connection $\Gamma_\mu$ as a sum of $\wh\Gamma_\mu$, and
the difference $\Gamma_\mu-\wh\Gamma_\mu$. This indicates that the
integrability analysis
of Eqn.\xemxe\ may be more transparent if we express the connection
$\Gamma_\mu$ as the sum of $\wt\Gamma_\mu$,
and $\Gamma_\mu-\wt\Gamma_\mu$, where $\wt\Gamma_\mu$  is a
connection with finite products (such as $c_\mu$ or $\bar c_\mu$).
Indirect evidence for the finiteness
of the field redefinition
is provided
by the perturbative finiteness of finite classical solutions of
string field theory [\mukherjisen ] which form the constant shift part
of the field redefinition.

\chapter{Discussion}

In this paper we have shown that given two nearby conformal field theories
CFT and CFT$'$, related to each other via a marginal deformation, and BV
string field theories formulated around each of these conformal field
theories, there is a field redefinition which relates the two master
actions and the antibrackets.
The constant shift involved in the
field redefinition is given by the classical solution in the string field
theory around CFT that represents the background given by CFT$'$. The
linear part of the field redefinition can be interpreted as a connection
in the space of conformal field theories, which differs from a canonical
connection by a term that can be expressed as an integral of a string
vertex over a certain region of the extended moduli space of punctured
Riemann surfaces. Finally, the non-linear part of the field redefinitions
can also be expressed as integrals of appropriate string vertices over
regions in the extended moduli space of punctured Riemann surfaces.

\subsection{Open String Field Theory.}
We have carried out our analysis for closed string field theory, but the
extension of our analysis to open string theories is
straightforward. In fact, a simpler field redefinition, one involving
a shift and a linear transformation, suffices to prove open string background
independence.
Recall that for open string theory the
interaction vertices $\bra{V^{(N)}}$ vanish for $N\ge 4$. Therefore,
if $\bra{\Gamma^{(3)}_\mu}$ can be shown to vanish, a consistent
solution of eqs.\emmfifteen\ is obtained by setting all the higher
$\bra{\Gamma_\mu^{(N)}}$'s to zero.
Proving that the redefinition need
not be nonlinear thus reduces
to showing that $\bra{\Gamma^{(3)}_\mu}$ vanishes.
On the other hand, $\bra{\Gamma^{(3)}_\mu}$ is to be determined from an
equation analogous to Eqn.\elalatwo, with no $\bra{V^{(4)}}$ term. Thus
all we need to show is that it is possible to choose
$\bra{\Delta\Gamma_\mu}$ in such a way that
${\bf S}(\bra{\Delta\Gamma_\mu}\bra{V^{(3)}}\s\rangle$  vanishes.
Note that for
open strings $\ket{\s_{12}}=\ket{R_{12}}$. Also, since the vertex
$\bra{V^{(3)}}$ has cyclic symmetry, but no exchange symmetry, {\bf S}
must explicitly symmetrize in the two external legs of $\bra{V^{(3)}}$.

Let us now analyze the quantity ${\bf
S}(\bra{\Delta\Gamma_\mu}\bra{V^{(3)}}\s\rangle$. In this expression
$\bra{V^{(3)}}$ denotes the Witten vertex where half of the
first string overlaps
with half of the second string, half of the second string overlaps with
half of the third string, and half of the third string overlaps with half
of the first string, with the strings 1, 2, 3 appearing in an
anticlockwise cyclic order. What about $\bra{\Delta\Gamma_\mu}$? It
satisfies an equation similar to \ethreeseven, except that the right hand
side of this equation must involve explicit symmetrization in the state
spaces 1 and 2 due to the lack of explicit exchange symmetry of the
vertices. A description of the vertex $\V'_3$ is given as follows;
the first and the second strings have a complete overlap, and the third string
is located at one of the common string endpoints,
with the strings 1, 2 and 3 being in
anticlockwise cyclic order.
The on-shell state $\ket{\wh\O_\mu}\equiv
\ket{c\O_\mu}$ is inserted at the third puncture, hence the final result
is insensitive to the choice of the coordinate system at the third
puncture.

We now consider an interpolating vertex
where a length $(1+t)/2$
of the first string coincides with the length $(1+t)/2$ of the second
string, a length $(1-t)/2$ of the second string coincides with a length
$(1-t)/2$ of the third string, and a length $(1-t)/2$ of the third string
coincides with the length $(1-t)/2$ of the first string.  The
first and the second strings are each of length one,
whereas the third string
is taken to be of length $(1-t)$, so that there is complete overlap of the
three strings.\foot{This, of course, is actually the description of a
Jenkins-Strebel quadratic differential.} Again, the
strings 1, 2 and 3, are in anticlockwise cyclic order. At $t=0$ this
describes Witten vertex, whereas at $t=1$ this describes $\V'_3$. Note
that the coordinate system on the third string becomes singular as $t\to
1$, since its length vanishes, but the result is insensitive to the choice
of the coordinate system on the third string. In fact, the above
description of the interpolating vertex can be taken as a specification of
the location of the punctures of the three strings and the coordinate
systems of the first and the second string, but not that of the third
string. If $\B_3$ denotes the region in $\P_3$\foot{Now $\P_n$ denotes
moduli space of a disk with $n$ punctures at the boundary of the disk.}
corresponding to the interpolating vertex, then we may write\foot{Note
that the forms $\bra{\Omega^{(k)N}}$ are now given in terms of correlation
functions of boundary operators in conformal field theory on the half
plane, and as a result has symmetry only under cyclic permutation of
the state space labels, but not under their exchange.}
$$\bra{(\Delta\Gamma_{\mu})_{12}}=\,-\,\int_{\B_3}(\bra{\,
\Omega^{(1)3}_{123}\,} - \bra{\,
\Omega^{(1)3}_{213}\,}\,)\,\ket{\wh\O_\mu}_3
\,,\eqn\econcone$$
This is an equation analogous to \ethreetwentyone, except that it has two
terms due to the lack of explicit symmetry of $\bra{V^{(3)}}$. The relative
$-$ sign between these two terms is due to the fact that the open string
master field is anticommuting (unlike the closed string master field which
is commuting).

We can now use this expression for $\bra{\Delta\Gamma_\mu}$ to compute
${\bf S}(\bra{\Delta\Gamma_\mu}\bra{V^{(3)}}\s\rangle$. $\int_{\B_3}$ simply
denotes an integral over $t$. We now see that for every value of $t$ the
contributions to ${\bf S}(\bra{\Delta\Gamma_\mu}\bra{V^{(3)}}\s\rangle$
cancel pairwise.
For example, the following pair of terms,
$$(I)\,=\, \bra{\,\Omega^{(1)3}_{3'34}(t)\,}\, \bra{\,\Omega^{(0)3}_{123''}
(t={1/2})\,}
\s_{3'3''}\rangle\ket{\wh\O_\mu}_4 \,,\eqn\econcthree$$
and,
$$(II)\,=\, \bra{\,\Omega^{(1)3}_{13'4}(t)\,}\, \bra{\,\Omega^{(0)3}_{233''}
(t={1/2})}
\s_{3'3''}\rangle\ket{\wh\O_\mu}_4 \,,\eqn\econctwo$$
yielding four strings with
the same cyclic order, cancel out since they correspond to identical
configurations. This can be seen diagramatically   ( $(I) = (II)$)
\par
\centerline{\vbox{\vskip .5cm
\hbox{\vbox{\hbox{~\hskip 1.85cm \vrule height2cm  \hskip .32cm
\vrule
height2cm }
\vskip -.68cm
\hbox{\vbox{\hrule width2cm} \hskip .2cm \vbox{\hrule width2cm}}
\vskip -.3cm
\hbox{\vbox{\hrule width2.7cm} \hskip .2cm \vbox{\hrule width1.3cm}}
\vskip -.07cm
\hbox{~\hskip 2.55cm \vrule height 1.3cm  \hskip .32cm \vrule
height1.3cm }
\vskip -2.8cm
\hbox{ ~\hskip 1cm 2 \hskip 2cm 1}
\vskip 1cm
\hbox{~\hskip 1cm 3  \hskip 2.4cm 4}
}
\hskip 2cm
\vbox{\hbox{~\hskip 1.85cm \vrule height2cm  \hskip .32cm \vrule
height2cm }
\vskip -.68cm
\hbox{\vbox{\hrule width2cm} \hskip .2cm \vbox{\hrule width2cm}}
\vskip -.3cm
\hbox{\vbox{\hrule width1.3cm} \hskip .2cm \vbox{\hrule width2.7cm}}
\vskip -.07cm
\hbox{~\hskip 1.15cm \vrule height 1.3cm  \hskip .32cm \vrule
height1.3cm }
\vskip -2.8cm
\hbox{ ~\hskip 1cm 3 \hskip 2cm 2}
\vskip 1cm
\hbox{~\hskip .5cm 4  \hskip 2.5cm 1}
} }\vskip -1.6cm \hbox{~\hskip 5.2cm =}\vskip 2.0cm }}
\par
This shows that it is possible to construct $\bra{\Delta\Gamma_\mu}$ in
such a way that a consistent field redefinition is obtained by setting
$\bra{\Gamma^{(N)}_\mu}=0$ for $N\ge 3$.  In plain english, this means that
it is possible to relate the actions of open string field theories
formulated around neighboring conformal field theories via a field
redefinition which only includes a shift and a linear transformation.

One can compare our result with what one expects in the purely cubic open
string field theory [\cubic]. In this formalism, the string field theory
action is given by the purely cubic term ${1\over
3!}\bra{V^{(3)}}\Psi\rangle \ket{\Psi}\ket{\Psi}$. A given background,
characterized by a BRST operator $Q$ corresponds to a specific classical
solution $Q_L\ket{\I}$, where $\ket{\I}$ is the identity operator of the
star product, satisfying $\bra{V^{(3)}_{123}}\I\rangle_3=\bra{R_{12}}$.
In this case, a shift in the background amounts to a change $\Delta Q$ in
the BRST charge, and hence, a simple shift $\Delta Q_L\ket{\I}$ in the string
field, without any further linear field redefinition. This shift, however,
is singular\foot{In the sense that string field products involving more
than one such field are typically ill-defined.}, since the state $\ket{\I}$
is a singular state. We expect
that the field redefinition that we have found is related to the one
induced by this simple shift by a (singular) gauge transformation.

\subsection{Other Directions for Closed String Field Theory}
The above discussion naturally raises the question as to whether it is
possible to find a formulation of closed string field theory analogous to
the purely cubic open string field theory. We do not have a definite
answer to this question. We note, however, a surprising fact.
The purely cubic closed string field theory action
$$S = {1\over 3!} \bra{V^{(3)}}\Psi\rangle\ket{\Psi} \ket{\Psi}\, ,
\eqn\pcub$$
is actually invariant under a gauge transformation
$$ \delta_\Lambda\,\ket{\Psi}=
\bra{V^{\prime(3)}_{123}}{\s_{1e}\rangle}\ket{\Psi}_2\ket{\Lambda}_3 =
\Psi \circ \Lambda \, ,
\eqn\econcfour$$
where $\ket{\Psi}$ is a classical string field  (a
ghost number two state in $\H\,$), and $\ket{\Lambda}$ is the gauge
transformation parameter (a ghost number one state in $\H\,$).
The product used in the above transfomation is the asymmetric product
where $\Psi$ is inserted on the second puncture of $\V'_3$ and
$\Lambda$ is inserted in the asymmetric puncture. Off-shell gauge invariance
follows from Eqn.\miden . The above action was never thought of as a possible
candidate for a closed string field action because it seemed
to have no gauge invariance. Is the above gauge invariance an indication that
it could, in fact, be a consistent classical action?
This is not clear to us. While
on-shell, this gauge invariance is equivalent to a gauge invariance generated
by the standard $\V_3$, the following properties seem to suggest complications.
First, in this case we would not have the standard BV relation between the
action and the gauge transformations. Second, the algebra of this gauge
transformations may need regularization; when performing two successive
gauge transformations the gauge parameters would collide. Third, thanks
to Eqn.\xiden, any action of the form
$$S \, = \,
\sum_{N=3}^\infty a_N \bra{V^{(N)}}\Psi\rangle_1 \cdots \ket{\Psi}_N
\eqn\econcfive$$
where the $a_N$'s are arbitrary coefficients, is invariant under the above
gauge transformation.

Another possibility for writing new closed string field theory actions
could involve the use of two string fields. Given that we now have
a vertex $\V'_3$ that naturally distinguishes one puncture from the
other two, it is tempting to couple a new string field through this
puncture. This brings us to the admittedly speculative possibility that
the theory-space connection, or some string field encoding such data, could
actually represent a dynamical string variable. In such formulation,
elimination of this connection through its field equations would
leave a `background' connection, along with a fully nonlinear action
for the string field. This `background connection' could play the role
of fixing the Riemann surface geometry that defines the string interactions.
Solving for the connection would amount to fixing the way string theory
would cut moduli space.

\subsection{Extending the Proof of Background Independence.}
Our proof of background independence
made important use of the fact that the interactions of the standard
classical closed string field theory are overlaps. Associated to
such vertices, the canonical connection $\wh\Gamma_\mu$ played
a prominent role in the analysis. It is
clearly possible to construct other string
field theories based on non-overlap type vertices (the simplest
example being a theory with stubs).  Although our analysis has not
been done for such theories, it was shown in a recent paper[\hatazwiebach]
that these different string field theories are related to the
standard one by canonical field
redefinitions, and hence this field redefinition, combined with the field
redefinition we have found in our paper, makes the result of the present
paper valid even for string field theories with non overlap vertices.

It would be more instructive, however, to apply our methods directly to these
theories. The main difference in this case is that
$D_\mu(\wh\Gamma)\bra{V^{(N)}}$ is no longer zero, but can be expressed as
an integral of $\bra{\Omega^{(0)N+1}}\wh\O_\mu\rangle$ over a certain
region of $\wh\P_{N+1}$. This will give rise to a new term on the right
hand side of Eqn.\etwox\ (and \elalatwo)  which has the same structure as
the other terms, and hence these equations can be solved in the same way.
Although we have not given a direct proof of existence of the solutions of
these modified equations, it is guaranteed by the results of
ref.[\hatazwiebach]. On the other hand, it is possible that a
connection different from $\wh\Gamma_\mu$
could be more appropriate to deal with such non-overlap theories.

A related question is whether we could have carried out our analysis
of the standard closed string field theory using
a reference connection $\wt\Gamma_\mu$ different from $\wh\Gamma_\mu$.
This
is not a merely academic question. As argued at the end
of last section, a connection different
from $\wh\Gamma_\mu$ may be useful to construct a manifestly finite
field redefinition relating two distant theories.
Again the main difference is that
$D_\mu(\wt\Gamma)\bra{V^{(N)}}$ would not be zero.
If we choose $\wt\Gamma$ to
be the any of the connections $\Gamma_D$ defined in ref.[\rangasonodazw]
we can again express $D_\mu(\wt\Gamma)\bra{V^{(N)}}$ as an
integral of $\bra{\Omega^{(0)N+1}}\wh\O_\mu\rangle$ over a certain region
of $\wh\P_{N+1}$, and the effect is to modify the right hand side of
Eqn.\etwox\ (and \elalatwo) by the addition of this term. In this case
$D_\mu(\,\wt\Gamma\,)Q$ also has a
form different from $D_\mu(\,\wh\Gamma\,)Q$,
and the result is to modify the right hand side of Eqn.\esevenaaa\ by the
replacement of $\bra{V^{\prime(3)}}$ by a different three string vertex.
Again, all the equations have the same structure as the ones we have
analyzed, and can be solved in an identical manner. The existence of the
solution of these equations is guaranteed by our result, together with the
fact that changing the connection amounts to a linear redefinition of the
string fields.

\subsection{Quantum Background Independence ?}
All of our analysis has been done in the context of
classical master action. How about the quantum theory?
Given that in the BV formulation of closed string field theory
one has a well defined quantum master action, the question of
quantum background independence is likely to be well defined.
A quantum theory, however, is defined by a BV supermanifold
$(M ,\omega ,d\mu )$, where
$M$ is the supermanifold, $\omega$ the symplectic form, and $d\mu$
a consistent volume element (leading to a nilpotent $\Delta$ operator),
together
with the master action $S$. It was found in ref.[\hatazwiebach], that
the symplectic diffeomorphisms relating theories using different string
vertices do not preserve the volume element $d\mu$ and the action $S$
separately, but do preserve $d\mu e^{2S}$ . This indicates that
the symplectic diffeomorphism implementing the physical requirement
of background independence also cannot preserve both the volume element
and the master action separately. Suppose we are comparing string field
theories
formulated on $\H_x$ and $\H_y$. Moreover, we have  volume elements
$d\mu_x$ and $d\mu_y$, respectively. Let $L_y$ be an arbitrary
lagrangian submanifold
of $\H_y$, and let $d\lambda_y$ be the measure induced on that submanifold
by the measure $d\mu_y$ on $\H_y$.
The physical requirement of background independence is that the
symplectic diffeomorphism
from $\H_x$ to $\H_y$ should map $[d\lambda_y\,e^{S_y}]$
to $[d\lambda_x\,e^{S_x}]$ where $d\lambda_x$ is the measure induced from
$d\mu_x$, on the
lagrangian submanifold obtained as the (inverse) image of $L_y$ under the
diffeomorphism. This is all that is required
physically. Actually such a result would follow from the possibly
stronger condition that the diffeomorphism takes
$[d\mu_y\, e^{2S_y}]$ to  $[d\mu_x \,e^{2S_x}]$, and we believe it is likely
that this stronger condition holds. As a technical point, we note that the
presence of higher loop tadpoles in the master action will probably force
us to modify even the constant part
$\bra{\Gamma^{(1)}}$ of the field redefinition from
its tree level value.

\ack
A. Sen acknowledges the hospitality of the Center for Theoretical Physics
at MIT.

\endpage
\singlespace
\refout
\end